\documentclass[12pt]{article}

\usepackage{amssymb}
\usepackage{amsmath}
\usepackage{amsfonts}
\usepackage{eurosym}
\usepackage{dsfont}
\usepackage{mathrsfs}
\usepackage{bbm}
\usepackage{bm}
\usepackage{fix-cm}
\usepackage{accents}

\usepackage{geometry}
\usepackage{setspace}
\usepackage{color}
\usepackage{enumitem}
\usepackage{soul}

\usepackage{float}
\usepackage{graphicx}
\usepackage{caption}
\usepackage{subcaption}
\usepackage{footmisc}
\usepackage{tablefootnote}

\usepackage{appendix}

\usepackage{tikz}
\usepackage{tikz-3dplot}
\usepackage{tikz-network}
\usetikzlibrary{backgrounds,automata,arrows,positioning,patterns,calc,shapes}
\usetikzlibrary{decorations.pathreplacing, decorations.markings}
\usetikzlibrary{positioning}

\usepackage[colorlinks=true, urlcolor=gray, linkcolor=blue,citecolor=purple]{hyperref}
\usepackage{natbib}

\usepackage{datetime}
\newdateformat{monthyeardate}{\monthname[\THEMONTH], \THEYEAR}

\newtheorem{theorem}{Theorem}
\newtheorem{proposition}{Proposition}
\newtheorem{lemma}{Lemma}

\newtheorem{assumption}{Assumption}
\newtheorem{definition}{Definition}
\newtheorem{remark}{Remark}
\newtheorem{example}{Example}

\allowdisplaybreaks

\DeclareMathOperator*{\argmax}{arg\,max}

\newcommand{\pfam}{\bm{\mathbbm{p}}}
\newcommand{\myd}{\mathcal{D}}
\newcommand{\myb}{\mathcal{B}}
\newcommand{\mye}{\mathcal{E}}
\newcommand{\mypsi}{\boldsymbol{\psi}}
\newcommand{\bH}{\mathbf{H}}
\newcommand{\ba}{\mathbf{a}}
\newcommand{\bp}{\mathbf{p}}
\newcommand{\bc}{\mathbf{c}}
\newcommand{\btheta}{\boldsymbol{\theta}}
\newcommand{\ubar}[1]{\underaccent{\bar}{#1}}

\begin{document}

\title{{\bf 
Price Regulation and Network Spillovers%
}\thanks{We are grateful to Mark Armstrong, Francis Bloch, Ying-Ju Chen, Sanjeev Goyal, Andrew Rhodes, 	Eduard Talamàs,  Tat-How Teh, Yiqing Xing, Yves Zenou,  Jidong Zhou, and various seminar and conference audiences for helpful comments.}
    }
\author{
	Chengqing Li\thanks{School of Economics and Management, Tsinghua University. lcq22@mails.tsinghua.edu.cn} \quad 
	Junjie Zhou\thanks{School of Economics and Management, Tsinghua University. zhoujj03001@gmail.com}}
\date{January 2026
}
\maketitle

\begin{abstract}

We study price regulation for a monopolist operating in networked markets with demand spillovers. Achieving efficiency requires price reductions proportional to consumers’ Katz–Bonacich centralities, which generally cannot be implemented by commonly used price regulations. Moreover, these regulations become asymptotically welfare neutral as spillovers grow. Nevertheless, some price regulations may still benefit consumers. In particular, average-price regulation robustly increases consumer surplus. By contrast, banning price discrimination increases consumer surplus only when more central consumers have higher intrinsic willingness to pay. 




\vspace{1ex}



\end{abstract}

\newpage

\section{Introduction}\label{sec.introduction}

A key feature of the digital economy is the prevalence of demand spillovers across different markets. While buyers are often segmented by geography or application, information diffusion and technological interoperability create strong demand interdependencies. For instance, in the consumer market, social media can transform a local trend, such as the "Labubu" series by Pop Mart, into a global phenomenon, where sales in one region stimulate demand in others. Similarly, in technology markets, a standard-essential patent may be licensed to various downstream producers and its usage in one device often enhances its value for others due to interoperability. In this context, a monopolist does not maximize profits in isolated markets but manages a network of interconnected demands where sales in one market generate positive externalities in others. Hence, the monopolist seeks to engage in third-degree price discrimination to extract maximum profit.\footnote{See, for example, \cite{candoganOptimalPricingNetworks2012}, \cite{blochPricingSocialNetworks2013}.}

In practice, however, the implementation of such price discrimination is frequently constrained by regulatory policies. For example, the EU Geo-blocking Regulation prohibits unjustified blocking of access to online interfaces, which effectively generates constraints on price discrimination due to potential arbitrage.\footnote{The Federal Network Agency of Germany clarifies that while traders may vary prices across regional websites, ``French customer must be able to order the goods at the same prices and under the same conditions as German customers'' if they access the German website. Consequently, if the price difference between the German and French websites becomes too large, the regulation enables consumer arbitrage, thus constraining the firm's pricing spread.} In the context of intellectual property, Standard Development Organizations often require members to license patents on Fair, Reasonable, and Non-Discriminatory (FRAND) terms, which place ceilings and fairness requirements on fees.\footnote{See, for example, \cite{contreras2017non} for detailed cases where violations of FRAND terms involve charging excessive or discriminatory royalties.} These regulations impose general constraints on the monopolist's pricing strategy, yet their welfare implications in the presence of demand spillovers remain understudied.

This paper applies a network approach to study price regulation in markets with demand spillovers. We characterize the monopolist's pricing strategy under regulations and examine whether different regulatory instruments can achieve Pareto-efficient outcomes. We also investigate whether price regulation necessarily benefits consumers despite always reducing the monopolist’s profit by imposing additional constraints.

Specifically, we consider a monopolist selling a product in $n$ markets, where demand spillovers across markets are represented by a network. The firm sets a price for each market subject to a feasible set of prices, which we refer to as a price regulation. We incorporate several commonly implemented forms of price regulation, including constraints on price differences, price floors and ceilings, and average-price regulation. Under price regulation, we show that the firm chooses the price vector within the feasible set that minimizes the “distance” from the unrestricted price, where distance is measured by a novel norm that reflects both the network structure and the intensity of demand spillovers.

Having characterized the firm’s pricing behavior, we turn to the welfare implications of price regulations. Our first criterion is efficiency, namely whether a given price regulation can attain the highest achievable consumer surplus for a given level of the firm’s profit loss. We first characterize the family of prices that induce efficient outcomes and show that it requires price reductions proportional to consumers’ Katz–Bonacich centralities. We then demonstrate that a price regulation is efficient if and only if it contains an efficient price and all other feasible prices under the regulation have a lower weighted average than this efficient level, where the weight vector is proportional to the Katz–Bonacich centrality weighted by the price reductions. However, under the three commonly implemented forms of price regulation considered in this paper, the equilibrium outcome is generically Pareto inefficient. Although a regulator may in principle select parameters that deliver efficiency, doing so requires complete information about consumers’ intrinsic values, the firm’s production costs, and the network structure, which is impractical in most real-world settings.

We next study the case of strong demand spillovers. Using the spectral decomposition of the network, we show that the firm’s objective depends only on the weighted average deviation of prices from the unrestricted benchmark, where the weights are proportional to the eigenvector centrality. This provides the firm with greater flexibility to avoid profit losses under regulation. As a result, commonly implemented price regulations tend to become welfare-neutral in the limit, in the sense that the relative changes in both profit and consumer surplus induced by regulation asymptotically vanish. However, the regulator can still select parameters that achieve efficient outcomes. Importantly, in the limit, achieving efficiency requires only knowledge of the network’s eigenvector centrality and the unrestricted price, which is substantially more practical than in the case of fixed spillovers.

Although commonly implemented price regulations generally fail to achieve efficiency, they may still benefit consumers. Even when relative welfare effects vanish, these regulations can generate large absolute increases in consumer surplus. We therefore examine whether price regulations benefit consumers under three important regulatory instruments. 
The first instrument imposes a fixed price that maximizes a weighted sum of firm profit and consumer surplus. We show that the resulting price is efficient. 
The second instrument is average-price regulation. We show that it always increases consumer surplus, but the magnitude of the gain depends on the average-price weight. The robustness of this regulation, measured by the ratio between the best- and worst-case surplus gains, is determined by the spectral properties of the network. 
Finally, we study the welfare effects of banning price discrimination. We show that consumer surplus increases when more central consumers have higher intrinsic values. In this case, price reductions for more central consumers stimulate demand through spillovers, and this effect dominates the price increases faced by less central markets. When the network is sufficiently symmetric, particularly when it consists of only two types of nodes such as core–periphery or complete bipartite networks, the surplus effect reduces to a comparison of average intrinsic values across more central and less central groups.

\vspace{12pt}

\noindent\textbf{Related literature}\hspace{1ex}
This paper complements the literature of optimal pricing with network externalities. Foundational works by \cite{candoganOptimalPricingNetworks2012} and \cite{blochPricingSocialNetworks2013} analyze the pricing strategies of a profit-maximizing monopoly in networked markets. Building on this framework, \citeauthor{chenCompetitivePricingStrategies2018} (\citeyear{chenCompetitivePricingStrategies2018}, \citeyear{chenImpactNetworkTopology2022}) extend the analysis to oligopolistic settings. Moving beyond the complete-information environments considered in these studies, \cite{fainmesserPricingNetworkEffects2016} analyze monopoly pricing under incomplete information, \cite{zhangOptimalNonlinearPricing2020} study nonlinear pricing schemes and \cite{shiScreeningNetworkInformation2025} investigate screening mechanisms in network settings. Most of these works focus on profit-maximizing prices or mechanisms that allow firms substantial freedom to price discriminate, with uniform pricing treated as a special case in \cite{candoganOptimalPricingNetworks2012} and \cite{fainmesserPricingNetworkEffects2016}. In contrast, our paper studies pricing behavior under more general forms of price regulation imposed as constraints on firms and analyzes their welfare implications.

This paper also contributes to the literature on regulating third-degree price discrimination by monopolists. Classic contributions examine whether banning third-degree price discrimination improves welfare (see, e.g., \citealp{robinson1969economics, schmalenseeOutputWelfareImplications1981, varianPriceDiscriminationSocial1985}). More recent work derives sharper sufficient conditions, based on properties of demand functions, for determining the welfare effects of price discrimination (\citealp{aguirreMonopolyPriceDiscrimination2010, cowan2016welfare}). These studies focus on independent markets. We extend this literature by incorporating demand spillovers across markets through a network structure, which alters the welfare effects of banning price discrimination.

Beyond banning price discrimination, several studies consider alternative forms of price regulation. \cite{armstrongWelfareEffectsPrice1991} analyze average-price constraints, while \cite{armstrongMultiproductPricingMade2018} study a Ramsey problem in multi-product settings. More recent work explores fairness-based regulations in pricing. \cite{cohenPriceDiscriminationFairness2022} examine multiple fairness criteria, including price fairness, demand fairness, and surplus fairness. \cite{yangFairnessRegulationPrices2024} study fairness constraints that limit price differences between two markets with demand spillovers. Our paper generalizes these works in two directions by considering a broader class of price regulations and by studying a network of $n$ interconnected markets.

Finally, this paper is related to the literature on network intervention. The seminal work of \cite{ballesterWhosWhoNetworks2006} studies structural intervention through the “key player” problem, which targets the network structure itself. Other work examines characteristic interventions that act on players’ attributes in network games, including \cite{demangeOptimalTargetingStrategies2017a} and \cite{galeottiTargetingInterventionsNetworks2020}. More recent studies consider tax/subsidy interventions applied to firms operating in networked markets (see, e.g., \citealp{galeotti2024robust, kohPricesSymmetries2025}). In contrast, our paper focuses on price regulation, which directly constrains the firm’s feasible action set rather than altering the network or subsidizing behavior.

The remainder of the paper is organized as follows. Section~\ref{sec.model} introduces the model. Section~\ref{sec.mainresult} first characterizes the monopolist’s optimal pricing strategy under price regulation and then studies the efficiency of price regulations in both fixed and large demand spillover cases. Section~\ref{sec.application} examines the implications of several important price regulation schemes for consumer surplus. Finally, Section~\ref{sec.conclusion} concludes.


\section{Model}\label{sec.model}

A monopolistic firm sells products in $n$ interconnected markets, $\mathcal{N} =\{1, 2,...,n\}$. In each market $i\in\mathcal{N}$, there is a representative consumer who can consume the product in any quantities $x_i$.
The utility function takes the following  form\footnote{See e.g., \cite{candoganOptimalPricingNetworks2012} and \cite{blochPricingSocialNetworks2013}}:
\begin{equation*}
	u_i(x_i,\mathbf{x}_{-i};\bp) = a_{i} x_{i} -\frac{1}{2}(x_{i})^2 + \delta \sum_{j=1}^{n}g_{ij}x_{i}x_{j} - p_{i} x_{i}.
\end{equation*}

The first two terms represent the intrinsic value from consuming the product itself, with decreasing marginal returns, while the last term captures the cost of consumption. In the third term, $\delta \geq 0$ measures the intensity of spillovers and we assume $g_{ij} \geq 0$. If $g_{ij} > 0$, there are positive spillovers from market $j$ to market $i$, otherwise there are no direct spillovers between the two markets. These spillovers can be represented by a $n\times n$  network adjacency matrix $\mathbf{G} = (g_{ij})$. We assume $g_{ii} = 0$, meaning there are no spillovers within a single market, and $g_{ij} = g_{ji}$, indicating that spillovers between two markets are symmetric. 

The marginal cost of producing product in market $i$ is denoted by $c_i$. We assume $a_i > c_i$ for all $i$. The firm chooses a price vector $\bp = (p_1, \ldots, p_n)^T$ to maximize its profit. However, the firm is subject to certain price regulations or restrictions and can only select prices from a feasible set $K\subseteq \mathbb{R}^n$.

\begin{assumption}\label{assump.k}
    The set of feasible prices $K$ is non-empty, closed, and convex.
\end{assumption}
The convexity of $K$ can be interpreted as allowing the firm to randomize among different prices. Moreover, this assumption is satisfied by three typical types of feasible sets which are illustrated in Figure \ref{fig.regulation}:
\begin{enumerate}
    \item Regulation on price difference:
    \begin{equation*}
        K = \myd_{\boldsymbol{\Delta}} := \left\{\bp \in \mathbb{R}^n: p_i - p_j \leq \Delta_{ij}, \forall i,j \in \mathcal{N} \right\}
    \end{equation*}
    where $\boldsymbol{\Delta} = (\Delta_{ij})$ is a  square matrix. These restrictions limit price differences across markets\footnote{In \cite{yangFairnessRegulationPrices2024}, such regulations stem from fairness concerns.}. Examples include:
    \begin{enumerate}
        \renewcommand{\labelenumi}{(\alph{enumi})}
        \item Uniform pricing: $\Delta_{ij} \equiv 0$ for all $i,j \in \mathcal{N}$ (\citealp{candoganOptimalPricingNetworks2012, chenCompetitivePricingStrategies2018}).
        
        \item Coarse pricing: given a partition of $\mathcal{N}$, regulators mandate uniform pricing within each block of the partition. This means $\Delta_{ij} = 0$ for all $i$, $j$ within the same block, generalizing uniform pricing.

        \item Price Parity Clause: some platforms or agents may require that the price in market~$i$ be no higher than in any other market, that is, $p_i-p_j \leq 0$ for any $j \neq i$.
    \end{enumerate}

    \item Regulations on price floors and ceilings: 
    \begin{equation*}
        K = \myb(\underline{\bp},\overline{\bp}) := \{\underline{\bp} \leq \bp \leq \overline{\bp}:  \mbox{where }   -\infty \leq \underline{p}_i\leq  \overline{p}_i \leq  +\infty, \forall i\}.
    \end{equation*}
    This type includes the following special cases:
    
    \begin{enumerate}
        \renewcommand{\labelenumi}{(\alph{enumi})}
        \item No price constraints: $K = \mathbb{R}^{n}$ (\citealp{candoganOptimalPricingNetworks2012,blochPricingSocialNetworks2013}). 
        \item Zero lower bound: $\underline{p}_i = 0$, $\forall i \in \mathcal{N}$ (\citealp{biscegliaFairGatekeepingDigital2024}).
        
        \item Fixed prices in some markets: given a subset of markets $\mathcal{S} \subseteq \mathcal{N}$, the regulator mandates the firm to set a fixed price vector $\tilde{\bp}_{\mathcal{S}}$ in these markets. Therefore, $\underline{p}_i = \overline{p}_i = \tilde{p}_i$ for $i \in \mathcal{S}$ and $\underline{p}_i = -\infty$, $\overline{p}_i = \infty$ for $i \in \mathcal{N} \backslash \mathcal{S}$.
        
    \end{enumerate}

    \item Regulation on average prices:
    \begin{equation*}
        K = \mye_{\btheta,M} := \{ \bp \in \mathbb{R}^n: \langle \btheta,\bp \rangle \leq M\}
    \end{equation*}
    where $M \in \mathbb{R}$ and $\btheta \in \Theta:= \{\btheta \in \mathbb{R}^n_+: \sum_{i=1}^n \theta_i=1\}$. This regulation ensures that the weighted average price does not exceed $M$ (\citealp{armstrongWelfareEffectsPrice1991}).
\end{enumerate}

\begin{figure}[ht]
    \centering

    \begin{subfigure}{0.3\textwidth}
    	\centering
        \begin{tikzpicture}[scale=0.9]
			
			\coordinate (D1) at ($(1,0.3)$);
			\coordinate (D2) at ($(D1)+1.8*(1,1)$);
			\coordinate (D3) at ($(D2)+0.8*(-1,1)$);
			\coordinate (D4) at ($(D1)+0.8*(-1,1)$);
			
			\fill[blue!20] (D1)--(D2)--(D3)--(D4)--cycle;
			

			\draw[blue,thick]
                (D1) -- (D2)
                node[midway, below, sloped, blue, font=\fontsize{10pt}{6pt}\selectfont]
                {$p_1 - p_2 = \Delta_{12}$};
			\draw[blue,thick] (D3)--(D4)
                node[midway, above, sloped, blue, font=\fontsize{10pt}{6pt}\selectfont]
                {$p_2 - p_1 = \Delta_{21}$};
			
		\end{tikzpicture}
    	\caption{$D_{\Delta}$}
        \label{fig.d}
    \end{subfigure}
    \hfill
    \begin{subfigure}{0.3\textwidth}
    	\centering
        \begin{tikzpicture}[scale=0.9]
            \coordinate (B1) at (0.7,0.9);
            \coordinate (B2) at (2.8,2.8);
            
            
            \fill[blue!20] (B1) rectangle (B2);
            \draw[blue,thick] (B1) rectangle (B2);
            

            \node at (B1) [anchor = north east, blue]{$\underline{p}$};
            \node at (B2) [anchor = south west, blue]{$\overline{p}$};
        \end{tikzpicture}
    	\caption{$\myb(\underline{\bp},\overline{\bp})$}
        \label{fig.b}
    \end{subfigure}
    \hfill
    \begin{subfigure}{0.3\textwidth}
    	\centering
        \begin{tikzpicture}[scale=0.9]
            
            \coordinate (v) at (1,-1.1);
            \coordinate (h) at (1.1,1);
            
            \coordinate (A1) at (0.9,3.2);
            \coordinate (A2) at ($(A1)+2.2*(v)$);
            \coordinate (A3) at ($(A2)-0.7*(h)$);
            \coordinate (A4) at ($(A1)-0.7*(h)$);
            
            \coordinate (A) at ($0.5*(A1)+0.5*(A2)$);
            \coordinate (AE) at ($(A)+0.5*(h)$);
            
            \fill[blue!20] (A1)--(A2)--(A3)--(A4)--cycle;
            
            
            \draw [-latex,blue] (A) -- (AE) ;
            \draw[thick,blue] (A1)--(A2)
                node[pos=1, above, sloped, blue, font=\fontsize{10pt}{6pt}\selectfont]
                {$\langle \btheta,\bp \rangle \leq M$};
            
            \node at (AE) [right, blue, font=\fontsize{10pt}{6pt}\selectfont]{$\btheta$};
        \end{tikzpicture}
    	\caption{$\mye_{\btheta,M}$}
        \label{fig.e}
    \end{subfigure}

    \caption{Three types of price regulations}
    \label{fig.regulation}
\end{figure}

The timing of the game is as follows. The multi-market monopolist first chooses a price vector $\bp$ from the feasible set $K$. After observing $\bp$, all consumers independently and simultaneously choose their consumption bundles. We analyze the subgame-perfect Nash equilibrium of this game. For any price vector $\bp$, let $\mathbf{x}(\bp)$ denote the equilibrium consumption profile. Then the firm selects its optimal price $\bp^*$ from $K$ by solving the following profit maximization problem:
\begin{equation*}
    \max\limits_{\bp \in K} \Pi(\bp) := (\bp-\bc)^T \mathbf{x}(\bp). 
\end{equation*}
The aggregate consumer surplus at price $\bp$ is given by:
\begin{equation*}
	V(\bp) := \sum_{i \in \mathcal{N}} u_i(\mathbf{x}(\bp);\bp ). 
\end{equation*}


\section{The (In)Efficiency of Price Regulations}\label{sec.mainresult}

In this section, we first characterize firm's optimal pricing strategy under price regulations. Section~\ref{sec.fix} then studies their efficiency properties under a fixed spillover $\delta$. Section~\ref{sec.large} provides a sharper characterization of equilibrium outcomes and their efficiency when spillovers become sufficiently large. Sections~\ref{sec.discussion} and \ref{sec.relationship} compare these two scenarios and provide additional discussion of the results.

Denote the largest eigenvalue of $\mathbf{G}$ as $\lambda_1(\mathbf{G})$. To guarantee the existence and uniqueness of equilibrium in the consumption stage, we have the following assumption.
\begin{assumption}[\citealp{ballesterWhosWhoNetworks2006}]\label{assump.spectrum}
    $0<\delta<\frac{1}{ \lambda_1(\mathbf{G})}$.
\end{assumption}

Given price $\bp$, the first-order condition for consumer $i$ gives
\begin{equation*}
    a_i - x_i + \delta \sum_{j=1}^{n}g_{ij}x_{j} - p_i = 0, \forall i\in\mathcal{N}\ \iff \
    \mathbf{x}(\bp) = \underbrace{[\mathbf{I}_n - \delta \mathbf{G}]^{-1}}_{\bH}(\ba-\bp).
\end{equation*}
Under Assumption~\ref{assump.spectrum}, $\bH$ is positive definite. We can define an inner product $\langle  \mathbf{z}, \mathbf{\hat{z}}  \rangle_{\mathbf{H}} := \mathbf{z}^T\mathbf{H}\mathbf{\hat{z}}$ for any vectors $\mathbf{z}, \hat{\mathbf{z}} \in \mathbb{R}^n$ and the induced norm $\lVert \cdot \rVert_{\mathbf{H}}$. Then the firm's profit equals
\begin{equation*}
    (\bp-\bc)^T \mathbf{x}(\bp)
    =(\bp-\bc)^T \mathbf{H}(\ba-\bp) =
    \Pi \left(\frac{\ba+\bc}{2} \right) -  \left\lVert\bp-\frac{\ba+\bc}{2}  \right\rVert^2_{\mathbf{H}}.
\end{equation*} 
Hence, the firm's profit maximization problem can be expressed as:
\begin{equation*}
    \max_{\bp \in K} \Pi(\bp) \iff \min_{\bp \in K} \left\lVert\bp-\frac{\ba+\bc}{2}  \right\rVert^2_{\mathbf{H}}.
\end{equation*}

\begin{lemma}\label{lemma.equi}
    If Assumptions~\ref{assump.k} and \ref{assump.spectrum} hold, the firm's optimal price $\bp^*$ is the unique projection of $\frac{\ba+\bc}{2}$ onto $K$ under the $\mathbf{H}$-norm.
\end{lemma}

In the benchmark cases where there are no price regulations ($K = \mathbb{R}^n$), the optimal price is exactly $\frac{\ba+\bc}{2}$ (see \citealp{candoganOptimalPricingNetworks2012,blochPricingSocialNetworks2013}), which we denote as $\bp^{ur}$. Hence, the firm's profit loss, $||\bp-\bp^{ur}||^2_{\mathbf{H}}$, is measured by the ``distance'' between the regulated price vector and the unrestricted benchmark $\bp^{ur}$. The associated indifference curves are ellipsoids in the price space.\footnote{When markets are independent ($\delta=0$), it becomes the standard $\ell^2$-norm, and the indifference curve becomes a circle.} To minimize profit loss, the firm chooses the price that makes the indifference curve just tangent to the boundary of the feasible set, as illustrated in Figure~\ref{fig.proj}.

\begin{figure}[H]
	\centering
    \begin{tikzpicture}[scale = 0.8]
        \pgfmathsetmacro{\sqrtthree}{sqrt(3)}
        
        \coordinate (pur) at (0,0);
        
        \coordinate (v1) at (1,\sqrtthree);
        
        \coordinate (v2) at (\sqrtthree,-1);
        
        \coordinate (w1) at ($(pur)+0.8*(v1)$);
        
        \coordinate (w2) at ($(pur)+1.3*(v2)$);		
        
        \coordinate (p) at ($(pur)-\sqrtthree*0.5*0.5*(v1)-0.5*(v2)$);


        \draw[ rotate=-30] (pur) ellipse (2cm and 1cm);

        \draw[fill=black] (pur) circle (2pt);
        
        
        \coordinate (a) at ($(p)+0.5*\sqrtthree*(v2)-0.25*(v1)$);
        \coordinate (b) at ($(p)-0.5*\sqrtthree*(v2)+0.25*(v1)$);
        \coordinate (c) at ($(p)-0.5*\sqrtthree*(v2)+0.25*(v1)-0.15*(v2)-0.3*\sqrtthree*(v1)$);
        \coordinate (d) at ($(p)+0.5*\sqrtthree*(v2)-0.25*(v1)-0.15*(v2)-0.3*\sqrtthree*(v1)$);
        \fill[blue!20] (a)--(b)--(c)--(d)--cycle; 
        
        \draw[blue,fill=blue] (p) circle (2pt);

        \draw[blue,thick] ($(p)+0.5*\sqrtthree*(v2)-0.25*(v1)$)--($(p)-0.5*\sqrtthree*(v2)+0.25*(v1)$);

        \node at (pur) [right] {$\bp^{ur}$};
        \node at (p) [left,blue] {$\bp^*$};
        \node at ($(p)-(0,0.8)$) [blue] {$K$};
        
    \end{tikzpicture}
	\caption{Firm's indifference curves and optimal pricing}
	\label{fig.proj}
\end{figure}

\begin{assumption}\label{assump.infeasible}
    The unrestricted price $\bp^{ur}=\frac{\ba+\bc}{2}$ is not in the feasible set $K$.
\end{assumption}

Given our focus on the welfare implications of price regulations, we impose this assumption to rule out uninteresting cases where the constraint has no effect. In practice, under average price regulation, this requires the regulator to set a sufficiently tight cap so that the average monopoly price exceeds the regulated level, i.e., $\boldsymbol{\theta}^T \bp^{ur} > M$.\footnote{For price difference regulation, this assumption implies a mandatory reduction in price dispersion relative to the unregulated benchmark. Similarly, for price floor and ceiling regulations, practical implementations typically involve setting caps below the monopoly price. All such practical cases satisfy this assumption.} Unless stated otherwise, Assumptions~\ref{assump.k}–\ref{assump.infeasible} hold throughout the paper.

\begin{example}\label{example.closedform}
    For some specific price regulations, we can derive a closed form solution for the firm's optimal price $\bp^*$.
    \begin{enumerate}
        \renewcommand{\labelenumi}{(\roman{enumi})}
        \item When $K$ is a singleton, the regulator can mandate the firm to choose any specific price.

        \item When $K=\mye_{\btheta,M}$, that is the average-price regulation.\footnote{The detailed derivation is provided in Appendix~\ref{appsec.proofpropaverage}.} Under Assumption \ref{assump.infeasible}, firm's optimal price is
        \begin{equation}\label{eq.paverage}
            \bp^*= \bp^{ur} - \frac{\btheta^T\bp^{ur}-M  }{\btheta^T\bH^{-1} \btheta} \bH^{-1} \btheta.
        \end{equation}
        
        \item When $K=\{s(1,\cdots, 1), s\in\mathbb{R}\}$, the regulator prohibits price discrimination, which is considered in \cite{candoganOptimalPricingNetworks2012}. The optimal uniform price is
        \begin{equation*}
            \bp^* = \frac{\mathbf{1}^T\mathbf{H}\bp^{ur}}{\mathbf{1}^T\mathbf{H}\mathbf{1}} \mathbf{1}.
        \end{equation*}
    \end{enumerate}

\end{example}


\subsection{Fixed spillovers}\label{sec.fix}
We now examine whether the commonly implemented price regulations lead to efficient outcomes given spillover intensity $\delta$. For any price vector $\bp$, the consumer surplus and firm’s profit can be written as:
\begin{align}
    &V(\bp) = \frac{1}{2} \mathbf{x}(\bp)^T\mathbf{x}(\bp) 
    = \frac{1}{2} (\ba-\bp)^T \bH^2 (\ba-\bp), \label{eq.cs} \\
    &\Pi(\bp) = \Pi(\bp^{ur}) - (\bp-\bp^{ur})^T \bH (\bp-\bp^{ur}), \label{eq.pi}
\end{align}
where the first identity follows from \cite{ballesterWhosWhoNetworks2006}, and the second is derived before. Instead of the absolute levels, we evaluate outcomes relative to the benchmark. Specifically, we define the following ratios\footnote{The results for fixed $\delta$ continues to hold when consumer surplus and profit are evaluated in absolute terms, $V(\bp)$ and $\Pi(\bp)$, rather than the ratios. However, both quantities depend on the intensity of spillovers $\delta$ and diverge as $\delta$ approaches the upper bound $1/\lambda_1$. To avoid this technical complication and present the limit results in a more coherent way, we evaluate consumer surplus and profit relative to their corresponding benchmark values, rather than in absolute values.}:
\begin{equation}\label{eq.ratio}
    R_{V}(\bp) := \frac{V(\bp)}{V(\bp^{ur})}, \quad
    R_{\Pi}(\bp) := \frac{\Pi(\bp)}{\Pi(\bp^{ur})}. 
\end{equation}

A price vector $\bp$ is said to be \emph{Pareto efficient} (or simply \emph{efficient}), if there is no $\bp' \in \mathbb{R}^n$ such that $R_V(\bp') \geq R_V(\bp)$ and $R_\Pi(\bp') \geq R_\Pi(\bp)$, with at least one inequality strict.\footnote{When prices in all markets are sufficiently low, for example, when $\bp < \bc$, the firm's profit may become negative. Although such a price vector may still be Pareto efficient, a firm earning negative profit would exit. Hence, throughout the paper we restrict attention to prices that yield nonnegative profit.} A price regulation~$K$ is called \emph{Pareto efficient} if its induced equilibrium price $\bp^*$ is Pareto efficient. Otherwise, the regulation is \emph{Pareto inefficient}.

As a first step, we characterize the set of efficient prices by considering the following problem
\begin{equation}\label{eq.intervention}
    \begin{gathered}
        \max\limits_{\bp\in \mathbb{R}^n} R_V(\bp) \\
        s.t. \ R_{\Pi}(\bp) \geq \tau.
    \end{gathered} 
\end{equation}
Here, $\tau$ can be interpreted as the firm's fixed operating cost. The regulator's objective is to maximize consumer surplus subject to the constraint that the firm generates sufficient profit to cover this cost.\footnote{This formulation differs from the classical Ramsey problem, where the objective is to maximize total surplus subject to a break-even constraint (\citealp{ramsey1927contribution,boiteux1956gestion}). We analyze a related problem, which maximizes a weighted sum of consumer surplus and profit, in Section~\ref{sec.fixp}.}

\begin{lemma}\label{lemma.pfamily}
	A price is Pareto efficient if and only if it is a solution to problem (\ref{eq.intervention}) for some $\tau \in [0,1]$. Moreover, the solution can be characterized as 
	\begin{equation}\label{eq.pfamily}
        \pfam(\eta) = 
        \underbrace{\frac{\ba+\bc}{2}}_{\bp^{ur}} - 
        \frac{\eta}{2-\eta} \underbrace{\left[\mathbf{I}_n-\frac{2\delta}{2-\eta}\mathbf{G}\right]^{-1}\frac{\ba-\bc}{2}}_{\mathbf{b}(\mathbf{G},\frac{2\delta}{2-\eta},\frac{\ba-\bc}{2})}, 
    \end{equation}
	where $\eta \geq 0$ is uniquely identified by $R_{\Pi}(\pfam(\eta)) = \tau$ and increases as $\tau$ decreases.
\end{lemma}

This lemma provides an alternative interpretation of efficient prices. As the form of problem~(\ref{eq.intervention}) indicates, an efficient price induces the highest attainable consumer surplus subject to a lower bound on the firm's profit.\footnote{Since the firm achieves its maximal profit at $\bp^{ur}$, the profit lower bound $\tau$ cannot exceed~1. Because we focus on prices yielding nonnegative profit, we also require $\tau \ge 0$.} Hence, there is no other price that can increase both consumer surplus and profit, with at least one strictly, because any such price would remain feasible and would contradict the optimality of the solution. This is precisely the notion of Pareto efficiency.

The efficient price can be expressed as the unrestricted price, $\bp^{ur}$, minus a discount proportional to the Katz-Bonacich centrality. The intuition is straightforward. Increasing consumer surplus requires lowering prices. At the same time, due to the presence of cross-market spillovers, larger price reductions in markets with higher centralities not only expand consumption within those markets but also amplify consumption in other markets via spillover effects. Consequently, the magnitude of price cuts is proportional to the Katz-Bonacich centrality. 

Furthermore, achieving the highest consumer surplus requires pushing the firm’s profit to its lower bound, so the constraint binds and thus determines the parameter $\eta$. As the profit lower bound $\tau$ decreases, larger price reductions become feasible, and $\eta$ correspondingly increases. When $\tau = 1$, we have $\eta = 0$ and the solution is the unrestricted price. When $\tau = 0$, we denote the highest value of $\eta$ by $\bar{\eta}$.

Let $\mathcal{P}:=\{\pfam(\eta): \eta \in [0,\bar{\eta}]\}$ denote the set of all efficient prices. The following lemma provides a necessary and sufficient condition for a price regulation to be Pareto efficient. 

\begin{lemma}\label{lemma.pareto}
    A price regulation $K$ is Pareto efficient if and only if there exists $\pfam \in \mathcal{P} \cap K$ such that $\langle \nabla R_{\Pi}(\pfam), \mathbf{z}-\pfam \rangle \leq 0$ for any $\mathbf{z} \in K$.
\end{lemma}

Based on the characterization of efficient prices, this lemma is straightforward. For a price regulation $K$ to be efficient, two conditions must be satisfied. First, at least one efficient price $\pfam$ is feasible in $K$. Second, this price must also be optimal for the firm. Since $R_{\Pi}$ is concave in $\bp$ and $K$ is convex, the optimality condition is equivalent to the variational inequality stated in Lemma~\ref{lemma.pareto}. This condition can be interpreted as requiring that the weighted average of any feasible price in $K$ does not exceed the efficient level, where the weights are given by $\nabla R_{\Pi}(\pfam)$. Moreover, note that 
\begin{equation*}
    \nabla R_{\Pi}(\pfam) \propto \bH (\bp^{ur} -\pfam),
\end{equation*}
which corresponds to K-B centrality weighted by price reductions. Geometrically, the second condition in Lemma~\ref{lemma.pareto} requires the entire feasible set to lie within the gray region illustrated in Figure~\ref{fig.pareto}.

\begin{figure}[ht]
    \centering

    \begin{tikzpicture}[scale=0.6]

        \def\ac{3}
        \def\a{4}
    
        \def\smallMajor{2.0}
        \def\smallMinor{1.22}
        \def\bigMajor{3.4}
        \def\bigMinor{1.84}
    
    
        \draw (\ac,0) ellipse[x radius=\smallMajor, y radius=\smallMinor, rotate=-26.565];
        \draw[magenta!30!red] (\a,0)  ellipse[x radius=\bigMajor, y radius=\bigMinor, rotate=-26.565];

        \coordinate (P) at (1.5, -0.39);
    
        \coordinate (d) at (0.756,0.654);
    
        \coordinate (t) at (0.654,-0.756);
    
        \fill[gray!20] ($(P)-3*(t)$) -- ($(P)+3*(t)$)--($(P)+3*(t)-1.5*(d)$)--($(P)-3*(t)-1.5*(d)$)--cycle;

        \coordinate (C1) at ($(P)  + 0.05*(t)$);
        \coordinate (C2) at ($(P)  - 0.05*(t)$);
        \coordinate (B1) at ($(P) - 0.2*(d) + 0.5*(t)$);
        \coordinate (B2) at ($(P) - 0.8*(d) + 1.1*(t)$);
        \coordinate (B3) at ($(P) - 1.5*(d) + 1.5*(t)$);
        \coordinate (B4) at ($(P) - 1.5*(d) - 1.5*(t)$);
        \coordinate (B5) at ($(P) - 0.8*(d) - 1.2*(t)$);
        \coordinate (B6) at ($(P) - 0.2*(d) - 0.6*(t)$);
        
        \begin{scope}
        \fill[blue!30, opacity=0.6]
            
            plot[smooth, tension=1] coordinates {
                (P) (C1)  (B1) (B2) (B3)
            }
            -- (B4)
            plot[smooth, tension=1] coordinates {
                (B4) (B5) (B6) (C2) (P)
            }
            --cycle;
        \end{scope}


        
        \draw[-latex,thick] ($(P)$) -- ($(P) + 1.5*(d)$);
    
        \draw[thick]  ($(P)-3*(t)$) -- ($(P)+3*(t)$);
    
        \fill (P) circle (3pt);

        \node at ($(P)+(0.1,0)$) [right] {$\pfam$};
        \node at ($(P) + 1.5*(d)-(0,0.2)$) [above, font=\fontsize{8pt}{6pt}\selectfont] {$\nabla R_{\Pi} (\pfam)$};
        \node at ($(P)+(3.2,0)$) [right, font=\fontsize{8pt}{6pt}\selectfont] {$R_{\Pi}$};
        \node at ($(P)+(5.5,0)$) [magenta!30!red, right, font=\fontsize{8pt}{6pt}\selectfont] {$R_{V}$};
        \node at ($(P)-0.75*(d)$) [blue, font=\fontsize{8pt}{6pt}\selectfont] {$K$};

    \end{tikzpicture}

    \caption{The black curve is the profit indifference curve and the red one is the surplus indifference curve. Pareto efficiency requires that the feasible set contains $\pfam$ and lies entirely within the shaded region. An illustrative feasible set is shown in blue.}
    \label{fig.pareto}
\end{figure}

Directly applying Lemma~\ref{lemma.pareto}, we obtain the following theorem, which characterizes the efficiency properties of commonly implemented price regulations.
\begin{theorem}\label{thm.pareto}
    For any $\delta$ satisfying Assumption \ref{assump.spectrum},
    \begin{enumerate}
        \renewcommand{\labelenumi}{(\roman{enumi})}
        \item $\myd_{\boldsymbol{\Delta}}$ is Pareto inefficient.

        \item $\myb(\underline{\bp},\overline{\bp})$ is Pareto efficient if and only if $\overline{\bp} \in \mathcal{P}$.

        \item $\mye_{\btheta,M}$  is Pareto efficient if and only if $\exists \pfam \in \mathcal{P}$ such that $\btheta \propto \nabla R_{\Pi}(\pfam)$ and $\langle \btheta, \pfam \rangle = M$.
    \end{enumerate}
\end{theorem}

Theorem \ref{thm.pareto} implies that all three commonly used price regulations are generally inefficient. The reason is straightforward: the set of efficient prices, $\mathcal{P}$, forms a one-dimensional curve in~$\mathbb{R}^n$ parametrized by~$\eta$, and the corresponding gradients of profit span another one-dimensional curve. Thus, when the regulator arbitrarily chooses a price bound~$\overline{\bp}$ or an averaging weights~$\btheta$, the equality required in Theorem~\ref{thm.pareto} generically does not hold. On the other hand, selecting parameters that do satisfy these conditions would require identifying at least one efficient price~$\pfam(\eta)$, which in turn demands complete information about the spillover intensity~$\delta$, the network structure~$\mathbf{G}$, consumers’ intrinsic values~$\ba$, and marginal costs~$\bc$. Such requirements are too demanding for a regulator in practice. Therefore, all three types of price regulations generically lead to inefficient outcomes.

Next, to understand why inefficiency arises when the conditions of Theorem~\ref{thm.pareto} fail, note that even if an efficient price is feasible, these regulations typically leave the firm with at least one feasible direction that raises profit at the expense of consumer surplus. Geometrically, the feasible set cannot be contained in the shaded region illustrated in Figure~\ref{fig.pareto}, which is necessary for efficiency.

Under the price-difference regulation $\myd_{\boldsymbol{\Delta}}$, relative prices are fixed but the firm may raise all prices uniformly. Such a shift strictly increases profit and reduces consumer surplus, making inefficiency inevitable. Under price floors and ceilings, $\myb(\underline{\bp},\overline{\bp})$, an efficient price is never optimal unless the upper bound exactly equals that price. Otherwise, the firm can increase price in at least one market, again increasing profit and reducing consumer surplus. Finally, under the average-price regulation $\mye_{\btheta,M}$, if the weight vector $\btheta$ is not proportional to the profit gradient at the efficient price, the firm can lower the price of a highly weighted market and offset it by raising prices elsewhere, leaving the weighted average unchanged. Because profit and consumer-surplus gradients have opposite signs at the efficient price, such an adjustment increases profit but decreases consumer surplus, implying that the resulting equilibrium cannot be efficient.

\subsection{Large spillovers}\label{sec.large}

As the intensity of spillovers becomes sufficiently large, the equilibrium outcomes have a much simpler form. In particular, we analyze the efficiency of price regulations as $\delta$ approaches the upper bound $1/\lambda_1$. The economic implications for this limiting regime are discussed in Section \ref{sec.discussion}.

Denote the limit of consumer surplus and profit by
\begin{equation*}
    \bar{R}_V(\bp) := \lim_{\delta \rightarrow 1/\lambda_1} R_V(\bp), \quad \bar{R}_{\Pi}(\bp) := \lim_{\delta \rightarrow 1/\lambda_1} R_{\Pi}(\bp).
\end{equation*}
To describe the limit outcomes in equilibrium, let $\bp^*(\delta)$ be the firm's optimal price under a given~$\delta$. The limiting equilibrium consumer surplus and profit are defined as:
\begin{equation*}
    \bar{R}^*_{V} := \lim_{\delta \rightarrow 1/\lambda_1} R_{V}(\bp^*(\delta)), \quad \bar{R}^*_{\Pi} := \lim_{\delta \rightarrow 1/\lambda_1} R_{\Pi}(\bp^*(\delta)).
\end{equation*}

Define an affine function $\mathscr{A}: \mathbb{R}^n \rightarrow \mathbb{R}$,
\begin{equation}\label{eq.a}
    \mathscr{A}(\bp) := 
    \frac{\langle \mathbf{w}_1, \bp-\bp^{ur} \rangle}{\langle \mathbf{w}_1, \frac{\ba-\bc}{2} \rangle},
\end{equation}
where $\mathbf{w}_1$ is the normalized eigenvector associated with the largest eigenvalue, also known as the eigenvector centrality.\footnote{We assume the network $\mathbf{G}$ is connected, then the normalized eigenvector corresponding to the largest eigenvalue is unique and positive. For background on the eigenvector centrality and its interpretations, see, e.g., \cite{bonacich1972factoring} and \cite{jackson2008social}.} The inner product $\langle \mathbf{w}_1, \bp\rangle$ can be interpreted as the weighted average of $\bp$ using $\mathbf{w}_1$ as the weights. (Henceforth, it will be abbreviated to average price.) Thus, $\mathscr{A}$ measures the average difference between a price vector $\bp$ and the benchmark $\bp^{ur}$.\footnote{This difference is normalized by $\langle \mathbf{w}_1, \frac{\ba-\bc}{2} \rangle = \langle \mathbf{w}_1, \ba-\bp^{ur} \rangle = -\langle \mathbf{w}_1, \bc-\bp^{ur} \rangle$, where $\ba$ and $\bc$ can be seen as the highest and lowest price the firm can charge: $\ba$ corresponds to consumers' intrinsic values and $\bc$ corresponds to marginal costs, and both lead to zero profit for the firm.} Denote the closure of any set $S$ by $\overline{S}$. We then state the following lemma.

\begin{lemma}\label{lemma.plimit}
~
    \begin{enumerate}
        \renewcommand{\labelenumi}{(\roman{enumi})}
        \item For any given $\bp$,
        \begin{equation*}
            \bar{R}_V(\bp) = (1-\mathscr{A}(\bp))^2, \quad
            \bar{R}_{\Pi}(\bp) = 1-\mathscr{A}(\bp)^2.
        \end{equation*}

        Then for any $\bp$ and $\bp'$, $\bar{R}_V(\bp)=\bar{R}_V(\bp')$ and $\bar{R}_{\Pi}(\bp) = \bar{R}_{\Pi}(\bp')$ if and only if $\mathscr{A}(\bp)=\mathscr{A}(\bp')$. 

        \item There exists a unique sufficient statistic 
        \begin{equation*}
            \bar{\mathscr{A}}^* := \argmax_{A \in \overline{\mathscr{A}(K)}} \{1-A^2\},
        \end{equation*}
        such that
        \begin{equation*}
            \bar{R}^*_{V}  = (1-\bar{\mathscr{A}}^*)^2, \quad
            \bar{R}^*_{\Pi}  = 1-(\bar{\mathscr{A}}^*)^2.
        \end{equation*}
    \end{enumerate}
\end{lemma}

The first part of the lemma fully characterizes the consumer surplus and profit in the limit and indicates that the average price difference serves as a sufficient statistic. The underlying intuition is that, as $\delta$ increases, the profit function places increasingly greater weight on the direction of the first eigenvector. Hence, any deviation in the $\mathbf{w}_1$ direction generates excessively high profit loss, which in turn forces the gradient at any $\bp$ to become more closely aligned with $\mathbf{w}_1$, as illustrated in Figure~\ref{fig.plimit}. In the limit, the gradient is parallel to $\mathbf{w}_1$, so any two prices with the same average level, i.e. $\mathscr{A}(\bp)=\mathscr{A}(\bp')$, yield the same profit ratio. The same argument applies to consumer surplus.

\begin{figure}[htbp]
	\centering
    \begin{subfigure}{0.48\textwidth}
        \centering
        \begin{tikzpicture}[scale=0.9]
            \coordinate (d1) at (1,2);
            \coordinate (d2) at (2,-1);
            \coordinate (pur) at (3,0);
            
            \def\smallMajor{2}
            \def\smallMinor{1.22}
        
            \draw (pur) ellipse[x radius=\smallMajor, y radius=\smallMinor, rotate=-26.57];

            \fill (pur) circle (1.5pt);
            \node at ($(pur)-(0,0.1)$) [below] {$\bp^{ur}$};
        
            \draw[dashed] ($(pur)-1.1*(d2)$) -- ($(pur)+1.1*(d2)$);
            
            \draw[thick,-latex] (pur) -- ($(pur)+0.4*(d1)$);
            \node at ($(pur)+0.3*(d1)$) [right] {$\mathbf{w}_1$};

            \coordinate (q) at (1.18, 0.23);
            \coordinate (v) at (1.14, 0.35);
            \coordinate (h) at (0.35,-1.14);

            \fill (q) circle (1.5pt);
            \node at (q) [below left] {$\bp$};
            \draw[thick,-latex] (q) -- ($(q)+0.8*(v)$);
            \node at ($(q)+0.6*(v)$) [above,font=\fontsize{10pt}{6pt}\selectfont] {$\nabla R_{\Pi}$};
    
        \end{tikzpicture}
        \caption{Moderate $\delta$}
    \end{subfigure}
    \hfill
    \begin{subfigure}{0.48\textwidth}
        \centering
        \begin{tikzpicture}[scale=0.9]
            \coordinate (d1) at (1,2);
            \coordinate (d2) at (2,-1);
            \coordinate (pur) at (3,0);

            \def\smallMajor{2.52}
            \def\smallMinor{0.84}

            \draw (pur) ellipse[x radius=\smallMajor, y radius=\smallMinor, rotate=-26.57];

            \fill (pur) circle (1.5pt);
            \node at ($(pur)-(0,0.1)$) [below] {$\bp^{ur}$};
        
            \draw[dashed] ($(pur)-1.3*(d2)$) -- ($(pur)+1.3*(d2)$);
            
            \draw[thick,-latex] (pur) -- ($(pur)+0.55*(d1)$);
            \node at ($(pur)+0.5*(d1)$) [right] {$\mathbf{w}_1$};
    
            \coordinate (q) at (1.18, 0.23);
            \coordinate (v) at (1.26,1.30);
            \coordinate (h) at (1.30,-1.26);
            \fill (q) circle (1.5pt);
            \node at (q) [below left] {$\bp$};
            \draw[thick,-latex] (q) -- ($(q)+0.6*(v)$);
            \node at ($(q)+0.5*(v)$) [right,font=\fontsize{10pt}{6pt}\selectfont] {$\nabla R_{\Pi}$};
    
        \end{tikzpicture}
        \caption{Large $\delta$}
    \end{subfigure}
    
	\caption{As $\delta$ increases, the firm’s indifference curve flattens in directions orthogonal to $\mathbf{w}_1$, and the gradient of profit for a given price $\bp$ becomes increasingly aligned with $\mathbf{w}_1$. In the limit, the gradient is parallel to $\mathbf{w}_1$, implying that all prices with the same $\mathbf{w}_1$–weighted average yield the same profit.}
	\label{fig.plimit}
\end{figure} 

The second part shows that $\bar{\mathscr{A}}^*$ can be interpreted as the solution to the firm's profit maximization problem in the limit. Accordingly, it characterizes the firm's limiting behavior and shows that the equilibrium consumer surplus and profit are fully determined by the sufficient statistic $\bar{\mathscr{A}}^*$.

Since increasing the firm’s profit requires reducing $|\bar{\mathscr{A}}^*|$, doing so lowers (raises) consumer surplus when $\bar{\mathscr{A}}^*$ is negative (positive). Therefore, a price regulation is called \emph{Pareto efficient} (\emph{inefficient}) in the limit if and only if $\bar{\mathscr{A}}^* < 0$ ($>0$). In particular, we refer to the case $\bar{\mathscr{A}}^* = 0$ as \emph{neutral}.\footnote{The neutral case is formally Pareto efficient, but we treat it separately because the distinction matters in the limit. When $\delta$ is fixed, neutrality occurs only at $\bp^{ur}$. However, when $\delta \to 1/\lambda_1$, any $\bp \in K$ with $\mathscr{A}(\bp)=0$ yields neutrality even if $\bp \neq \bp^{ur}$. Thus, with sufficiently strong spillovers, price regulations may have negligible effects.}



\begin{theorem}\label{thm.neutral}
    As $\delta \rightarrow 1/\lambda_1$,
    \begin{enumerate}
        \renewcommand{\labelenumi}{(\roman{enumi})}
        \item $\myd_{\boldsymbol{\Delta}}$ is neutral.

        \item $\myb(\underline{\bp},\overline{\bp})$ is neutral if and only if $\mathscr{A}(\underline{\bp}) \leq 0 \leq \mathscr{A}(\overline{\bp})$. Otherwise, it is efficient when $\mathscr{A}(\overline{\bp}) < 0$ and inefficient when $0 < \mathscr{A}(\underline{\bp})$.

        \item $\mye_{\btheta,M}$ is neutral if and only if $\btheta \not\propto \mathbf{w}_1$. Otherwise, it is efficient.
    \end{enumerate}
            
            
            

\end{theorem}

The neutrality in Theorem~\ref{thm.neutral} directly follows from Lemma~\ref{lemma.plimit}, which implies that whenever the feasible set contains a price whose average level matches that of the unrestricted price, such a price induces no profit loss for the firm in the limit. Consequently, the firm will select among these prices, which makes the regulation neutral in the limit. Such scenario is illustrated in Figure~\ref{fig.limit} for the three commonly implemented price regulations.

\begin{figure}[ht]
	\centering
    \begin{subfigure}{0.32\textwidth}
        \centering
        \begin{tikzpicture}[scale = 1.2]

            \coordinate (d1) at (1,2);
            \coordinate (d2) at (2,-1);
            \coordinate (pur) at (3,0);

            \coordinate (D1) at ($(1.5,-0.5)$);
			\coordinate (D2) at ($(D1)+1.8*(1,1)$);
			\coordinate (D3) at ($(D2)+0.6*(-1,1)$);
			\coordinate (D4) at ($(D1)+0.6*(-1,1)$);
			
			\fill[blue!20] (D1)--(D2)--(D3)--(D4)--cycle;

			\draw[blue,thick] (D1)--(D2);
			\draw[blue,thick] (D3)--(D4);

            \fill (pur) circle (1.5pt);
            \node at ($(pur)-(0,0.1)$) [below] {$\bp^{ur}$};
        
            \draw[dashed] ($(pur)-(d2)$) -- ($(pur)+0.5*(d2)$);
            
            \draw[thick,-latex] (pur) -- ($(pur)+0.4*(d1)$);
            \node at ($(pur)+0.4*(d1)$) [right] {$\mathbf{w}_1$};

        \end{tikzpicture}
        \caption{$\myd_{\boldsymbol{\Delta}}$}
		\label{fig.dlim}
    \end{subfigure}
    \hfill
	\begin{subfigure}{0.32\textwidth}
		\centering
        \begin{tikzpicture}[scale = 1.2]

            \coordinate (d1) at (1,2);
            \coordinate (d2) at (2,-1);
            \coordinate (pur) at (3,0);

            \coordinate (B1) at (1.2,0);
            \coordinate (B2) at (2.5,1.2);
            
            \fill[blue!20] (B1) rectangle (B2);
            \draw[blue,thick] (B1) rectangle (B2);
        
            \fill (pur) circle (1.5pt);
            \node at ($(pur)-(0,0.1)$) [below] {$\bp^{ur}$};
        
            \draw[dashed] ($(pur)-(d2)$) -- ($(pur)+0.5*(d2)$);
            
            \draw[thick,-latex] (pur) -- ($(pur)+0.4*(d1)$);
            \node at ($(pur)+0.4*(d1)$) [right] {$\mathbf{w}_1$};
    
        \end{tikzpicture}
		\caption{$\myb(\underline{\bp},\overline{\bp})$}
		\label{fig.blim}
	\end{subfigure}
    \hfill
    \begin{subfigure}{0.32\textwidth}
		\centering
        \begin{tikzpicture}[scale = 1.2]

            \coordinate (d1) at (1,2);
            \coordinate (d2) at (2,-1);
            \coordinate (pur) at (3,0);

            \coordinate (v) at (1,-2);
            \coordinate (h) at (2,1);
            \coordinate (A1) at (1.7,1.2);
            \coordinate (A2) at ($(A1)+(v)$);
            \coordinate (A3) at ($(A2)-0.4*(h)$);
            \coordinate (A4) at ($(A1)-0.4*(h)$);
            
            \coordinate (A) at ($0.5*(A1)+0.5*(A2)$);
            \coordinate (AE) at ($(A)+0.5*(h)$);
            
            \fill[blue!20] (A1)--(A2)--(A3)--(A4)--cycle;
            
            \draw[blue,thick] (A1)--(A2);

            \fill (pur) circle (1.5pt);
            \node at ($(pur)-(0,0.1)$) [below] {$\bp^{ur}$};
        
            \draw[dashed] ($(pur)-(d2)$) -- ($(pur)+0.5*(d2)$);
            
            \draw[thick,-latex] (pur) -- ($(pur)+0.4*(d1)$);
            \node at ($(pur)+0.4*(d1)$) [right] {$\mathbf{w}_1$};
    
            \draw[thick,-latex,blue] ($0.8*(A1)+0.2*(A2)$) -- ($0.8*(A1)+0.2*(A2)+0.3*(h)$);
            \node at ($0.8*(A1)+0.2*(A2)+0.3*(h)$) [right,blue] {$\btheta$};
    
        \end{tikzpicture}
		\caption{$\mye_{\btheta, M}$}
		\label{fig.elim}
	\end{subfigure}

	\caption{The dashed hyperplane consists of all price vectors whose average level equals that of $\bp^{ur}$. Hence, whenever the feasible set intersects this hyperplane, the limiting optimal price lies on this intersection, and the regulation becomes neutral in the limit.}
	\label{fig.limit}
\end{figure} 

For $\myd_{\boldsymbol{\Delta}}$, the firm can freely increase prices by the same amount across all markets. Hence, there must exist a feasible price with the same average level as the unrestricted one, and this regulation is neutral in the limit.

For regulations with price floors and ceilings, $\myb(\underline{\bp},\overline{\bp})$, neutrality in the limit is achieved whenever the average of the lower bound lies below the unrestricted average and the average of the upper bound lies above it. Otherwise, if the upper bound is low enough so that the average of all feasible prices falls below the unrestricted level, the limiting outcome becomes efficient, as illustrated in Figure~\ref{fig.blimefficient}. Conversely, the limit outcome becomes inefficient when the lower bound is sufficiently high.

For regulations on the average price, $\mye_{\btheta, M}$, if the weight $\btheta$ is not parallel to $\mathbf{w}_1$, the firm can adjust prices along the direction orthogonal to $\btheta$, which leaves the $\btheta$-weighted average unchanged while modifying the $\mathbf{w}_1$-weighted average. This allows the firm to reach a price with the same average level as $\bp^{ur}$, making the regulation approximately neutral. On the other hand, if $\btheta$ is parallel to $\mathbf{w}_1$, which generically does not occur, the regulator can set the bound low enough such that every feasible price delivers a strictly lower average level than the unrestricted one, i.e., $\mathscr{A}(\bp) < 0$ for all $\bp \in K$, as shown in Figure~\ref{fig.elimefficient}. In this case, the limit outcome is efficient. 

\begin{figure}[ht]
	\centering

    \begin{subfigure}{0.45\textwidth}
		\centering
        \begin{tikzpicture}[scale = 1.2]

            \coordinate (d1) at (1,2);
            \coordinate (d2) at (2,-1);
            \coordinate (pur) at (3,0);

            \coordinate (B1) at (1.2,-1);
            \coordinate (B2) at (2.7,0);
            
            \fill[blue!20] (B1) rectangle (B2);
            \draw[blue,thick] (B1) rectangle (B2);
        
            \fill (pur) circle (1.5pt);
            \node at ($(pur)+(0,0.1)$) [right] {$\bp^{ur}$};
        
            \draw[dashed] ($(pur)-0.7*(d2)$) -- ($(pur)+0.7*(d2)$);
            
            \draw[thick,-latex] (pur) -- ($(pur)+0.4*(d1)$);
            \node at ($(pur)+0.4*(d1)$) [right] {$\mathbf{w}_1$};
    
        \end{tikzpicture}
		\caption{$\myb(\underline{\bp},\overline{\bp})$ with $\mathscr{A}(\overline{\bp}) < 0$}
		\label{fig.blimefficient}
	\end{subfigure}
    \hfill
    \begin{subfigure}{0.45\textwidth}
		\centering
        \begin{tikzpicture}[scale = 1.2]

            \coordinate (d1) at (1,2);
            \coordinate (d2) at (2,-1);
            \coordinate (pur) at (3,0);


            \coordinate (A1) at ($(pur)-0.1*(d1)-0.7*(d2)$);
            \coordinate (A2) at ($(pur)-0.1*(d1)+0.7*(d2)$);
            \coordinate (A3) at ($(A2)-0.4*(d1)$);
            \coordinate (A4) at ($(A1)-0.4*(d1)$);
            
            \coordinate (A) at ($0.5*(A1)+0.5*(A2)$);
            \coordinate (AE) at ($(A)+0.5*(d1)$);
            
            \fill[blue!20] (A1)--(A2)--(A3)--(A4)--cycle;
            
            \draw[blue,thick] (A1)--(A2);

            \fill (pur) circle (1.5pt);
            \node at ($(pur)+(0,0.1)$) [right] {$\bp^{ur}$};
        
            \draw[dashed] ($(pur)-0.7*(d2)$) -- ($(pur)+0.7*(d2)$);
            
            \draw[thick,-latex] (pur) -- ($(pur)+0.4*(d1)$);
            \node at ($(pur)+0.4*(d1)$) [right] {$\mathbf{w}_1$};

            \draw[thick,-latex,blue] ($0.7*(A1)+0.3*(A2)$) -- ($0.7*(A1)+0.3*(A2)+0.4*(d1)$);
            \node at ($0.7*(A1)+0.3*(A2)+0.4*(d1)$) [right,blue] {$\btheta$};

        \end{tikzpicture}
		\caption{$\mye_{\btheta, M}$ with $\btheta = \mathbf{w}_1$}
		\label{fig.elimefficient}
	\end{subfigure}

	\caption{Some specifications deliver efficient outcomes in the limit}
	\label{fig.limitefficient}
\end{figure}


\subsection{Remarks and discussion}\label{sec.discussion}

\textbf{Comparison of the fixed and large spillovers}\hspace{1ex}
Table~\ref{table.summary} summarizes the main efficiency results for three commonly implemented price regulations, both for a fixed spillover intensity and in the limit as spillovers become sufficiently strong.

\begin{table}[ht]
    \centering
    \renewcommand{\arraystretch}{1}
    \setlength{\tabcolsep}{10pt}
    \begin{tabular}{c|l|l}
        \hline\hline
        Price regulation 
        & \multicolumn{1}{c|}{Outcome given $\delta$}
        & \multicolumn{1}{c}{Outcome in the limit}
        \\
        \hline\hline
        
        $\myd_{\boldsymbol{\Delta}}$
        & Inefficient
        & Neutral
        \\
        \hline

        $\myb(\underline{\bp},\overline{\bp})$ 
        & Inefficient, generically
        & Neutral, under mild conditions
        \\
        \hline
        
        $\mye_{\btheta,M}$ 
        & Inefficient, generically
        & Neutral, generically
        \\
        \hline\hline
    \end{tabular}
    \caption{Summary of outcomes under different price regulations}
    \label{table.summary}
\end{table}


Overall, for a fixed spillover intensity, these regulations are highly likely to be inefficient, which means there exist other prices that improve both consumer surplus and firm profit. As the spillover intensity increases, as shown in Lemma~\ref{lemma.plimit}, the firm only cares about deviations of the average price from the unrestricted benchmark, which substantially enlarges the set of prices that can be implemented without incurring profit losses in equilibrium. Consequently, the impact of price regulation tends to become neutral: both surplus and profit converge to their unregulated levels, so that the welfare effects of regulation vanish in the limit.

However, these results also reveal that a price regulation which is inefficient for a fixed $\delta$ may become neutral (a special case of efficiency) or even efficient in the limit. To explain the mechanism, the shaded area in Figure~\ref{fig.fixfeasible} depicts the feasible set of $(R_V(\bp), R_\Pi(\bp))$ for a fixed $\delta$.\footnote{The blue curve represents the Pareto frontier induced by problem~(\ref{eq.intervention}). The red curve is the lower frontier, obtained by considering the minimization analogue of problem~(\ref{eq.intervention}).}
In this case, the feasible set is genuinely two-dimensional, and an equilibrium outcome may lie strictly inside the set, implying inefficiency. In contrast, as $\delta \to 1/\lambda_1$, surplus and profit collapse onto a one-dimensional curve indexed by the sufficient statistic $\mathscr{A}$, as shown in Figure~\ref{fig.largefeasible}. For any price vector $\bp$, the outcome $(R_V(\bp), R_\Pi(\bp))$ converges to a point on this curve. As a result, an equilibrium that is inefficient for a fixed $\delta$ (point $E$ in Figure~\ref{fig.relation}) may converge to a neutral or efficient outcome in the limit (point $\bar{E}$ in Figure~\ref{fig.largefeasible}).



\begin{figure}[ht]
	\centering
	\begin{subfigure}{0.47\textwidth}
		\centering
        \begin{tikzpicture}[scale=1.5]

            \fill[gray!50] (0,0)--(2.56,0)--
            plot[domain=0:1,smooth, variable=\t]({0.25*(2+1.2*\t)^2},{0.25*4*(1-\t^2)})--
            plot[domain=0:1,smooth, variable=\t]({(1-\t)^2},{0.25*4*(1-\t^2)})
            --cycle;
            \draw[domain=0:1,blue, smooth, variable=\t, samples=200,thick] 
            plot({0.25*(2+1.2*\t)^2},{0.25*4*(1-\t^2)});
             \draw[domain=0:1,red, smooth, variable=\t, samples=200,thick] 
            plot({(1-\t)^2},{0.25*4*(1-\t^2)});

            \draw[->] (-0.5, 0) -- (3, 0) node[right,font=\fontsize{10pt}{0}\selectfont] {$R_{V}$};
            \draw[->] (0, -0.5) -- (0, 2) node[above,font=\fontsize{10pt}{0}\selectfont] {$R_{\Pi}$};
    

            \coordinate (e) at (1.35,0.5);
            \fill (e) circle (1pt);
            \node at (e) [above] {$E$};
            \draw[dashed] (0,0.5) -- (2.03,0.5);
            \draw[dashed] (e) -- (1.35,0);
            \draw[dashed] (2.03,0.5) -- (2.03,0);
            \node at (0,0.5) [left, font=\fontsize{8pt}{6pt}\selectfont] {$R^*_\Pi (\delta)$};
            \node at (1.35,0) [below, font=\fontsize{8pt}{6pt}\selectfont] {$R_V^* (\delta)$};
            \node at (2.03,0) [below, font=\fontsize{8pt}{6pt}\selectfont] {$R_V^+ (\delta)$};

            \draw[
                decorate,
                decoration={brace, raise=2pt}
            ]
            (1.35,0) -- (2.03,0)
            node[midway, above=3pt, font=\fontsize{8pt}{6pt}\selectfont]
            {$Gap(\delta)$};

        \end{tikzpicture}
		\caption{}
		\label{fig.fixfeasible}
	\end{subfigure}
	\hfill
	\begin{subfigure}{0.47\textwidth}
		\centering
        \begin{tikzpicture}[scale=1.5]

            \draw[domain=0:1,blue, smooth, variable=\t, samples=200,thick] 
            plot({0.25*(2+1*\t)^2},{0.25*4*(1-\t^2)});
            \draw[domain=0:1,red, smooth, variable=\t, samples=200,thick] 
            plot({(1-\t^3)^2},{(1-\t^5)});

            \draw[->] (-0.5, 0) -- (3, 0) node[right,font=\fontsize{10pt}{0}\selectfont] {$\bar{R}_{V}$};
            \draw[->] (0, -0.5) -- (0, 2) node[above,font=\fontsize{10pt}{0}\selectfont] {$\bar{R}_{\Pi}$};
    
            \draw[dashed] (0,1)--(1,1);
            \draw[dashed] (1,0)--(1,1);
            \node at (0,1) [left] {$1$};
            \node at (1,0) [below] {$1$};

            \node at (0,0.5) [left,red] {$\mathscr{A} < 0$};
            \node at (2,0.5) [right,blue] {$\mathscr{A} > 0$};

            \fill (1,1) circle (1pt);
            \node at (1,1) [above right] {$\bar{E}$ (neutral, $\mathscr{A}=0$)};

            \fill (1.35,0.5) circle (1pt);
            \node at (1.35,0.5) [right] {$E$};

            \draw[
                dashed,
                thick,
                postaction={
                    decorate,
                    decoration={
                        markings,
                        mark=at position 0.6 with {\arrow{latex}}
                    }
                }
            ]
            (1.35,0.5) -- (1,1);
            
        \end{tikzpicture}
		\caption{}
		\label{fig.largefeasible}
	\end{subfigure}
    \caption{Panel~\ref{fig.fixfeasible} illustrates the feasible set of consumer surplus and profit for a fixed spillover intensity $\delta$, shown by the gray region. Panel~\ref{fig.largefeasible} depicts the limiting outcomes as $\delta \to 1/\lambda_1$, shown by a one-dimensional curve indexed by $\mathscr{A}$.}
	\label{fig.relation}
\end{figure}


Formally, for any given price regulation $K$ and fixed $\delta$, let $R_\Pi^*(\delta)$ denote the equilibrium profit. There exists a corresponding consumer surplus level on the Pareto frontier, denoted by $R_V^+(\delta)$, that delivers the same profit.\footnote{This value is obtained as the maximal value of problem~(\ref{eq.intervention}) with $\tau = R_\Pi^*(\delta)$.} The efficiency loss induced by the regulation can therefore be measured by the gap between the equilibrium and Pareto-efficient surplus,
\begin{equation*}
	Gap(\delta) := R^+_V(\delta) - R_V^*(\delta) \geq 0,
\end{equation*}
as illustrated in Figure~\ref{fig.fixfeasible}.
\begin{lemma}\label{lemma.gap}
	The price regulation is efficient or neutral in the limit if and only if
	\begin{equation*}
		\lim_{\delta \rightarrow 1/\lambda_1} Gap(\delta) = 0.
	\end{equation*}
    In particular, it is neutral if, in addition, $\lim_{\delta \rightarrow 1/\lambda_1}R_V^*(\delta) = 1$.
\end{lemma}

This lemma formally indicates that, even if a regulation is inefficient for a fixed spillover intensity, the efficiency gap may vanish as $\delta$ approaches its upper bound. In this case, the regulation becomes neutral or even efficient in the limit.

\vspace{12pt}

\noindent\textbf{Large spillovers made simple}\hspace{1ex}
The previous results and discussion shows that, in the limit, the analysis of equilibrium outcomes can be reduced to a single index $\mathscr{A}$. To clarify how this simplification arises and to highlight its economic implications, we adopt the spectral decomposition approach. Specifically, the symmetric real matrix $\mathbf{G}$ admits the following decomposition:
\begin{equation*}
    \mathbf{G} = \mathbf{W}\mathbf{\Lambda}\mathbf{W}^T,
\end{equation*}
where $\mathbf{\Lambda}$ is a diagonal matrix whose diagonal elements $\Lambda_{ii} = \lambda_i$ are the eigenvalues of $\mathbf{G}$ ordered as $\lambda_1 \geq \ldots \geq \lambda_n$. Also, $\mathbf{W}$ is an orthogonal matrix, whose $i$th column $\mathbf{w}_i$ is the corresponding eigenvector associated with $\lambda_i$ which is normalized as $||\mathbf{w}_i||_2 = 1$.

Applying this decomposition to the demand, the Slutsky matrix can be written as
\begin{equation*}
    \frac{\partial \mathbf{x}}{\partial \bp} = -\bH = - \frac{1}{1-\delta\lambda_1} \mathbf{w}_1 (\mathbf{w}_1)^T - \sum_{i\geq 2} \frac{1}{1-\delta\lambda_i} \mathbf{w}_i (\mathbf{w}_i)^T.
\end{equation*}
When the spillovers are sufficiently large, the term associated with the first eigenvector dominates. As a result, the demand system can be approximated arbitrarily well by the first principal component, even without full knowledge of the network structure. In \cite{galeotti2024robust}, matrices satisfying such a structure admit a recoverable form, and they show the existence of robust interventions under which one unit of subsidy can robustly increase aggregate welfare by approximately two units. Their approach relies on a generalized technique analogous to recovering the principal eigenvector $\mathbf{w}_1$. In our setting, by contrast, the network is connected and deterministic. As a result, the limiting structure is fully determined by $\mathbf{w}_1$, yielding a particularly simple characterization of equilibrium consumer surplus and profit in the limit.

Beyond simplifying the characterization of equilibrium outcomes, this approximation also substantially reduces the regulator’s information requirements. As shown in Theorem~\ref{thm.pareto}, when the spillover intensity is moderate, achieving efficiency requires detailed information: the regulator must know the entire network structure across markets, consumers’ intrinsic values, and the firm’s production costs. This information requirement becomes much weaker when spillovers are sufficiently strong. In the limit, selecting parameters that lead to efficiency only requires two pieces of information: the eigenvector centrality $\mathbf{w}_1$ and the unrestricted price $\bp^{ur}$. This is much more practical than the fixed $\delta$ case.\footnote{Several studies show that Aggregated Relational Data, which are easier to collect than complete network data, can be used to estimate network statistics such as eigenvector centrality. For example, see \cite{brezaUsingAggregatedRelational2020} and \cite{brezaConsistentlyEstimatingNetwork2023}.}

\subsection{Relationship to \cite{galeottiTargetingInterventionsNetworks2020}}\label{sec.relationship}


Our framework closely relates to the literature on characteristic interventions in networks, particularly the seminal work of \cite{galeottiTargetingInterventionsNetworks2020}. They study a utilitarian planner who maximizes aggregate individual surplus by modifying agents' intrinsic values, $\ba$, subject to a budget constraint on the $\ell^2$-norm of the intervention vector. Our efficient price problem~(\ref{eq.intervention}) presents a direct analogy, where the planner adjusts prices to maximize aggregate consumer surplus. In our setting, the ``cost'' of intervention is the reduction in the firm's profit. Consequently, the lower bound on firm profit acts as a constraint analogous to the budget constraint in \cite{galeottiTargetingInterventionsNetworks2020}. Due to network spillovers, adjusting the price in one market affects the demand and profit across the entire network. Thus, in contrast to their $\ell^2$-norm cost, our cost structure induces cross-market dependencies and is not separable. However, it still admits simultaneous diagonalization. We exploit this property to apply spectral decomposition techniques similar to theirs, deriving Lemma~\ref{lemma.pfamily}.

Furthermore, the firm's profit-maximizing problem under price regulation shares structural similarities with the intervention problem but differs fundamentally in incentives. Here, the firm acts as the intervener, using personalized prices to influence consumption and maximize profit. The price regulation serves as a generalized ``budget''. Notably, unlike a benevolent planner, the firm's objective is not aligned with consumer surplus. Thus, our analysis focuses on the surplus-profit trade-off and efficiency rather than surplus alone. Using simultaneous diagonalizability, we apply spectral decomposition to both consumer surplus and profit. While we achieve a clean characterization in the large $\delta$ limit, like the "simple" intervention under a large budget in \cite{galeottiTargetingInterventionsNetworks2020}, the underlying mechanisms differ. In their setting, a large budget makes the status quo intrinsic values negligible, simplifying the solution. In contrast, increasing $\delta$ in our model alters the shape of the indifference curves, inducing the firm to place greater weight on the leading eigenvector, as illustrated in Figure~\ref{fig.plimit}. Consequently, the equilibrium outcomes in the limit are fully determined by eigenvector centrality.

\section{Effects on Consumer Surplus}\label{sec.application}


The previous analysis evaluates price regulations through the lens of efficiency, which captures the joint welfare of the firm and consumers. Inefficiency, however, only indicates that a regulation fails to attain the maximal feasible consumer surplus given the induced loss in firm's profit. Nevertheless, such regulations may still increase consumer surplus, making them appealing to a regulator who prioritizes consumer welfare. Moreover, although these regulations generically have negligible effects on the relative consumer surplus and profit in the limit, the changes in absolute levels may still be substantial. Together, these considerations motivate a closer examination of whether these price regulations ultimately benefit consumers.\footnote{Technically, equations~(\ref{eq.cs}) and~(\ref{eq.pi}) show that as spillover grows, consumer surplus diverges to infinity faster than profit. Consequently, in markets with large spillovers, increasing consumer surplus is asymptotically equivalent to increasing total surplus, thereby improving efficiency. We also briefly address the case of moderate or small spillovers while primarily focusing on the asymptotic properties.}

In this section, we focus on the effects on consumer surplus and derive more policy implications for several special yet important price regulations that belong to the commonly implemented classes discussed before.

\subsection{Fixing prices}\label{sec.fixp}
As a first step, we consider the most stringent form of regulation, in which the regulator imposes fixed prices $\bp$ across all markets. To assess the implications for consumer surplus, we allow the regulator to place different priorities on consumer surplus relative to firm profit. Specifically, the regulator solves a Ramsey problem that maximizes a weighted sum of firm profit and consumer surplus, subject to a non-negativity constraint on profit:
\begin{equation}\label{eq.ramsey}
    \begin{gathered}
        \max\limits_{\bp \in \mathbb{R}^n} \Pi(\bp) + \alpha V(\bp)\\
        s.t. \ \Pi(\bp) \geq 0
    \end{gathered}
\end{equation}
where $\alpha \geq 0$ is the Pareto weight on consumer surplus.

\begin{proposition}\label{prop.ramsey}
    For any $\alpha \in [0,\bar{\eta}]$, the solution to the Ramsey problem~(\ref{eq.ramsey}) is exactly the efficient price $\pfam(\alpha)$. For $\alpha > \bar{\eta}$, the solution is $\pfam(\bar{\eta})$.  
\end{proposition}

The regulator who considers problem~(\ref{eq.intervention}) only focuses on consumer surplus, but subject to a lower bound on the firm’s profit. In contrast, in the Ramsey problem, the regulator’s priority placed on consumer surplus is determined by the parameter $\alpha$. Proposition~\ref{prop.ramsey} establishes an equivalence between these two problems. To see this, note that the Ramsey objective is equivalent to
\begin{equation*}
    R_\Pi(\bp) + \frac{\alpha V(\bp^{ur})}{\Pi(\bp^{ur})} R_V(\bp),
\end{equation*}
Hence, any pair $(R_V,R_\Pi)$ lying on the same line with slope $-\frac{\alpha V^{ur}}{\Pi^{ur}}$ yields the same value of this objective. As shown in Figure~\ref{fig.frontier}, maximization requires this line to be tangent to the Pareto frontier, which is characterized by solving the efficient price problem~(\ref{eq.intervention}).


\begin{figure}[ht]
	\centering
	
    \begin{tikzpicture}[scale = 2.5]
        \draw[->] (-0.2, 0) -- (2.7, 0) node[right,font=\fontsize{10pt}{0}\selectfont] {$R_{V}$};
        \draw[->] (0, -0.2) -- (0, 1.4) node[above,font=\fontsize{10pt}{0}\selectfont] {$R_{\Pi}$};
        
        \draw[domain=0:1,blue, smooth, variable=\t, samples=200,thick] 
        plot({0.25*(2+1.2*\t)^2},{0.25*4*(1-\t^2)});

        \coordinate (p) at (1.69,0.75);
        \fill[black] (p) circle (1pt);
        \node at ($(p)+(0.1,0.1)$) [right,font=\fontsize{10pt}{0}\selectfont]{slope = $-\alpha V^{ur}/\Pi^{ur}$};
        
        \draw[thick] ($(p)-0.4*(1.56,-1)$)--($(p)+0.4*(1.56,-1)$);

        \draw[dashed] (0,1)--(1,1);
        \draw[dashed] (1,0)--(1,1);
        \draw[dashed] (0,0.75)--(p);
        \draw[dashed] (1.69,0)--(p);
        \node at (0,1) [left] {$1$};
        \node at (1,0) [below] {$1$};
        \node at (0,0.75) [left] {$\tau$};
        \node at (1.69,0) [below] {$R_V(\tau)$};

    \end{tikzpicture}

	\caption{Pareto frontier and Ramsey pricing}
	\label{fig.frontier}
\end{figure}

Finally, Proposition~\ref{prop.ramsey} assigns an economic interpretation to the parameter $\eta$ associated with the efficient price. When the Pareto weight $\alpha$ is moderate, these two parameters are equivalent. As $\alpha$ increases, the optimal allocation shifts toward higher consumer surplus and lower firm's profit. Once $\alpha$ exceeds $\bar{\eta}$, the profit constraint binds at zero, and the optimal price is set at its lowest feasible level for all consumers, thus achieving the maximum possible consumer surplus.



\subsection{Average-price regulation}\label{sec.average}

We now analyze the regulation on average prices, $\mye_{\btheta,M}$. Recall that, under Assumption~\ref{assump.infeasible}, the equilibrium price vector is given by equation~(\ref{eq.paverage}). The resulting consumption vector under regulation can be written as
\begin{align*}
    \mathbf{x}(\bp^*) = \bH(\ba-\bp^*) = \mathbf{x}(\bp^{ur}) + \frac{\btheta^T\bp^{ur}-M   }{\btheta^T\bH^{-1} \btheta} \btheta.
\end{align*}

Assumptions~\ref{assump.spectrum} and~\ref{assump.infeasible} imply that $\btheta^\top \bH^{-1} \btheta > 0$ and $\btheta^\top \bp^{ur} - M > 0$. Since the weight vector $\btheta$ is nonnegative, it follows that consumption weakly increases in every market under the average-price regulation. Consequently, the equilibrium utility of each consumer weakly increases, implying that aggregate consumer surplus strictly increases.


\begin{proposition}\label{prop.average}
    The average-price regulation strictly increases consumer surplus.
\end{proposition}

Notably, Proposition~\ref{prop.average} does not depend on the particular specification of $\ba$, $\bc$, $\delta$, or the network $\mathbf{G}$. Hence, average-price regulation is robust in the sense that, whenever the unrestricted price is infeasible, any admissible choice of weights and bound strictly improves consumer welfare.

\begin{remark}\label{remark.average}
    Although consumption increases in all markets, prices need not decrease in every market. To see this, note that the vector $\bH^{-1}\btheta$ may contain negative elements.
    For example, consider a star network with three nodes, where node~1 is the center, that is, $g_{1j} = g_{j1} = 1$ for $j = 2,3$, and all other entries are zero. Let $\delta = 0.5$, which satisfies Assumption~\ref{assump.spectrum}, and let $\btheta = (0.8, 0.1, 0.1)^\top$. Then $\bH^{-1}\btheta = (0.7, -0.3, -0.3)^\top$, implying that prices in markets~2 and~3 increase under regulation. Nevertheless, the price reduction in market~1 increases its own demand, which in turn raises demand in markets~2 and~3 through spillover effects. As a result, consumption increases in all markets despite price increases in some of them.

    By contrast, without spillovers, that is when $\delta = 0$, prices decrease and consumption increases in all markets. This comparison shows that, in the presence of spillovers, the effect of price regulation on individual prices may be non-monotone, even though consumption and consumer surplus strictly increase.
\end{remark}

Although average-price regulation always raises consumer surplus, the magnitude of this improvement can vary substantially. Consequently, selecting a suboptimal weight may lead to a significant reduction in surplus gains. We therefore analyze the robustness of average-price regulation. For clarity, we focus on connected, unweighted networks, i.e., adjacency matrices $\mathbf{G}$ with entry 0 or 1. 

As implied by Theorem~\ref{thm.neutral}, when $\delta \to 1/\lambda_1$, efficiency is achieved if and only if $\btheta\propto \mathbf{w}_1$. Hence, a regulator has an incentive to select a weight vector that is closely aligned with $\mathbf{w}_1$. In practice, however, the regulator may not observe $\mathbf{w}_1$ perfectly and instead relies on an estimator whose correlation with $\mathbf{w}_1$ is $k<1$.\footnote{The correlation between two vectors is defined as $corr(\mathbf{z},\mathbf{z}') := \frac{\langle \mathbf{z} , \mathbf{z}' \rangle}{||\mathbf{z}||  ||\mathbf{z}'||}$.} We therefore assume that the regulator chooses among weight vectors with a fixed correlation $k$ with $\mathbf{w}_1$. To ensure comparability across instruments, we normalize $\btheta^\top\btheta=1$ and fix the tightness of the constraint via $\btheta^\top\bp^{ur}-M \equiv C>0$. The set of feasible regulatory parameters is therefore
\begin{equation*}
    \mathcal{F} := \{(\btheta,M) \in \mathbb{R}^n_{+} \times \mathbb{R}: \btheta^T \btheta \equiv 1, \btheta^T\bp^{ur}-M \equiv C, corr(\btheta, \mathbf{w}_1) = k \}.
\end{equation*}

The percentage increase in consumer surplus is $R_V(\bp^*(\delta))-1$. However, Theorem~\ref{thm.neutral} implies that $\lim_{\delta \rightarrow 1/\lambda_1}R_V(\bp^*(\delta))-1=0$ for any regulation in $\mathcal{F}$. Therefore, we consider the first-order approximation and define
\begin{equation*}
    I(\btheta,M,\mathbf{G}) := \lim\limits_{\delta \rightarrow 1/\lambda_1}\frac{R_V(\bp^*(\delta))-1}{1-\delta\lambda_1}.
\end{equation*}
Under this definition,
\begin{equation*}
    R_V(\bp^*(\delta))-1 = I(\btheta,M,\mathbf{G}) \times (1-\delta\lambda_1) + \mathcal{O} \left(\frac{1}{\lambda_1} -\delta  \right)^2.
\end{equation*}
Thus, when $\delta$ is large enough, the surplus gain from the regulation is fully determined by $I(\btheta,M,\mathbf{G})$.

\begin{proposition}\label{prop.averagesensitivity}
    The best-to-worst ratio of average-price regulation is given by
    \begin{equation*}
        \xi(\mathbf{G}) := \frac{\max_{(\btheta,M) \in \mathcal{F}}
        I(\btheta,M,\mathbf{G})}{\min_{(\btheta,M) \in \mathcal{F}}
        I(\btheta,M,\mathbf{G})}
        =
        \frac{\lambda_1-\lambda_n}{\lambda_1-\lambda_2}.
    \end{equation*}
\end{proposition}

The index $\xi(\mathbf{G})$ measures the robustness of average-price regulation given network $\mathbf{G}$. Note that a well-specified regulator with complete information chooses $(\btheta, M)$ to maximize the surplus improvement index $I(\btheta, M, \mathbf{G})$, attaining the maximal consumer surplus gain. By contrast, a misspecified regulator, for example one who holds an incorrect belief about the network structure, may select a suboptimal pair $(\btheta, M)$ and, in the worst case, attain the minimal surplus improvement. The index $\xi(\mathbf{G})$, defined as the ratio of the best-case to the worst-case value of $I(\btheta, M, \mathbf{G})$, captures the magnitude of this risk. A larger value of $\xi(\mathbf{G})$ indicates that, although average-price regulation continues to improve consumer surplus even under misspecification, the resulting gains may be substantially smaller than those achieved under the optimal choice of $(\btheta, M)$. 

\begin{example}\label{example.averagenetwork}
Since $\xi(\mathbf{G})$ is determined entirely by the spectrum of the network, we evaluate it for several canonical networks.\footnote{Closed-form spectral characterizations for these networks are provided in \cite{brouwer2011spectra}.}
    \begin{enumerate}
        \renewcommand{\labelenumi}{(\roman{enumi})}
        \item Complete networks: $\xi(\mathbf{G})=1$. In this case, average-price regulation exhibits perfect robustness. Symmetry ensures that every admissible weight vector yields exactly the same surplus improvement, so the regulator cannot make mistakes when choosing $\btheta$.
        
        \item Complete bipartite networks: $\xi(\mathbf{G}) = 2$. Here, average-price regulation has moderate sensitivity. Misallocating weights across the two partitions can reduce the surplus gain by half relative to the optimal allocation.
        
        \item Path/cycle networks: $\xi(\mathbf{G}) = \mathcal{O}(n^2)$. The index diverges as $n$ grows. These networks have large diameter and weak connectivity, which yields an asymptotically vanishing spectral gap. Hence, deviations from the optimal weights can generate very large reductions in surplus gains. Regulation in such environments requires considerable precision.
    \end{enumerate}
\end{example}

\begin{remark}
    While the average regulation always increases consumer surplus, we also investigate its impact on total surplus. We maintain the normalization of average weights, $\btheta^T \btheta \equiv 1$, and define the tightness of the regulation as $C := \btheta^T\bp^{ur}-M$. Using equations~(\ref{eq.paverage})-(\ref{eq.pi}), the change in total surplus is given by
    \begin{equation*}
        \Delta V+ \Delta \Pi = C \frac{\btheta^T \bH(\ba-\bp^{ur})}{1-\delta\btheta^T\mathbf{G} \btheta} + \frac{C^2}{(1-\delta\btheta^T\mathbf{G} \btheta)^2} \left(\delta\btheta^T\mathbf{G} \btheta - \frac{1}{2} \right).
    \end{equation*}
    Note that $\btheta^T\mathbf{G} \btheta \leq \lambda_1$, with equality holding if and only if $\btheta = \mathbf{w}_1$. Therefore, when spillovers or network connectivity are sufficiently large (specifically, $\delta \lambda_1 \geq \frac{1}{2}$), average price regulation always increases total surplus. Moreover, a strictly tighter constraint (i.e., a larger $C$) induces a greater welfare gain.

    In contrast, if spillovers are small or network connectivity is weak such that $\delta \lambda_1 < \frac{1}{2}$, the result is non-monotonic. There exists a threshold $\hat{C}$ such that for any $C > \hat{C}$, tightening the constraint reduces the surplus gain, and may even result in a lower total surplus compared to the unregulated benchmark.
\end{remark}

\subsection{Banning price discrimination}\label{sec.uniform}

There is a large literature on the welfare effects of third-degree price discrimination (see, for example, \citealp{schmalenseeOutputWelfareImplications1981,varianPriceDiscriminationSocial1985, armstrongWelfareEffectsPrice1991, aguirreMonopolyPriceDiscrimination2010}). These studies mainly examine settings in which markets are independent. In this section, we investigate when banning price discrimination can benefit consumers in the presence of large demand spillovers across markets.

For any $\mathbf{z} \in \mathbb{R}^n$, define the demeaned vector as $\tilde{\mathbf{z}} := \mathbf{z} - \frac{\langle \mathbf{z}, \mathbf{1}  \rangle}{n} \mathbf{1}$. We impose the following assumption throughout this section.

\begin{assumption}\label{assump.uniform}
    $\bc = \mathbf{0}$, $\tilde{\ba} \neq \mathbf{0}$ and $\mathbf{G}$ is not a regular graph.
\end{assumption}

The assumption of $\bc = \mathbf{0}$ is just a normalization by assuming that production costs are identical across markets.\footnote{For heterogeneous costs, the main results hold by replacing intrinsic values $\ba$ with $\bp^{ur}$, or equivalently, $(\ba+\bc)/2$, the cost-adjusted intrinsic values.} The second argument ensures that the unrestricted prices are not uniform across all consumers.\footnote{Note that $\tilde{\mathbf{a}}=\mathbf{0}$ if and only if $  \mathbf{a}$ is proportional to $ \mathbf{1}$. Then this assumption assures that $\bp^{ur} = \ba/2$ is not identical among all markets.} Otherwise, there is no scope for banning price discrimination. The third point implies that $\tilde{\mathbf{w}}_1 \neq \mathbf{0}$, which ensures sufficient heterogeneity in the network structure, particularly in terms of centrality.\footnote{The discussion of regular graph can be found in Appendix \ref{appsec.regular}.}

Define an index of the network structure:\footnote{When $\mathbf{G}$ is a regular graph, $\langle \mathbf{w}_i , \mathbf{1} \rangle \equiv 0$ for any $i \geq 2$ and thus $\mypsi(\mathbf{G}) \equiv \mathbf{0}$. In this case, this vector becomes ineffective for welfare analysis. This provides another reason for assuming that $\mathbf{G}$ is not regular. }
\begin{equation}\label{eq.psi}
    \mypsi(\mathbf{G}) := \lim\limits_{\delta \rightarrow 1/\lambda_1}
    [\mathbf{w}_1\mathbf{1}^T\mathbf{H}-\mathbf{H}\mathbf{1}\mathbf{w}_1^T]\mathbf{1}
    =
    \left(\sum_{i=2}^{n} \frac{\langle \mathbf{w}_i , \mathbf{1} \rangle^2}{1- \lambda_i/\lambda_1} \right) \mathbf{w}_1
        - \sum_{i=2}^{n} \frac{\langle \mathbf{w}_i , \mathbf{1} \rangle \langle \mathbf{w}_1 , \mathbf{1} \rangle}{1- \lambda_i/\lambda_1} \mathbf{w}_i. 
\end{equation}

\begin{proposition}\label{thm.uniform}
    When the regulator bans price discrimination:
    \begin{enumerate}
        \renewcommand{\labelenumi}{(\roman{enumi})}
        \item If $corr(\mypsi(\mathbf{G}),\ba)>0$, there exists $\bar{\delta} < 1/\lambda_1$ such that banning price discrimination harms the firm but benefits consumers for $\delta \in (\bar{\delta}, 1/\lambda_1)$.
        
        \item If $corr(\mypsi(\mathbf{G}),\ba)<0$, there exists $\bar{\bar{\delta}} < 1/\lambda_1$ such that banning price discrimination harms both the firm and consumers for $\delta \in (\bar{\bar{\delta}}, 1/\lambda_1)$.
    \end{enumerate}
\end{proposition}


Proposition~\ref{thm.uniform} indicates that whether banning price discrimination benefits consumers is determined by the correlation between the sufficient statistic of network structure, $\mypsi(\mathbf{G})$, and consumers' intrinsic willingness to pay, $\ba$. This correlation is highly parameter dependent. Specifically, if banning price discrimination benefits consumers under intrinsic value vector $\ba$, for $\tilde{\ba}' = - \tilde{\ba}$, which reverses the order of intrinsic values and keeps the variation unchanged, it can be shown that banning price regulation will harm consumers.\footnote{Indeed, one can verify that $\langle \mypsi,\mathbf{1} \rangle=0$. Also, $\tilde{\ba}' = - \tilde{\ba}$ is equivalent to $\ba + \ba' \propto \mathbf{1}$, which implies $\langle \mypsi, \ba'\rangle = -\langle \mypsi,\ba\rangle<0$. Then the correlation $corr(\mypsi, \ba')$ will also be negative.} This observation implies that the ``order'' of the intrinsic value is important in determining the welfare implication.

\begin{proposition}\label{prop.uniform}
    There exists $\epsilon > 0$ such that
    \begin{enumerate}
        \renewcommand{\labelenumi}{(\roman{enumi})}
        \item For any $\ba$ satisfying $corr(\tilde{\ba}, \tilde{\mathbf{w}}_1) > 1-\epsilon$, there exists $\bar{\delta} < 1/\lambda_1$ such that banning price discrimination harms the firm but benefits consumers for all $\delta \in (\bar{\delta}, 1/\lambda_1)$.
        
        \item For any $\ba$ satisfying $corr(\tilde{\ba}, \tilde{\mathbf{w}}_1) < -1+\epsilon$, there exists $\bar{\bar{\delta}} < 1/\lambda_1$ such that banning price discrimination harms both the firm and consumers for all $\delta \in (\bar{\bar{\delta}}, 1/\lambda_1)$.
    \end{enumerate}
\end{proposition} 

This proposition provides a sharper sufficient condition under which banning price discrimination is surplus improving. The intuition behind this correlation-based condition is that increasing consumer surplus requires larger price reductions for more central consumers. A higher correlation between demeaned intrinsic values and eigenvector centrality indicates that central players are charged excessively high prices under discrimination. Consequently, banning price discrimination reduces prices for central consumers. Although prices for less central consumers increase, spillover effects dominate, so that aggregate consumer surplus still increases. By contrast, when the correlation is sufficiently negative, banning price discrimination raises prices for central consumers, which in turn reduces consumer surplus.



Furthermore, we next show that when the network structure exhibits sufficient symmetry, determining whether banning price discrimination benefits consumers becomes even simpler.
\begin{definition}\label{def.twotype}
    An undirected graph $\mathbf{G} = (V,E)$ is said to have two types of nodes if the vertex set $V$ can be partitioned into two disjoint subsets, $V = V_1 \cup V_2$ with $V_1 \cap V_2 = \emptyset$, such that:
    \begin{enumerate}
        \renewcommand{\labelenumi}{(\roman{enumi})}
        \item For any graph automorphism $\sigma \in Aut(\mathbf{G})$, $\sigma(V_1)=V_1$, $\sigma(V_2)=V_2$.

        \item For any $i,j \in V_k$, $k=1,2$, there exists a graph automorphism $\sigma \in Aut(\mathbf{G})$ such that $\sigma(i)=j$ and $\sigma(j)=i$.
    \end{enumerate}
\end{definition}

This definition implies that any two nodes within the same subset are structurally symmetric in the network. To illustrate, we provide two representative examples: the core-periphery network like Figure \ref{fig.cp} and the complete bipartite network like Figure \ref{fig.cb}. In the core-periphery network, the core nodes (red triangle) are symmetric and form the set $V_1 = V_{\text{core}} = \{1,2,3\}$, while the periphery nodes (blue circle) are also symmetric and form $V_2 = V_{\text{periphery}} = \{4,5,6,7,8,9\}$. In the complete bipartite graph, the two independent sets naturally correspond to $V_1$ and $V_2$. The core-periphery structure captures the phenomenon known as the Law of the Few \citep{galeottiLawFew2010}, in which a large number of individuals (periphery nodes) interact with a small, influential subset (core nodes). In contrast, the complete bipartite graph is a stylized representation of two-sided markets \citep{rochetTwosidedMarketsProgress2006, armstrongCompetitionTwosidedMarkets2006}.
\begin{figure}[ht]
    \centering
    \begin{subfigure}{0.4\textwidth}
        \centering
        \resizebox{0.8\textwidth}{!}{%
            \begin{tikzpicture}
                \pgfmathsetmacro{\sqrtthree}{sqrt(3)}
                
                \coordinate (p1) at (-\sqrtthree,1);
                \coordinate (p2) at (\sqrtthree,1);
                \coordinate (p3) at (0,-2);
                \coordinate (p4) at ($2*(p1)-(p2)$);
                \coordinate (p5) at ($2*(p1)-(p3)$);
                \coordinate (p6) at ($2*(p2)-(p3)$);
                \coordinate (p7) at ($2*(p2)-(p1)$);
                \coordinate (p8) at ($2*(p3)-(p1)$);
                \coordinate (p9) at ($2*(p3)-(p2)$);
                
                \draw[thick]  (p1) -- (p2);
                \draw[thick]  (p1) -- (p3);
                \draw[thick]  (p3) -- (p2);
                \draw[thick]  (p4) -- (p1);
                \draw[thick]  (p5) -- (p1);
                \draw[thick]  (p6) -- (p2);
                \draw[thick]  (p7) -- (p2);
                \draw[thick]  (p8) -- (p3);
                \draw[thick]  (p9) -- (p3);
                
                \fill[red] ($(p1)+(0,0.3)$)--($(p1)+(-0.15*\sqrtthree,-0.15)$)--($(p1)+(0.15*\sqrtthree,-0.15)$)--cycle;
                \fill[red] ($(p2)+(0,0.3)$)--($(p2)+(-0.15*\sqrtthree,-0.15)$)--($(p2)+(0.15*\sqrtthree,-0.15)$)--cycle;
                \fill[red] ($(p3)+(0,0.3)$)--($(p3)+(-0.15*\sqrtthree,-0.15)$)--($(p3)+(0.15*\sqrtthree,-0.15)$)--cycle;

                \draw[blue,fill=blue] (p4) circle (4pt);
                \draw[blue,fill=blue] (p5) circle (4pt);
                \draw[blue,fill=blue] (p6) circle (4pt);
                \draw[blue,fill=blue] (p7) circle (4pt);
                \draw[blue,fill=blue] (p8) circle (4pt);
                \draw[blue,fill=blue] (p9) circle (4pt);

                \node[font=\bfseries\Large] at ($(p1)+(-0.2,0)$) [above left] {1};
                \node[font=\bfseries\Large] at ($(p4)+(-0.1,0)$) [left] {4};
                \node[font=\bfseries\Large] at (p5) [above left] {5};
                \node[font=\bfseries\Large] at ($(p2)+(0.2,0)$) [above right] {2};
                \node[font=\bfseries\Large] at (p6) [above right] {6};
                \node[font=\bfseries\Large] at ($(p7)+(0.1,0)$) [right] {7};
                \node[font=\bfseries\Large] at ($(p3)+(0,-0.2)$) [below] {3};
                \node[font=\bfseries\Large] at (p8) [below right] {8};
                \node[font=\bfseries\Large] at (p9) [below left] {9};
                
            \end{tikzpicture}
        }%
        \caption{Core-periphery network}
        \label{fig.cp}
    \end{subfigure}
    \hfill
    \begin{subfigure}{0.4\textwidth}
        \centering
        \resizebox{0.8\textwidth}{!}{%
            \begin{tikzpicture}
                \pgfmathsetmacro{\sqrtthree}{sqrt(3)}
                
                \coordinate (p2) at (-0.5,3);
                \coordinate (p3) at (0.5,3);
                \coordinate (p5) at (-2,0);
                \coordinate (p6) at (-1,0);
                \coordinate (p7) at (0,0);
                \coordinate (p8) at (1,0);
                \coordinate (p9) at (2,0);
                
                
                \draw  (p2) -- (p5);
                \draw  (p2) -- (p6);
                \draw  (p2) -- (p7);
                \draw  (p2) -- (p8);
                \draw  (p2) -- (p9);
                
                \draw  (p3) -- (p5);
                \draw  (p3) -- (p6);
                \draw  (p3) -- (p7);
                \draw  (p3) -- (p8);
                \draw  (p3) -- (p9);
                
                
                \fill[red] ($(p2)+(0,0.15)$)--($(p2)+(-0.075*\sqrtthree,-0.075)$)--($(p2)+(0.075*\sqrtthree,-0.075)$)--cycle;
                \fill[red] ($(p3)+(0,0.15)$)--($(p3)+(-0.075*\sqrtthree,-0.075)$)--($(p3)+(0.075*\sqrtthree,-0.075)$)--cycle;
                \draw[blue,fill=blue] (p5) circle (2pt);
                \draw[blue,fill=blue] (p6) circle (2pt);
                \draw[blue,fill=blue] (p7) circle (2pt);
                \draw[blue,fill=blue] (p8) circle (2pt);
                \draw[blue,fill=blue] (p9) circle (2pt);

                \node[font=\bfseries\small] at ($(p2)+(0,0.1)$) [above] {1};
                \node[font=\bfseries\small] at ($(p3)+(0,0.1)$) [above] {2};
                \node[font=\bfseries\small] at (p5) [below] {3};
                \node[font=\bfseries\small] at (p6) [below] {4};
                \node[font=\bfseries\small] at (p7) [below] {5};
                \node[font=\bfseries\small] at (p8) [below] {6};
                \node[font=\bfseries\small] at (p9) [below] {7};
                
            \end{tikzpicture}
        }%
        \caption{Complete bipartite network}
        \label{fig.cb}
    \end{subfigure}
    \caption{Two examples of network with two types of nodes}
    \label{fig.twotype}
\end{figure}

Given this network symmetry, we can show that the eigenvector centrality is identical within each type, that is, $\mathbf{w}_1(i) = \mathbf{w}_1(j) \equiv \mathbf{w}_1^{(k)}$ for any $i, j \in V_k$ ($k = 1, 2$). Without loss of generality, we assume $\mathbf{w}_1^{(1)} > \mathbf{w}_1^{(2)}$.\footnote{Equality cannot hold since $\mathbf{G}$ is not regular.} The average intrinsic value for each type $k$ is defined as $\bar{a}(V_k) := \frac{1}{|V_k|}\sum_{i \in V_k} a_i$.

\begin{proposition}\label{prop.twotype}
    Let $\mathbf{G} = (V, E)$ be a non-regular graph with two types of nodes.
    \begin{enumerate}
        \renewcommand{\labelenumi}{(\roman{enumi})}
        \item If $\bar{a}(V_1) > \bar{a}(V_2)$, there exists $\bar{\delta} < 1/\lambda_1$ such that banning price discrimination harms the firm but benefits consumers for $\delta \in (\bar{\delta}, 1/\lambda_1)$.
        
        \item If $\bar{a}(V_1) < \bar{a}(V_2)$, there exists $\bar{\bar{\delta}} < 1/\lambda_1$ such that banning price discrimination harms both the firm and consumers for $\delta \in (\bar{\bar{\delta}}, 1/\lambda_1)$.
    \end{enumerate}
\end{proposition}

With such network symmetry, the intuition that ``more central consumers should receive higher discounts'' reduces to a simple comparison of the average intrinsic values across the two types. If the more central type has a higher average intrinsic value, or equivalently, a higher average price without regulation, then banning price discrimination benefits consumers if the spillovers are strong enough. Otherwise, the ban tends to reduce welfare for both consumers and the firm. Importantly, these welfare implications depend only on the average intrinsic values across types rather than on their within-type heterogeneity. In other words, redistributing values among consumers of the same type does not affect the outcome as long as the average remains unchanged.

\begin{remark}\label{remark.uniform}
    The previous analysis focuses on the case where spillovers are sufficiently strong. On the other hand, the welfare implications of uniform pricing without demand spillovers have been extensively studied \citep{schmalenseeOutputWelfareImplications1981, varianPriceDiscriminationSocial1985, aguirreMonopolyPriceDiscrimination2010}. Here, we revisit these results as a special case of our model with $\delta = 0$: under Assumption \ref{assump.uniform},
    \begin{equation*}
        \bp^0 = \frac{\mathbf{1}^T\bp^{ur}}{n} \mathbf{1}
        = \frac{\overline{a}}{2} \mathbf{1},
    \end{equation*}
    where $\overline{a} = \mathbf{1}^T \ba/n$ is the average intrinsic value. Since the average price remains unchanged at $\overline{a}/2$, total consumption does not change, implying that at $\delta=0$,  banning price discrimination strictly increases consumer surplus (see, for instance, \citealp{varianPriceDiscriminationSocial1985}). Moreover, the total surplus also increases.\footnote{Indeed, at $\delta=0$, the change in consumer surplus is
    \begin{align*}
        V(\bp^{0}) - V(\bp^{ur})
        =& \frac{1}{2}||\ba-\bp^{0}||^2 - \frac{1}{2}||\ba-\bp^{ur}||^2
        = \frac{3}{8} (||\ba||^2 - n\overline{a}^2)
        = \frac{3(n-1)}{8} \text{Var}[\ba] >0.
    \end{align*}
    Similarly, the change in total surplus is given by $\frac{1(n-1)}{8} \text{Var}[\ba] >0$.
    }
    By continuity, there exists $\hat{\delta} >0$ such that banning price discrimination harms the firm but benefits consumers for $\delta \in [0,\hat{\delta})$.
\end{remark}

Finally, we provide a numerical example to illustrate the main results presented before.
\begin{example}\label{example.uniform}
    Consider a core-periphery network (as in Figure \ref{fig.cp}) with $n = 9$, $\bc = \mathbf{0}$ and $a_i = \theta_k$ if $i \in V_k$ for $V_1 = \{1,2,3\}$ representing the \emph{core} and $V_2 = \{4,5,6,7,8,9\}$ representing the \emph{periphery}. Therefore, the core nodes and the periphery nodes have identical intrinsic values respectively. We set $(\theta_1, \theta_2) = (20,10)$ for case (i) and $(\theta_1, \theta_2) = (10,20)$ for case (ii).
    
    \begin{figure}[ht]
        \centering
        \par
        \subfloat[Case (i): $(\theta_1, \theta_2) = (20,10)$]{
            \includegraphics[width=0.45\textwidth]{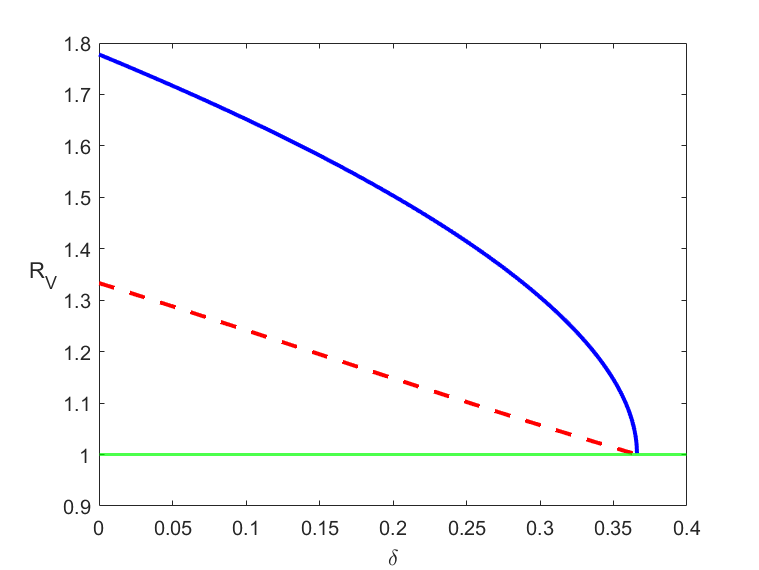}
            \label{fig.uniform}
        }\hfill
        \subfloat[Case (ii): $(\theta_1, \theta_2) = (10,20)$]{
            \includegraphics[width=0.45\textwidth]{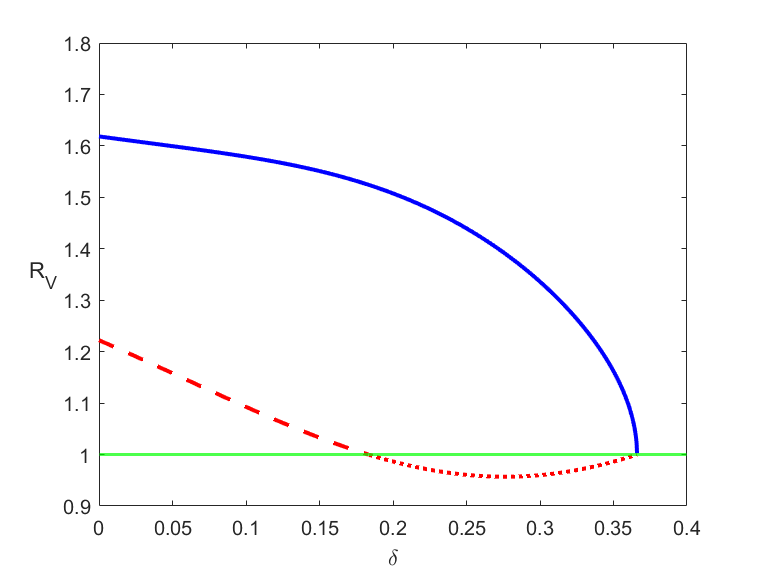}
            \label{fig.uniform2}
        }
        \caption{Changes of $R_V^*$ (\textcolor{red}{Red curve}) and $R_V^+$ (\textcolor{blue}{Blue curve}) as $\delta$ varies from $0$ to $1/\lambda_1$  }
        \label{fig.example}
    \end{figure}
    
    In this example, $\lambda_1 = 1+\sqrt{3}$ and thus $1/\lambda_1 \approx 0.366$. As shown in Figure \ref{fig.example}, the blue curve lies above the red curve for all $\delta < 0.366$, consistent with Theorem \ref{thm.pareto}, which states that such price regulation is Pareto inefficient for any fixed $\delta < 0.366$. When $\delta \rightarrow 0.366$, the surplus ratio approaches 1. This aligns with Theorem \ref{thm.neutral}, which asserts that uniform pricing is neutral. Moreover, the two curves converge at $\delta=0.366$, as implied by Lemma~\ref{lemma.gap}. These align with results in previous sections.
    
    Also note that $\mathbf{w}_1^{(1)} = 0.513 > 0.188 = \mathbf{w}_1^{(2)}$, and that $\overline{a}(V_k) = \theta_k$ for $k = 1, 2$. Then, according to Proposition~\ref{prop.twotype}, if the regulator prohibits price discrimination, consumers benefit in case (i) ($\theta_1 > \theta_2$), but are worse off in case (ii) ($\theta_1 < \theta_2$), provided that $\delta$ is sufficiently large. For small $\delta$, banning price discrimination always benefits consumers as stated in Remark \ref{remark.uniform}.

    In Appendix \ref{appsec.pdiff}, we extend this analysis by conducting numerical simulations for price difference regulation, which nests banning price discrimination as a special case where the permitted difference is zero. By varying the allowable price gap, we examine how the stringency of regulation impacts consumer surplus. A parallel analysis for complete bipartite networks is provided in Appendix~\ref{appsec.bipartite}.
\end{example}


\section{Conclusion}\label{sec.conclusion}

This paper studies the welfare effect of price regulations for a monopolist operating across multiple markets with demand spillovers. We show that achieving efficiency requires price reduction proportional to the K-B centralities. However, at a fixed spillover intensity, commonly used price regulations generically fail to implement efficient outcomes. Moreover, as the intensity of spillovers increases, these price regulations become generically welfare neutral in the limit, in the sense that both the relative profit loss and the relative change in consumer surplus converge to zero. Despite these negative results, there exist specific parameterizations of price regulation that achieve efficiency. In the fixed-spillover case, such efficiency requires complete information about consumers’ intrinsic values, the firm’s production costs, and the network structure. In contrast, when spillovers are strong, the information requirements are substantially reduced, as efficiency can be attained using only knowledge of the network’s eigenvector centrality and unrestricted price, which is more feasible in empirical applications.

We further analyze whether some commonly used price regulations can benefit consumers. We find that the solution of Ramsey pricing problem is efficient. Moreover, we show that average-price regulation robustly increases consumer surplus, but the dispersion of surplus gains across different average-price weights depends on the network structure. We also find that banning price discrimination can raise consumer surplus under strong spillovers when more central consumers have higher willingness to pay. In networks with two types of nodes such as core–periphery or complete bipartite networks, the welfare implications simplify to a comparison of average intrinsic values between more central and less central groups.

Finally, several topics remain for future research. One extension is to study price regulation in competitive environments, where strategic interactions among firms may alter the welfare implications of regulation. Another avenue is to analyze the design and robustness of price regulation under incomplete information, where regulators may rely on limited knowledge of consumers' attributes, firm's cost structure, and network structure.


\newpage
\appendix 


\renewcommand{\thetheorem}{\thesection.\arabic{theorem}}
\renewcommand{\theproposition}{\thesection.\arabic{proposition}}
\renewcommand{\thelemma}{\thesection.\arabic{lemma}}
\renewcommand{\thecorollary}{\thesection.\arabic{corollary}}
\renewcommand{\theclaim}{\thesection.\arabic{claim}}
\renewcommand{\theassumption}{\thesection.\arabic{assumption}}
\renewcommand{\thedefinition}{\thesection.\arabic{definition}}
\renewcommand{\theremark}{\thesection.\arabic{remark}}
\renewcommand{\theequation}{\thesection.\arabic{equation}}
\renewcommand{\thefigure}{\thesection.\arabic{figure}}
\counterwithin{theorem}{section}
\counterwithin{proposition}{section}
\counterwithin{lemma}{section}
\counterwithin{corollary}{section}
\counterwithin{claim}{section}
\counterwithin{assumption}{section}
\counterwithin{definition}{section}
\counterwithin{remark}{section}
\counterwithin{equation}{section}
\counterwithin{figure}{section}

\begin{center}
    {\bf\huge Appendix}
\end{center}

\singlespacing

\section{Proofs}\label{appsec.proof}

\subsection{Proof of Lemma \ref{lemma.equi}}\label{appsec.prooflemmaequi}
Under Assumption \ref{assump.spectrum}, $\mathbb{R}^n$ with the $\mathbf{H}$-inner product is a Hilbert space. Moreover, $K$ is a non-empty, closed and convex set under Assumption \ref{assump.k}. Then the solution of the minimization problem
\begin{equation*}
    \min_{\bp \in K} \left\lVert\bp-\frac{\ba+\bc}{2}  \right\rVert^2_{\mathbf{H}},
\end{equation*}
exists and is unique, which is the projection of $\bp^{ur}$ onto $K$ under the $\mathbf{H}$-inner product.\hfill{$\square$}


\subsection{Proof of Lemma \ref{lemma.pfamily}}\label{appsec.proofpfamily}
We first show that a price $\pfam$ is Pareto efficient if and only if it solves problem~(\ref{eq.intervention}). Suppose $\pfam$ is efficient. Consider the regulator's problem with $\tau = R_{\Pi}(\pfam)$. For any $R_{\Pi}(\bp) \geq R_{\Pi}(\pfam)$, efficiency implies $R_{V}(\pfam) \geq R_{V}(\bp)$. Hence, $\pfam$ solves problem~(\ref{eq.intervention}) for this particular $\tau$. Conversely, suppose $\pfam$ is a solution. As we will show below, the solution to problem~(\ref{eq.intervention}) is unique. Therefore, for any $\bp$ satisfying $R_{\Pi}(\bp) \geq R_{\Pi}(\pfam) \geq \tau$, the uniqueness of the maximizer implies $R_{V}(\pfam) > R_{V}(\bp)$. Hence, $\pfam$ must be Pareto efficient.

Then we adopt the spectral decomposition approach to solve the problem. For any vector $\mathbf{z}$, we define $\hat{z}_i := \langle \mathbf{w}_i, \mathbf{z} \rangle$. Let $\phi_i := \frac{1-\delta\lambda_1}{1-\delta\lambda_i}$. Since $\mathbf{G}$ is connected, it follows that $0 \leq \phi_n \leq \ldots \leq \phi_2 < \phi_1 = 1$. To simplify the notation, we define $\mathbf{y}:=\bp-\bp^{ur}$. Using spectral decomposition, we can reformulate problem (\ref{eq.intervention}) as the following equivalent problem:
\begin{equation}\label{appeq.intervention2}
    \begin{gathered}
        \max \sum_{i=1}^{n} \phi_i^2 (\frac{\hat{a}_i-\hat{c}_i}{2}-\hat{y}_i )^2\\
        s.t. \ \sum_{i=1}^{n} \phi_i \hat{y}_i^2 \leq (1-\tau) \sum_{i=1}^{n} \phi_i(\frac{\hat{a}_i - \hat{c}_i}{2})^2.
    \end{gathered} 
\end{equation}
The Lagrangian is defined as 
\begin{equation*}
    L(\mathbf{\hat{y}},\mu) = \sum_{i=1}^{n} \phi_i^2 (\frac{\hat{a}_i-\hat{c}_i}{2}-\hat{y}_i )^2 + \mu \left[(1-\tau) \sum_{i=1}^{n} \phi_i(\frac{\hat{a}_i - \hat{c}_i}{2})^2- \sum_{i=1}^{n} \phi_i \hat{y}_i^2\right].
\end{equation*}

Suppose $\mathbf{\hat{y}}$ is the solution, we first show that the constraint must be binding. Otherwise, since $\tau \in [0,1]$, there exists some $i$ such that $(\hat{y}_i)^2 < (\frac{\hat{a}_i - \hat{c}_i}{2})^2$. Consider the perturbed solution defined by $\hat{y}_i^{**} = \hat{y}_i - \epsilon sgn(\frac{\hat{a}_i-\hat{c}_i}{2})$ and $\hat{y}_j^{**} = \hat{y}_j$ for $j \neq i$. For $\epsilon$ small enough, $\mathbf{\hat{y}}^{**}$ remains feasible and
\begin{align*}
    (\frac{\hat{a}_i-\hat{c}_i}{2}-\hat{y}_i^{**} )^2 =&
    (\frac{\hat{a}_i-\hat{c}_i}{2}-\hat{y}_i +\epsilon sgn(\frac{\hat{a}_i-\hat{c}_i}{2})  )^2\\
    =& (\frac{\hat{a}_i-\hat{c}_i}{2}-\hat{y}_i)^2 + \epsilon^2 + 2  \epsilon \lvert \frac{\hat{a}_i-\hat{c}_i}{2}-\hat{y}_i \rvert\\
    > & (\frac{\hat{a}_i-\hat{c}_i}{2}-\hat{y}_i)^2.
\end{align*}
The second equality follows from the fact that $sgn(\frac{\hat{a}_i-\hat{c}_i}{2}-\hat{y}_i)=sgn(\frac{\hat{a}_i-\hat{c}_i}{2})$ since $(\hat{y}_i)^2 < (\frac{\hat{a}_i - \hat{c}_i}{2})^2$. Thus, $\mathbf{\hat{y}}^{**}$ yields a strictly higher value for the objective function, contradicting the optimality of $\mathbf{\hat{y}}$. Therefore, the constraint must be binding, and the solution satisfies the following KKT conditions:
\begin{align}
	&2 \phi_i^2 (\hat{y}_i - \frac{\hat{a}_i-\hat{c}_i}{2})-2\mu \phi_i \hat{y}_i = 0, \ i=1,\ldots,n  \label{appeq.foc}\\\
	&(1-\tau) \sum_{i=1}^{n} \phi_i(\frac{\hat{a}_i - \hat{c}_i}{2})^2- \sum_{i=1}^{n} \phi_i \hat{y}_i^2=0. \label{appeq.bind}
\end{align}
Equation (\ref{appeq.foc}) can be rewritten as 
\begin{equation}\label{appeq.yhat}
	\hat{y}_i = \frac{ \phi_i}{ \phi_i - \mu}\frac{\hat{a}_i - \hat{c}_i}{2}, \ i=1,\ldots,n.
\end{equation}

Note that the sign of $\hat{y}_i$ should the same as $- \frac{\hat{a}_i-\hat{c}_i}{2}$. This holds because the objective is to maximize the distance between $\hat{y}_i$ and $\frac{\hat{a}_i-\hat{c}_i}{2}$. As a consequence, we have $\mu > \phi_i$ for all $i$, that is, $\mu > \max_{i} \phi_i = \phi_1 = 1$. Now we pin down the value of $\mu$. Define
\begin{equation*}
	f(\mu) := \sum_{i=1}^{n} \phi_i \hat{y}_i^2 = \sum_{i=1}^{n} \phi_i(\frac{\hat{a}_i - \hat{c}_i}{2})^2 (\frac{\phi_i}{\mu - \phi_i})^2,
\end{equation*}
which is decreasing in $\mu > \phi_i = 1$ and the value ranges from $0$ to $\infty$. Then for any $\tau \in [0,1]$, there is a unique solution in $(1,\infty]$ for equation (\ref{appeq.bind}).\footnote{Here we allow $\mu=\infty$ which is the solution for $\tau=1$.} 

Define $\eta = (2-2\delta \lambda_1)/\mu \in [0,2-2\delta \lambda_1)$, then equation (\ref{appeq.yhat}) can be rewritten as
\begin{equation*}
	\hat{y}_i = - \frac{\eta}{2-2\delta\lambda_i - \eta} \frac{\hat{a}_i - \hat{c}_i}{2}, \ i=1,\ldots,n.
\end{equation*}
which can be expressed in matrix form for $\bp$:
\begin{equation}\label{appeq.p}
	\pfam(\eta) = 
	\frac{\ba+\bc}{2} - 
	\frac{\eta}{2-\eta} [\mathbf{I}_n-\frac{2\delta}{2-\eta}\mathbf{G}]^{-1}\frac{\ba-\bc}{2}
\end{equation}
Then equation (\ref{appeq.bind}) is equivalent to
\begin{equation}\label{appeq.bind2}
	R_{\Pi}(\pfam(\eta)) = \tau,
\end{equation}
where the solution is unique for $\tau \in [0,1]$ and $\eta \in [0,2-2\delta \lambda_1)$.
\hfill{$\square$}


\subsection{Proof of Lemma \ref{lemma.pareto}}\label{appsec.prooflemmapareto}
The firm's profit can be written as
\begin{equation*}
    (\bp-\bc)^T \mathbf{x}(\bp)
    =(\bp-\bc)^T \mathbf{H}(\ba-\bp),
\end{equation*}
which is concave in $\bp$ since $\bH$ is positive definite. Consequently, the profit ratio $R_{\Pi}(\bp)$ is also concave. For a differentiable concave function on a convex set $K$, a price $\bp^*$ is a maximizer if and only if it satisfies the first order optimality condition:
\begin{equation}\label{eq.vi}
    \langle \nabla R_{\Pi}(\bp^*), \bp-\bp^* \rangle \leq 0 \quad  \forall \bp \in K,
\end{equation}
as shown in Section 4.2.3 of \cite{boyd2004convex}.

Therefore, if an efficient price $\pfam$ lies in $K$ and satisfies (\ref{eq.vi}), then it will be the equilibrium price and the regulation is indeed efficient. Conversely, if a regulation is efficient, then firm's optimal price $\bp^*$ must satisfies (\ref{eq.vi}) and coincide with some efficient price $\pfam$.
\hfill$\square$


\subsection{Proof of Theorem \ref{thm.pareto}}\label{appsec.proofthmpareto}

For any $\pfam(\eta)$, we have
\begin{equation*}
    \nabla R_{\Pi}(\pfam(\eta)) \propto \nabla \Pi(\pfam(\eta))
    = -\bH(\pfam(\eta)-\bp^{ur})
    =\frac{\eta}{2-\eta} \bH [\mathbf{I}_n-\frac{2\delta}{2-\eta}\mathbf{G}]^{-1}\frac{\ba-\bc}{2}.
\end{equation*}
In Appendix~\ref{appsec.proofpfamily}, we have shown that $\eta \in [0,2-2\delta \lambda_1)$, which implies $1-\frac{2\delta}{2-\eta}\lambda_1 >0$. By Theorem III of \cite{debreuNonnegativeSquareMatrices1953}, all entries of $\bH = [\mathbf{I}_n - \delta \mathbf{G}]^{-1}$ and $[\mathbf{I}_n-\frac{2\delta}{2-\eta}\mathbf{G}]^{-1}$ are nonnegative. Then it follows that $\nabla R_{\Pi}(\pfam(\eta)) > \mathbf{0}$, that is, all entries of $\nabla R_{\Pi}(\pfam(\eta))$ are nonnegative and $\nabla R_{\Pi}(\pfam(\eta)) \neq \mathbf{0}$. This holds for any $\pfam \in \mathcal{P}$.

\vspace{12pt}

\noindent\textbf{For $\myd_{\boldsymbol{\Delta}}$.}  Fix $\hat{\bp} \in \myd_{\boldsymbol{\Delta}}$, for any $\pfam \in \mathcal{P}$, there exists some $t \in \mathbb{R}$ such that 
\begin{equation*}
    \langle \nabla R_{\Pi}(\pfam), \hat{\bp}+t \mathbf{1} \rangle
    = \langle \nabla R_{\Pi}(\pfam), \hat{\bp} \rangle +t  \langle \nabla R_{\Pi}(\pfam), \mathbf{1} \rangle > \langle \nabla R_{\Pi}(\pfam), \pfam \rangle.
\end{equation*}
The strict inequality comes from the fact that $\nabla R_{\Pi}(\pfam) > \mathbf{0}$, implying $\langle \nabla R_{\Pi}(\pfam), \mathbf{1} \rangle > 0$. However, $\hat{\bp}+t \mathbf{1}$ also belongs to $\myd_{\boldsymbol{\Delta}}$ since adding a common constant $t$ does not affect price differences across markets. This violates the second condition in Lemma \ref{lemma.pareto}. Therefore, $\myd_{\boldsymbol{\Delta}}$ is always Pareto inefficient.

\vspace{12pt}

\noindent\textbf{For $\myb(\underline{\bp},\overline{\bp})$.} Suppose $\exists \pfam \in \mathcal{P}$ such that $\overline{\bp} = \pfam$. Then, for any $\bp \in \myb(\underline{\bp},\overline{\bp})$,
\begin{equation*}
    \langle \nabla R_{\Pi}(\pfam), \bp \rangle
    \leq \langle \nabla R_{\Pi}(\pfam), \overline{\bp} \rangle
    =  \langle \nabla R_{\Pi}(\pfam), \pfam \rangle,
\end{equation*}
where the inequality follows from $\nabla R_{\Pi}(\pfam) > \mathbf{0}$.
By Lemma \ref{lemma.pareto}, this implies that $\myb(\underline{\bp},\overline{\bp})$ is Pareto efficient.

Conversely, if $\myb(\underline{\bp},\overline{\bp})$ is Pareto efficient, by Lemma \ref{lemma.pareto}, there exists some $\pfam \in \mathcal{P}$ such that $\pfam \in K$ and 
\[
\langle \nabla R_{\Pi}(\pfam), \overline{\bp} \rangle \leq  \langle \nabla R_{\Pi}(\pfam), \pfam \rangle.
\]
Since $\pfam \in \myb(\underline{\bp},\overline{\bp})$, then $\pfam \leq \overline{\bp}$. Hence, the equality must hold and we have $\pfam = \overline{\bp}$.

In summary, $\myb(\underline{\bp},\overline{\bp})$ is Pareto efficient if and only if there exists some $\pfam \in \mathcal{P}$ such that $\overline{\bp}=\pfam$, i.e. $\overline{\bp} \in \mathcal{P}$. However, 
the set of Pareto efficient prices $\pfam$ forms a one-dimensional smooth curve in $\mathbb{R}^n$. Consequently, the equality $\overline{\bp} = \pfam$ holds only for a measure-zero set of parameter values. Therefore, this condition is generically violated, and $\myb(\underline{\bp},\overline{\bp})$ is generically Pareto inefficient.

\vspace{12pt}

\noindent\textbf{For $\mye_{\btheta, M}$.}
If $\btheta \propto \nabla R_{\Pi}(\pfam)$ and $\langle \btheta, \pfam \rangle = M$. Then $\pfam \in \mye_{\btheta, M}$ and
\begin{equation*}
    \langle \btheta,\bp \rangle \leq M \implies 
    \langle \btheta,\bp \rangle \leq \langle \btheta, \pfam \rangle \implies 
    \langle \btheta,\bp-\pfam \rangle \leq 0 \implies 
    \langle \nabla R_{\Pi}(\pfam),\bp-\pfam \rangle \leq 0
\end{equation*}
By Lemma~\ref{lemma.pareto}, $\mye_{\btheta, M}$ is Pareto efficient.


Conversely, suppose $\mye_{\btheta, M}$ is Pareto efficient, then there exists $\pfam \in \mathcal{P}$ such that $\pfam \in \mye_{\btheta, M}$. If $\btheta \not\propto \nabla R_{\Pi}(\pfam)$, then there exists some $\hat{\mathbf{z}} \in \mathbb{R}^n$ such that $\langle \btheta,  \hat{\mathbf{z}} \rangle = 0$ but $\langle\nabla R_{\Pi}(\pfam),  \hat{\mathbf{z}} \rangle \neq 0$. Consequently, there exists some $t \in \mathbb{R}$ such that 
\begin{align*}
    & \langle \btheta,  \pfam + t\hat{\mathbf{z}} \rangle = \langle \btheta,  \pfam \rangle + t \langle \btheta, \hat{\mathbf{z}} \rangle = \langle \btheta,  \pfam \rangle \leq M,\\
    & \langle \nabla R_{\Pi}(\pfam),  \pfam + t\hat{\mathbf{z}} \rangle = \langle \nabla R_{\Pi}(\pfam),  \pfam \rangle + t \langle \nabla R_{\Pi}(\pfam), \hat{\mathbf{z}} \rangle > \langle \nabla R_{\Pi}(\pfam),  \pfam \rangle.
\end{align*}
The first inequality implies $\pfam + t\hat{\mathbf{z}} \in \mye_{\btheta, M}$ but the second one implies that $\pfam + t\hat{\mathbf{z}}$ does not satisfy the second condition in Lemma~\ref{lemma.pareto}. This leads to a contradiction to the efficiency. Hence, we must have $\btheta = k \nabla R_{\Pi}(\pfam)$ for some $k >0$ as both vectors are positive. Moreover, fix $\bp \in \mye_{\btheta, M}$ such that $\langle \btheta,  \bp \rangle = M$, we have
\begin{align*}
    M = \langle \btheta,  \bp \rangle =  k \langle  \nabla R_{\Pi}(\pfam),  \bp \rangle \leq k \langle \nabla R_{\Pi}(\pfam),  \pfam \rangle = \langle \btheta,  \pfam \rangle
\end{align*}
where the inequality follows the efficiency of $\mye_{\btheta, M}$ and Lemma~\ref{lemma.pareto}. Moreover, $\pfam \in \mye_{\btheta, M}$ implies $\langle \btheta,  \pfam \rangle \leq M$. Then these two inequalities imply that  $\langle \btheta,  \pfam \rangle = M$.

In summary, $\mye_{\btheta, M}$ is Pareto efficient if and only if $\exists \pfam \in \mathcal{P}$ such that $\btheta \propto \nabla R_{\Pi}(\pfam)$ and $\langle \btheta, \pfam \rangle = M$. This requires that their normal vectors must be proportional and the associated scalar terms must scale accordingly. That is, there must exist some $\exists k > 0$ such that
\begin{equation*}
    (\btheta, M) = k \left(\nabla R_{\Pi}(\pfam),  \langle \nabla R_{\Pi}(\pfam),  \pfam \rangle  \right).
\end{equation*}
Since the right-hand side spans only a one-dimensional curve in $\mathbb{R}^n_+ \times \mathbb{R}$, this proportionality condition holds only on a measure-zero subset. Therefore, $\mye_{\btheta,M}$ is generically Pareto inefficient.
\hfill$\square$


\subsection{Proof of Lemma \ref{lemma.plimit}}\label{appsec.prooflemmaplimit}
\textbf{Part I.} The consumer surplus and firm's profit ratios can be written as:
\begin{align*}
    &R_V(\bp) = \frac{(\ba-\bp)^T \bH^2 (\ba-\bp)}{(\ba-\bp^{ur})^T \bH^2 (\ba-\bp^{ur})} = \frac{(\ba-\bp)^T \bH^2 (\ba-\bp)}{(\frac{\ba-\bc}{2})^T \bH^2 (\frac{\ba-\bc}{2})},\\
    &R_\Pi(\bp) = 1-\frac{(\bp-\bp^{ur})^T \bH (\bp-\bp^{ur})}{\Pi(\bp^{ur})}
    = 1-\frac{(\bp-\bp^{ur})^T \bH (\bp-\bp^{ur})}{(\frac{\ba-\bc}{2})^T \bH (\frac{\ba-\bc}{2})}.
\end{align*}
Apply the spectral decomposition, we have
\begin{equation}\label{appeq.spectralr}
    \begin{aligned}
    	& R_{V}(\bp) =
    	\frac{\langle \mathbf{w}_1, \ba-\bp \rangle^2+ \sum_{i=2}^{n}  (\frac{1-\delta \lambda_1}{1-\delta \lambda_i})^2 \langle \mathbf{w}_i , \ba-\bp \rangle^2}{\langle \mathbf{w}_1, \frac{\ba-\bc}{2} \rangle^2+ \sum_{i=2}^{n}  (\frac{1-\delta \lambda_1}{1-\delta \lambda_i})^2 \langle \mathbf{w}_i , \frac{\ba-\bc}{2} \rangle^2},\\
    	& R_{\Pi}(\bp) = 1-
    	\frac{\langle \mathbf{w}_1, \bp-\bp^{ur} \rangle^2+ \sum_{i=2}^{n}  \frac{1-\delta \lambda_1}{1-\delta \lambda_i} \langle \mathbf{w}_i , \bp-\bp^{ur} \rangle^2}{\langle \mathbf{w}_1, \frac{\ba-\bc}{2} \rangle^2+ \sum_{i=2}^{n}  \frac{1-\delta \lambda_1}{1-\delta \lambda_i} \langle \mathbf{w}_i , \frac{\ba-\bc}{2} \rangle^2}.
    \end{aligned}
\end{equation}

Since we assume the network $\mathbf{G}$ is connected, then the largest eigenvalue is unique. Hence, as $\delta \rightarrow 1/\lambda_1$, we have $1-\delta\lambda_1 \rightarrow 0$ and $1-\delta\lambda_i \rightarrow 1-\lambda_i/\lambda_1 > 0$. Then we have
\begin{align*}
    &\bar{R}_V(\bp) := \lim_{\delta \rightarrow 1/\lambda_1} R_V(\bp) = 
    \frac{\langle \mathbf{w}_1, \ba-\bp \rangle^2}{\langle \mathbf{w}_1, \frac{\ba-\bc}{2} \rangle^2} = (1-\mathscr{A}(\bp))^2,\\
    &\bar{R}_{\Pi}(\bp) := \lim_{\delta \rightarrow 1/\lambda_1} R_{\Pi}(\bp)
    =1- \frac{\langle \mathbf{w}_1, \bp-\bp^{ur} \rangle^2}{\langle \mathbf{w}_1, \frac{\ba-\bc}{2} \rangle^2} = 1-\mathscr{A}(\bp)^2.
\end{align*}
This expression implies that if $\mathscr{A}(\bp) = \mathscr{A}(\bp')$, then $\bar{R}_V(\bp)=\bar{R}_V(\bp')$ and $\bar{R}_{\Pi}(\bp) = \bar{R}_{\Pi}(\bp')$.

\vspace{12pt}

\noindent\textbf{Part II.} According to the spectral decomposition, the firm's profit maximization problem is equivalent to
\begin{equation}\label{appeq.opdecomp}
	\min\limits_{\bp \in K} 
    f(\bp,\delta) := 
    \underbrace{\frac{\langle \mathbf{w}_1, \bp-\bp^{ur} \rangle^2 }{\langle \mathbf{w}_1, \frac{\ba-\bc}{2} \rangle^2}}_{\mathscr{A}(\bp)^2}
	+ \sum_{i=2}^n \frac{1-\delta \lambda_1}{1-\delta \lambda_i}
    \frac{\langle \mathbf{w}_i, \bp-\bp^{ur} \rangle^2}{\langle \mathbf{w}_1, \frac{\ba-\bc}{2} \rangle^2}.
\end{equation}
Denote the solution for any given $\delta$ as $\bp^*(\delta)$.

\begin{lemma}\label{applemma.limita}
    If Assumptions~\ref{assump.k} and \ref{assump.spectrum} hold,
    \begin{equation*}
         \lim_{\delta \to 1/\lambda_1} \mathscr{A}(\bp^*(\delta)) = \argmax_{A \in \overline{\mathscr{A}(K)}} \{1-A^2\}
         =:\bar{\mathscr{A}}^* .
    \end{equation*}
\end{lemma}
\emph{Proof.}
The closure of the image of $K$ under $\mathscr{A}$ is defined as
\begin{equation*}
    \overline{\mathscr{A}(K)} :=\text{cl}\left(  \{\mathscr{A}(\bp): \bp \in K\}  \right).
\end{equation*}
Since $\mathscr{A}$ is linear and $K$ is convex, the image $\mathscr{A}(K)$ is an interval in $\mathbb{R}$. Hence, its closure must be a closed interval, denote $\overline{\mathscr{A}(K)} = [\ubar{b},\bar{b}]$. The maximizer of $1-A^2$ over this interval is then given by
\begin{equation}\label{appeq.limita}
    \bar{\mathscr{A}}^*  = 
    \begin{cases}
        \ubar{b}, \quad & \ubar{b}>0\\
        \bar{b}, \quad & \bar{b}<0\\
        0, \quad & 0 \in [\ubar{b},\bar{b}]
    \end{cases}
\end{equation}
Hence, $\bar{\mathscr{A}}^*$ is well defined.

Fix $\bp \in K$, by the definition of $f$ and $\bp^*(\delta)$, we have
\begin{equation*}
    \mathscr{A}(\bp^*(\delta))^2 \leq f(\bp^*(\delta),\delta) \leq f(\bp,\delta) \leq f(\bp,0), \ \forall \delta \in [0,1/\lambda_1).
\end{equation*}
Therefore, the family $\{\mathscr{A}(\bp^*(\delta)): \delta \in [0,1/\lambda_1)\}$ is bounded. Consider any convergent subsequence, denote its limit by $A^{\infty}$. In what follows we show that $A^{\infty} = \bar{\mathscr{A}}^*$. With a slight abuse of notation, we henceforth assume that $\mathscr{A}(\bp^*(\delta))$ itself converges.

First, note that $\mathscr{A}(\bp^*(\delta)) \in \mathscr{A}(K) \subseteq \overline{\mathscr{A}(K)}$ for any $\delta$, then we have $A^{\infty} \in [\ubar{b},\bar{b}]$. If $A^{\infty} \neq \bar{\mathscr{A}}^*$, by equation~(\ref{appeq.limita}) we have $|A^{\infty}| >  |\bar{\mathscr{A}}^*|$. There exists $\tilde{\bp} \in K$ such that $\lvert \mathscr{A}(\tilde{\bp})\rvert < |A^{\infty}|$. For any $\epsilon > 0$, define
\begin{equation*}
    \overline{\delta} := \frac{1}{\lambda_1}
    \left[1-\frac{\langle \mathbf{w}_1, \frac{\ba-\bc}{2} \rangle^2(1-\lambda_2/\lambda_1)\epsilon}{||\tilde{\bp}-\bp^{ur}||^2_2}
    \right]< \frac{1}{\lambda_1}.
\end{equation*}
For $\overline{\delta} < \delta < 1/\lambda_1$, we have
\begin{align*}
    \mathscr{A}(\bp^*(\delta))^2 \leq&
    f(\bp^*(\delta),\delta) \leq 
    f(\tilde{\bp},\delta) 
    = 
    \mathscr{A}(\tilde{\bp})^2
	+ \sum_{i=2}^n \frac{1-\delta \lambda_1}{1-\delta \lambda_i}
    \frac{\langle \mathbf{w}_i, \tilde{\bp}-\bp^{ur} \rangle^2}{\langle \mathbf{w}_1, \frac{\ba-\bc}{2} \rangle^2}\\
    \leq & 
    \mathscr{A}(\tilde{\bp})^2
    + \frac{1-\delta \lambda_1}{1-\lambda_2/\lambda_1} 
    \frac{||\tilde{\bp}-\bp^{ur}||^2_2}{\langle \mathbf{w}_1, \frac{\ba-\bc}{2} \rangle^2}\\
    \leq&
    \mathscr{A}(\tilde{\bp})^2 + \epsilon
\end{align*}
The inequality in the second line follows from $\lambda_2 \geq \lambda_i$ for $i \geq 2$ and $\sum_{i=2}^n \langle \mathbf{w}_i, \tilde{\bp}-\bp^{ur} \rangle^2 \leq \sum_{i=1}^n \langle \mathbf{w}_i, \tilde{\bp}-\bp^{ur} \rangle^2 = ||\tilde{\bp}-\bp^{ur}||^2_2$. Taking limit on the left-hand side we get $\mathscr{A}(\tilde{\bp})^2 +\epsilon\geq (A^{\infty})^2$ for any $\epsilon > 0$. This implies $\mathscr{A}(\tilde{\bp})^2 \geq (A^{\infty})^2$, which contradicts with $\lvert \mathscr{A}(\tilde{\bp})\rvert < |A^{\infty}|$. Therefore, we must have $A^{\infty} = \bar{\mathscr{A}}^*$.
\hfill $\square$

Now we derive the limits of the consumer surplus and profit ratios. For any $\epsilon >0$, there exists $\hat{\bp} \in K$ such that $\mathscr{A}(\hat{\bp})^2 \leq (\bar{\mathscr{A}}^*)^2 + \epsilon/2$. By the same approach as before, there exists $\hat{\delta}$ such that for any $\hat{\delta} < \delta < 1/\lambda_1$,
\begin{equation*}
    \mathscr{A}(\bp^*(\delta))^2 \leq
    f(\bp^*(\delta),\delta) \leq 
    f(\hat{\bp},\delta) \leq \mathscr{A}(\hat{\bp})^2 + \frac{\epsilon}{2} \leq (\bar{\mathscr{A}}^*)^2 + \epsilon.
\end{equation*}
This implies $\lim_{\delta \rightarrow1/\lambda_1} f(\bp^*(\delta),\delta) = (\bar{\mathscr{A}}^*)^2$. Then we have
\begin{align*}
    \overline{R}^{*}_{\Pi} := & \lim\limits_{\delta \rightarrow 1/\lambda_1} R_{\Pi}(\bp^{*}(\delta))\\
    = & 1-\lim\limits_{\delta \rightarrow 1/\lambda_1}
    \frac{\langle \mathbf{w}_1 , \bp^{*}(\delta)-\bp^{ur} \rangle^2+\sum_{i=2}^{n} \frac{1-\delta \lambda_1}{1-\delta \lambda_i} \langle \mathbf{w}_i , \bp^{*}(\delta)-\bp^{ur} \rangle^2 }
    {\langle \mathbf{w}_1 , \frac{\mathbf{a}-\mathbf{c}}{2} \rangle^2+\sum_{i=2}^{n} \frac{1-\delta \lambda_1}{1-\delta \lambda_i} \langle \mathbf{w}_i , \frac{\mathbf{a}-\mathbf{c}}{2} \rangle^2 }\\
    =& 1-\lim\limits_{\delta \rightarrow 1/\lambda_1}
    \frac{ f(\bp^*(\delta),\delta) }
    {1+\sum_{i=2}^{n} \frac{1-\delta \lambda_1}{1-\delta \lambda_i} \langle \mathbf{w}_i , \frac{\mathbf{a}-\mathbf{c}}{2} \rangle^2 / \langle \mathbf{w}_1 , \frac{\mathbf{a}-\mathbf{c}}{2} \rangle^2}
    \\
    = & 1-(\bar{\mathscr{A}}^*)^2.
\end{align*}

The previous convergence results for $\mathscr{A}(\bp^*(\delta))^2$ and $f(\bp^*(\delta),\delta)$ also imply
\begin{equation}\label{appeq.wilimit}
    \lim_{\delta \rightarrow 1/\lambda_1} \frac{1-\delta \lambda_1}{1-\delta \lambda_i}\langle \mathbf{w}_i, \bp^*(\delta)-\bp^{ur} \rangle^2 = 0, \ i \geq 2.
\end{equation}
Then for $i \geq 2$, we have
\begin{align*}
    & \lim_{\delta \rightarrow 1/\lambda_1} (\frac{1-\delta \lambda_1}{1-\delta \lambda_i})^2 \langle \mathbf{w}_i, \bp^*(\delta)-\bp^{ur} \rangle^2 = 0,\\
    & \lim_{\delta \rightarrow 1/\lambda_1} (\frac{1-\delta \lambda_1}{1-\delta \lambda_i})^2 \langle \mathbf{w}_i, \bp^*(\delta)-\bp^{ur} \rangle = 0,
\end{align*}
where the second equality follows from $|\langle \mathbf{w}_i, \bp^*(\delta)-\bp^{ur} \rangle| \leq \max\{1, \langle \mathbf{w}_i, \bp^*(\delta)-\bp^{ur} \rangle^2\}$. Then we obtain that for $i \geq 2$, 
\begin{align*}
	&\lim_{\delta \rightarrow 1/\lambda_1}
	(\frac{1-\delta \lambda_1}{1-\delta \lambda_i})^2 \langle \mathbf{w}_i , \mathbf{a}-\bp^{*}(\delta) \rangle^2\\
	=& \lim_{\delta \rightarrow 1/\lambda_1}
	(\frac{1-\delta \lambda_1}{1-\delta \lambda_i})^2 \left[\langle \mathbf{w}_i , \bp^{*}(\delta)-\bp^{ur} \rangle^2+\langle \mathbf{w}_i , \frac{\mathbf{a}-\mathbf{c}}{2} \rangle^2 - 2\langle \mathbf{w}_i , \bp^{*}(\delta)-\bp^{ur} \rangle \langle \mathbf{w}_i , \frac{\mathbf{a}-\mathbf{c}}{2} \rangle \right]=0.
\end{align*}
Therefore, we have
\begin{align*}
    \overline{R}^{*}_{V} := & \lim\limits_{\delta \rightarrow 1/\lambda_1} R_{V}(\bp^{*}(\delta)) \\
    = & \lim\limits_{\delta \rightarrow 1/\lambda_1}  
    \frac{\langle \mathbf{w}_1 , \mathbf{a}-\bp^{*}(\delta) \rangle^2+ \sum_{i=2}^{n} (\frac{1-\delta \lambda_1}{1-\delta \lambda_i})^2 \langle \mathbf{w}_i , \mathbf{a}-\bp^{*}(\delta) \rangle^2}
    {\langle \mathbf{w}_1 , \frac{\mathbf{a}-\mathbf{c}}{2} \rangle^2+ \sum_{i=2}^{n} (\frac{1-\delta \lambda_1}{1-\delta \lambda_i})^2 \langle \mathbf{w}_i , \frac{\mathbf{a}-\mathbf{c}}{2} \rangle^2}\\
    = & \lim\limits_{\delta \rightarrow 1/\lambda_1}  
    \frac{\langle \mathbf{w}_1 , \mathbf{a}-\bp^{*}(\delta) \rangle^2}
    {\langle \mathbf{w}_1 , \frac{\mathbf{a}-\mathbf{c}}{2} \rangle^2}\\
    =& \lim\limits_{\delta \rightarrow 1/\lambda_1} 
    \left[1-\mathscr{A}(\bp^{*}(\delta))  \right]^2
    = (1-\bar{\mathscr{A}}^*)^2.
    \tag*{\text{$\square$}}
\end{align*}




\subsection{Proof of Theorem \ref{thm.neutral}}\label{appsec.proofthmneutral}

For $\myd_{\boldsymbol{\Delta}}$, and any feasible price vector $\bp \in \myd_{\boldsymbol{\Delta}}$, we have $\bp+t \mathbf{1} \in \myd_{\boldsymbol{\Delta}}$ for any $t \in \mathbb{R}$. Note that $\langle \mathbf{w}_1, \mathbf{1} \rangle \neq 0$ since $\mathbf{w}_1 > \mathbf{0}$. Therefore, there exists $t$ such that $\langle \mathbf{w}_1, \bp+t \mathbf{1} \rangle = \langle \mathbf{w}_1, \bp^{ur} \rangle$ and thus $\mathscr{A}(\bp+t \mathbf{1}) = 0 \in \mathscr{A}(\myd_{\boldsymbol{\Delta}})$. According to equation~(\ref{appeq.limita}), $\bar{\mathscr{A}}^*=0$ and $\myd_{\boldsymbol{\Delta}}$ is always neutral.

For $\myb(\underline{\bp},\overline{\bp})$, note that all elements of $\mathbf{w}_1$ are positive. Then $\mathscr{A}(\bp) \leq \mathscr{A}(\bp')$ if $\bp \leq \bp'$. Therefore, $\mathscr{A}(\myb(\underline{\bp},\overline{\bp})) = [\mathscr{A}(\underline{\bp}),\mathscr{A}(\overline{\bp})]$. According to equation~(\ref{appeq.limita}), $\myb(\underline{\bp},\overline{\bp})$ is neutral if and only if $0 \in [\mathscr{A}(\underline{\bp}),\mathscr{A}(\overline{\bp})]$. Otherwise, it is efficient (inefficient) if $\mathscr{A}(\overline{\bp}) < 0$ ($\mathscr{A}(\underline{\bp}) > 0$).

For $\mye_{\btheta, M}$, if $\btheta \not\propto \mathbf{w}_1$, then $\exists \hat{\mathbf{z}} \in \mathbb{R}^n$ such that $\langle \btheta,  \hat{\mathbf{z}} \rangle = 0$ and $\langle \mathbf{w}_1,  \hat{\mathbf{z}} \rangle \neq 0$. Then for any $\bp \in \mye_{\btheta, M}$, $\exists t \in \mathbb{R}$ such that $\bp + t \hat{\mathbf{z}} \in \mye_{\btheta, M}$ and
\begin{align*}
	& \langle \mathbf{w}_1,  \bp + t \hat{\mathbf{z}} \rangle = \langle \mathbf{w}_1,  \bp \rangle + t \langle \mathbf{w}_1, \hat{\mathbf{z}} \rangle = \langle \mathbf{w}_1,  \bp^{ur} \rangle.
\end{align*}
This implies $\mathscr{A}(\bp+t \hat{\mathbf{z}}) = 0 \in \mathscr{A}(\mye_{\btheta, M})$. Then $\mye_{\btheta, M}$ is neutral if $\btheta \not\propto \mathbf{w}_1$, which generically holds. Otherwise, suppose that $\btheta \propto \mathbf{w}1$. Since $\bp^{ur}$ is infeasible under Assumption~\ref{assump.infeasible}, for any $\bp \in \mye_{\btheta, M}$ we have
\begin{equation*}
    \mathscr{A}(\bp) = \frac{\langle \mathbf{w}_1, \bp-\bp^{ur} \rangle}{\langle \mathbf{w}_1, \frac{\ba-\bc}{2} \rangle} 
    = \frac{\langle \btheta, \bp-\bp^{ur} \rangle}{\langle \btheta, \frac{\ba-\bc}{2} \rangle}
    \leq \frac{M-\langle \btheta, \bp^{ur} \rangle}{\langle \btheta, \frac{\ba-\bc}{2} \rangle} <0.
\end{equation*}
It follows that $\overline{\mathscr{A}(\mye_{\btheta, M})} \subseteq (\infty,0)$ and hence $\bar{\mathscr{A}}^*<0$, which implies that $\mye_{\btheta, M}$ is efficient.
\hfill $\square$


\subsection{Proof of Lemma \ref{lemma.gap}}\label{appsec.prooflemmagap}
Rewrite problem~(\ref{eq.intervention}) and define the value function as
\begin{equation}\label{appeq.intervention}
    \begin{gathered}
        g(\tau,\delta) := \max\limits_{\bp\in \mathbb{R}^n} R_V(\bp,\delta) \\
        s.t. \ R_{\Pi}(\bp,\delta) \geq \tau.
    \end{gathered} 
\end{equation}
The spectral decomposition representation of the ratios in equation~(\ref{appeq.spectralr}) implies that they are continuous in $(\bp,\delta)$. By Berge's maximum theorem, $g(\tau,\delta)$ is continuous and note that $g(\tau,1/\lambda_1) = (1+\sqrt{1-\tau})^2$. Then we have
\begin{align*}
    \lim_{\delta \rightarrow 1/\lambda_1} Gap(\delta) =&
    \lim_{\delta \rightarrow 1/\lambda_1} R^+_V(\delta) - R_V(\delta)\\
    =&
    \lim_{\delta \rightarrow 1/\lambda_1} g(R_{\Pi}(\bp^*(\delta)),\delta) - R_V(\bp^*(\delta)) \tag*{(\text{by definition})} \\
    =& \left(1+\sqrt{1-\bar{R}^*_{\Pi}} \right)^2 - \bar{R}^*_{V} \tag*{(\text{by continuity)}}\\
    =& \left(1+\sqrt{(\bar{\mathscr{A}}^*)^2} \right)^2 - (1-\bar{\mathscr{A}}^*)^2 \tag*{(\text{by Lemma~\ref{lemma.plimit}})}\\
    =&
    \begin{cases}
        0 \quad & \bar{\mathscr{A}}^* \leq 0\\
        4\bar{\mathscr{A}}^* \quad & \bar{\mathscr{A}}^* > 0
    \end{cases}
\end{align*}
Therefore, $\lim_{\delta \rightarrow 1/\lambda_1} Gap(\delta) = 0$ if and only if $\bar{\mathscr{A}}^* \leq 0$, that is, the price regulation is efficient or neutral. Moreover, neutrality arises when $\bar{\mathscr{A}}^*=0$, i.e. $\bar{R}^*_{V} = 1$.
\hfill$\square$

\subsection{Proof of Proposition \ref{prop.ramsey}}\label{appsec.proofpropramsey}

According to equations~(\ref{eq.cs}) and~(\ref{eq.pi}), the objective can be written as
\begin{equation}\label{appeq.ramsey}
    -(\bp-\bp^{ur})\bH(\bp-\bp^{ur}) + \frac{\alpha}{2} (\ba-\bp)\bH^2(\ba-\bp).
\end{equation}
We first ignore the constraint $R_{\Pi}(\bp) \geq 0$ and derive the first-order condition:
\begin{align*}
    & -2\bH(\bp-\bp^{ur})+\alpha \bH^2(\bp-\ba) = 0\\
    \iff & [\alpha\bH^2-2\bH]\bp = [\alpha\bH^2-2\bH]\bp^{ur} + \alpha\bH^2 (\ba-\bp^{ur}).
\end{align*}
Note that
\begin{equation*}
    \alpha\bH^2-2\bH = -\bH^2[(2-\alpha)\mathbf{I}_n-2\delta\mathbf{G}].
\end{equation*}
As shown in Appendix~\ref{appsec.proofpfamily}, $\bar{\eta} < 2-2\delta\lambda_1$. Hence, when $\alpha \in [0, \bar{\eta}]$, this matrix is negative definite. Therefore, the objective in~\eqref{appeq.ramsey} is strictly concave, and the unique maximizer is characterized by the first-order condition, which can be simplified as:
\begin{equation*}
    \bp = \bp^{ur} - \frac{\alpha}{2-\alpha} \left[\mathbf{I}_n-\frac{2\delta}{(2-\alpha)}\mathbf{G} \right]^{-1} \frac{\ba-\bc}{2} = \pfam(\alpha).
\end{equation*}
Moreover, since $R_{\Pi}(\pfam(\alpha))\ge0$, the constraint in problem~(\ref{eq.ramsey}) is also satisfied. Therefore, when $\alpha \in [0, \bar{\eta}]$, the solution is exactly $\pfam(\alpha)$.

Next, consider $\alpha>\bar{\eta}$ and let $\hat{\bp}$ denote the solution to problem~\eqref{eq.ramsey}. Optimality implies
\begin{align*}
    & \Pi(\hat{\bp}) + \alpha V(\hat{\bp}) \geq \Pi(\pfam(\bar{\eta})) + \alpha V(\pfam(\bar{\eta})),\\
    & \Pi(\pfam(\bar{\eta})) + \bar{\eta} V(\pfam(\bar{\eta})) \geq \Pi(\hat{\bp}) + \bar{\eta} V(\hat{\bp}),
\end{align*}
which together imply
\begin{equation*}
    (\alpha-\bar{\eta}) [\Pi(\pfam(\bar{\eta})) - \Pi(\hat{\bp})] \geq 0.
\end{equation*}
Since $\Pi(\pfam(\bar{\eta})) = 0$, it follows that $\Pi(\hat{\bp}) \leq 0$. Combined with the feasibility constraint $\Pi(\hat{\bp}) \geq 0$, we have $\Pi(\hat{\bp}) = 0$. Consequently, $V(\hat{\bp}) = V(\pfam(\bar{\eta}))$. The uniqueness of the solution to problem~(\ref{eq.intervention}) then implies $\hat{\bp} = \pfam(\bar{\eta})$.
\hfill$\square$


\subsection{Proof of Proposition \ref{prop.average}}\label{appsec.proofpropaverage}
We first provide a closed form of the firm's optimal pricing under some linear-equality constraints. Formally, consider the problem
\begin{equation}\label{appeq.lineareq}
    \begin{gathered}
        \min\limits_{\bp} (\bp-\bp^{ur})^T \bH (\bp-\bp^{ur})\\
        s.t. \ \mathbf{V}\bp = \mathbf{y}.
    \end{gathered}
\end{equation}
Here, $\mathbf{V}$ is an $m \times n$ ($m\leq n$) matrix  with full row rank and $\mathbf{y}$ is an $m \times 1$ vector.
\begin{lemma}\label{applemma.closedform}
	The solution of problem~(\ref{appeq.lineareq}) can be written as
	\begin{equation}\label{eq.closedformp}
		\begin{aligned}
			\bp 
			=  \bp^{ur} - \bH^{-1} \mathbf{V}' \left[\mathbf{V} \bH^{-1} \mathbf{V}'\right]^{-1}
			(\mathbf{V} \bp^{ur} - \mathbf{y}).
		\end{aligned}
	\end{equation}
\end{lemma}
\emph{Proof.} Define the Lagrangian as:
\begin{equation*}
	\mathcal{L} =  (\bp-\bp^{ur})^T \bH (\bp-\bp^{ur})
	+ \boldsymbol{\lambda}' (\mathbf{V}\bp - \mathbf{y}).
\end{equation*}
Since both the objective function and the feasible set are convex, the standard KKT conditions are sufficient for the solution:
\begin{equation}\label{appeq.foc1}
	2 \bH(\bp-\bp^{ur}) +  \mathbf{V}' \boldsymbol{\lambda} = \mathbf{0},
\end{equation}
\begin{equation}\label{appeq.foc2}
	\mathbf{V}\bp - \mathbf{y} = 0.
\end{equation}
Equation (\ref{appeq.foc1}) is equivalent to
\begin{equation}\label{appeq.foc3}
	\bp = \bp^{ur} - \frac{1}{2} \bH^{-1} \mathbf{V}' \boldsymbol{\lambda}.
\end{equation}
Multiply equation (\ref{appeq.foc3}) by $\mathbf{V}$ and use equation (\ref{appeq.foc2}), we get 
\begin{equation*}
	\mathbf{y} = \mathbf{V} \bp^{ur} - \frac{1}{2} \mathbf{V} \bH^{-1} \mathbf{V}' \boldsymbol{\lambda}.
\end{equation*}

Since $\mathbf{V}$ has full row rank and $\bH^{-1}$ is positive definite under Assumption \ref{assump.spectrum}, then $\mathbf{V} \bH^{-1} \mathbf{V}'$ is invertible and we get
\begin{equation*}
	\boldsymbol{\lambda} = 2[\mathbf{V} \bH^{-1} \mathbf{V}']^{-1} 
	(\mathbf{V} \bp^{ur} - \mathbf{y}).
\end{equation*}
Plug this equation into equation (\ref{appeq.foc3}) and we get
\begin{equation*}
	\bp = \bp^{ur} - \bH^{-1} \mathbf{V}' \left[\mathbf{V} \bH^{-1} \mathbf{V}'\right]^{-1}
    (\mathbf{V} \bp^{ur} - \mathbf{y}).
    \tag*{\text{$\square$}}
\end{equation*}

Now we examine the specific price regulation, $\mye_{\btheta,M}$. Denote the firm's optimal price as $\bp^*$, which cannot be equal to $\bp^{ur}$ under Assumption~\ref{assump.infeasible}. If $\langle \btheta, \bp^* \rangle < M$, then there exists $t \in (0,1)$ such that $\bp^*+t(\bp^{ur}-\bp^*) \in \mye_{\btheta,M}$. However,
\begin{equation*}
    \left\lVert \bp^*+t(\bp^{ur}-\bp^*)-\bp^{ur}  \right\rVert_{\bH} 
    = (1-t) \left\lVert \bp^*-\bp^{ur}  \right\rVert_{\bH} 
    < \left\lVert \bp^*-\bp^{ur}  \right\rVert_{\bH} ,
\end{equation*}
which contradicts Lemma~\ref{lemma.equi}. Then we must have $\langle \btheta, \bp^* \rangle = M$. Hence, the feasible set $\mye_{\btheta,M}$ can be without loss reduced to a linear equality constraint $\langle \btheta, \bp \rangle = M$. According to Lemma~\ref{applemma.closedform}, the firm's optimal price equals
\begin{equation*}
    \bp^*= \bp^{ur} - \frac{\btheta^T\bp^{ur}-M  }{\btheta^T\bH^{-1} \btheta} \bH^{-1} \btheta.
\end{equation*}

Given this price, we can calculate the demand:
\begin{equation*}
	\mathbf{x}^* = \bH(\mathbf{a}-\mathbf{p}^*)
	= \mathbf{x}^{ur} + \frac{\btheta^T\bp^{ur}-M  }{\btheta^T\bH^{-1} \btheta}  \btheta.
\end{equation*}
The infeasibility of $\bp^{ur}$ implies $\btheta^T\bp^{ur}-M>0$. Under Assumption~\ref{assump.spectrum}, the matrix $\bH$ is positive definite, which implies $\btheta^T\bH^{-1} \btheta>0$. Since $\btheta > \mathbf{0}$, it follows that $x^*_i \geq x^{ur}$ for all $i \in \mathcal{N}$. Consequently, the equilibrium utility $u^*_i = \frac{1}{2} (x^*_i)^2 \geq \frac{1}{2} (x^{ur}_i)^2 = u^{ur}_i$ with strict inequality for at least one consumer. Therefore, aggregate consumer surplus strictly increases: $V^* = \sum_{i \in \mathcal{N}} u^*_i > \sum_{i \in \mathcal{N}} u^{ur}_i=V^{ur}$.
\hfill$\square$


\subsection{Proof of Proposition \ref{prop.averagesensitivity}}\label{appsec.proofpropaveragesen}

For any $(\btheta,M) \in \mathcal{F}$, the consumer surplus satisfies
\begin{align*}
    V(\bp^*) =& \frac{1}{2} (\mathbf{x}^*)^T\mathbf{x}^*
    = V(\bp^{ur})
    +\frac{\btheta^T\bp^{ur}-M}{\btheta^T\bH^{-1} \btheta}\btheta^T \bH (\ba-\bp^{ur})
    + \frac{1}{2} \frac{(\btheta^T\bp^{ur}-M )^2  }{(\btheta^T\bH^{-1} \btheta)^2} \btheta^T\btheta\\
    =&
    V(\bp^{ur})
    +\frac{C }{\btheta^T\bH^{-1} \btheta}\btheta^T \bH (\ba-\bp^{ur})
    + \frac{1}{2} \frac{C^2  }{(\btheta^T\bH^{-1} \btheta)^2},
\end{align*}
where the second equality uses the normalization $\btheta^\top\bp^{ur}-M=C$ and $\btheta^\top\btheta=1$.

Applying the spectral decomposition of $\bH=[\mathbf{I}_n-\delta\mathbf{G}]^{-1}$, we obtain
\begin{align*}
    &\lim\limits_{\delta \rightarrow 1/\lambda_1} \btheta^T\bH^{-1} \btheta
    = \lim\limits_{\delta \rightarrow 1/\lambda_1} \sum_{i=1}^n (1-\delta\lambda_i) \langle \mathbf{w}_i, \btheta \rangle^2
    = \sum_{i=2}^n (1-\lambda_i/\lambda_1) \langle \mathbf{w}_i, \btheta \rangle^2,\\
    &\lim\limits_{\delta \rightarrow 1/\lambda_1} (1-\delta\lambda_1)\btheta^T \bH (\ba-\bp^{ur})
    = \lim\limits_{\delta \rightarrow 1/\lambda_1} \sum_{i=1}^n \frac{1-\delta\lambda_1}{1-\delta\lambda_i} \langle \mathbf{w}_i, \btheta \rangle \langle \mathbf{w}_i, \ba-\bp^{ur} \rangle
    = \langle \mathbf{w}_1, \btheta \rangle \langle \mathbf{w}_1, \ba-\bp^{ur} \rangle,\\
    &\lim\limits_{\delta \rightarrow 1/\lambda_1} (1-\delta\lambda_1)^2 V(\bp^{ur}) = \lim\limits_{\delta \rightarrow 1/\lambda_1} \frac{1}{2}\sum_{i=1}^n \frac{(1-\delta\lambda_1)^2}{(1-\delta\lambda_i)^2} \langle \mathbf{w}_i, \ba-\bp^{ur} \rangle^2
    = \frac{1}{2}\langle \mathbf{w}_1, \ba-\bp^{ur} \rangle^2.
\end{align*}

It follows that
\begin{equation*}
    I(\btheta,M,\mathbf{G}) := \lim\limits_{\delta \rightarrow 1/\lambda_1}\frac{R_V(\bp^*(\delta))-1}{1-\delta\lambda_1}
    = \frac{2C}{\langle \mathbf{w}_1, \ba-\bp^{ur} \rangle}\frac{\langle \mathbf{w}_1, \btheta \rangle}{\sum_{i=2}^n (1-\lambda_i/\lambda_1) \langle \mathbf{w}_i, \btheta \rangle^2}.
\end{equation*}
Under the normalization $\lVert \btheta \rVert = \lVert \mathbf{w}_1 \rVert = 1$, then $\langle \mathbf{w}_1, \btheta \rangle = corr(\mathbf{w}_1, \btheta) = k$ and $\sum_{i=2}^n\langle \mathbf{w}_i, \btheta \rangle^2 = 1-k^2$. Since $\lambda_2 \geq \ldots \geq \lambda_n$, this implies
\begin{equation*}
    \frac{2C}{\langle \mathbf{w}_1, \ba-\bp^{ur} \rangle}\frac{k}{(1-\lambda_n/\lambda_1) (1-k^2)} 
    \leq I(\btheta,M,\mathbf{G}) \leq 
    \frac{2C}{\langle \mathbf{w}_1, \ba-\bp^{ur} \rangle}\frac{k}{(1-\lambda_2/\lambda_1) (1-k^2)}.
\end{equation*}
The upper and lower bounds are attained by $k \mathbf{w}_1 + \sqrt{1-k^2} \mathbf{w}_2$ and $k \mathbf{w}_1 + \sqrt{1-k^2} \mathbf{w}_n$ respectively. For $k$ sufficiently large, both vectors are nonnegative and hence feasible.

Therefore,
\begin{equation*}
    \xi(\mathbf{G}) := \frac{\max_{(\btheta,M) \in \mathcal{F}}
    I(\btheta,M,\mathbf{G})}{\min_{(\btheta,M) \in \mathcal{F}}
    I(\btheta,M,\mathbf{G})}
    =
    \frac{\lambda_1-\lambda_n}{\lambda_1-\lambda_2}.
    \tag*{\text{$\square$}}
\end{equation*}


\subsection{Proof of Proposition \ref{thm.uniform}}
According to Lemma \ref{lemma.equi}, the firm chooses a price $p^0 \in \mathbb{R}$ to minimize 
\begin{equation*}
    (p^0 \mathbf{1} - \bp^{ur})^T\mathbf{H}(p^0 \mathbf{1} -\bp^{ur}).
\end{equation*}
Since the object is convex, the price can be characterized through the first order condition:
\begin{equation*}
    2 \mathbf{1}^T\mathbf{H}\mathbf{1}p^0 - 2\mathbf{1}^T\mathbf{H}\bp^{ur}
    \implies
    \bp^0(\delta) := p^0 \mathbf{1}   =\frac{\mathbf{1}^T\mathbf{H}\bp^{ur}}{\mathbf{1}^T\mathbf{H}\mathbf{1}} \mathbf{1}.
\end{equation*}

Note that $\bp^{ur}=\ba/2$ under Assumption \ref{assump.uniform}, then we have
\begin{align*}
    \mathscr{A}(\bp^{0}(\delta)) 
    = \frac{\langle \mathbf{w}_1, \bp^0(\delta)-\bp^{ur} \rangle}{\langle \mathbf{w}_1, \ba/2 \rangle}
    = \frac{\frac{\mathbf{1}^T\mathbf{H}\bp^{ur}}{\mathbf{1}^T\mathbf{H}\mathbf{1}} \mathbf{1}^T \mathbf{w}_1 - \ba^T\mathbf{w}_1/2}{\langle \mathbf{w}_1, \ba/2 \rangle}
    = - \frac{\mathbf{1}^T[\mathbf{H}\mathbf{1}\mathbf{w}_1^T-\mathbf{w}_1\mathbf{1}^T\mathbf{H}]\ba}{\langle \mathbf{w}_1, \ba \rangle \mathbf{1}^T\mathbf{H}\mathbf{1}}.
\end{align*}
By spectral decomposition
\begin{align*}
    \lim\limits_{\delta \rightarrow 1/\lambda_1} \mathbf{1}^T[\mathbf{H}\mathbf{1}\mathbf{w}_1^T-\mathbf{w}_1\mathbf{1}^T\mathbf{H}] =& 
    \lim\limits_{\delta \rightarrow 1/\lambda_1} \sum_{i=1}^n \left[  
    \frac{\langle\mathbf{w}_i,\mathbf{1}\rangle^2 }{1-\delta\lambda_i} \mathbf{w}_1^T-
    \frac{\langle\mathbf{w}_i,\mathbf{1}\rangle \langle\mathbf{w}_1,\mathbf{1}\rangle}{1-\delta\lambda_i} \mathbf{w}_i^T
    \right]\\
    =& (\sum_{i=2}^{n} \frac{\langle \mathbf{w}_i , \mathbf{1} \rangle^2}{1- \lambda_i/\lambda_1}) \mathbf{w}_1^T
    - \sum_{i=2}^{n} \frac{\langle \mathbf{w}_i , \mathbf{1} \rangle \langle \mathbf{w}_1 , \mathbf{1} \rangle}{1- \lambda_i/\lambda_1} \mathbf{w}_i^T.
\end{align*}
The limit is well-defined and is in fact $\mypsi^T$. Furthermore,
\begin{align*}
    \frac{1}{\mathbf{1}^T\mathbf{H}\mathbf{1}} = 
    \frac{1}{\langle\mathbf{w}_1,\mathbf{1}\rangle^2 + \sum_{i=2}^n\frac{(1-\delta\lambda_1)\langle\mathbf{w}_i,\mathbf{1}\rangle^2 }{1-\delta\lambda_i}}(1-\delta\lambda_1) =
    \frac{1}{\langle\mathbf{w}_1,\mathbf{1}\rangle^2}(1-\delta\lambda_1) + 
    \mathcal{O}\left( \frac{1}{\lambda_1} - \delta\right)^2
\end{align*}
These two equations implies
\begin{equation*}
    \mathscr{A}(\bp^{0}(\delta)) 
    = 0-\frac{\lambda_1\langle \mypsi(\mathbf{G}), \ba \rangle}{\langle \mathbf{w}_1, \ba \rangle \langle \mathbf{w}_1, \mathbf{1} \rangle^2} \left(\frac{1}{\lambda_1}-\delta \right)+
    \mathcal{O}\left( \frac{1}{\lambda_1}-\delta \right)^2
\end{equation*}

Moreover, the spectral decomposition gives:
\begin{align*}
	R_{V}(\bp^{0}(\delta)) 
	= & 
	\frac{\langle \mathbf{w}_1 , \ba-\bp^{0}(\delta) \rangle^2+ \sum_{i=2}^{n} (\frac{1-\delta \lambda_1}{1-\delta \lambda_i})^2 \langle \mathbf{w}_i , \ba-\bp^{0}(\delta) \rangle^2}
	{\langle \mathbf{w}_1 , \frac{\ba-\bc}{2} \rangle^2+ \sum_{i=2}^{n} (\frac{1-\delta \lambda_1}{1-\delta \lambda_i})^2 \langle \mathbf{w}_i , \frac{\ba-\bc}{2} \rangle^2}\\
	= & 
	(1-\mathscr{A}(\bp^{0}(\delta)))^2\\
	& +
	\frac{ \sum_{i=2}^{n} (\frac{1-\delta \lambda_1}{1-\delta \lambda_i})^2 [\langle \mathbf{w}_i , \ba-\bp^{0}(\delta) \rangle^2\langle \mathbf{w}_1 , \frac{\ba-\bc}{2} \rangle^2 - \langle \mathbf{w}_i , \frac{\ba-\bc}{2} \rangle^2\langle \mathbf{w}_1 , \ba-\bp^{0}(\delta) \rangle^2]}
	{[\langle \mathbf{w}_1 , \frac{\ba-\bc}{2} \rangle^2+ \sum_{i=2}^{n} (\frac{1-\delta \lambda_1}{1-\delta \lambda_i})^2 \langle \mathbf{w}_i , \frac{\ba-\bc}{2} \rangle^2]\langle \mathbf{w}_1 , \frac{\ba-\bc}{2} \rangle^2}\\
	=&
	(1-\mathscr{A}(\bp^{0}(\delta)))^2 + \mathcal{O}\left( \frac{1}{\lambda_1}-\delta \right)^2.
\end{align*}
The third equality holds since $\bp^{0}(\delta)$ is bounded. Plug the first order approximation of $\mathscr{A}(\bp^{0}(\delta))$ into this equation, we have
\begin{equation*}
    R_{V}(\bp^{0}(\delta)) = 1 + \frac{2\lambda_1\langle \mypsi(\mathbf{G}), \ba \rangle}{\langle \mathbf{w}_1, \ba \rangle \langle \mathbf{w}_1, \mathbf{1} \rangle^2}\left(\frac{1}{\lambda_1}-\delta \right)+
    \mathcal{O}\left( \frac{1}{\lambda_1}-\delta \right)^2.
\end{equation*}

Therefore, if $corr(\mypsi(\mathbf{G}),\ba) > 0$ (resp. $<0$), there exists $\bar{\delta}  < 1/\lambda_1$ (resp. $\bar{\bar{\delta}} < 1/\lambda_1$) such that $R_{V}(\bp^{0}(\delta)) > 1$ (resp. $<1$) for $\delta \in (\bar{\delta}, 1/\lambda_1)$ (resp. $\delta \in (\bar{\bar{\delta}}, 1/\lambda_1)$). Consumers gain from the prohibition of price discrimination in the first case, but are harmed in the second.
\hfill $\square$


\subsection{Proof of Proposition \ref{prop.uniform}}
Firstly, observe that
\begin{align*}
    \langle \mypsi,\mathbf{1} \rangle = \lim\limits_{\delta \rightarrow 1/\lambda_1}\mathbf{1}^T[\mathbf{w}_1\mathbf{1}^T\mathbf{H}-\mathbf{H}\mathbf{1}\mathbf{w}_1^T]\mathbf{1}
    = \lim\limits_{\delta \rightarrow 1/\lambda_1}[\mathbf{1}^T\mathbf{w}_1\mathbf{1}^T\mathbf{H}\mathbf{1}-\mathbf{1}^T\mathbf{H}\mathbf{1}\mathbf{w}_1^T\mathbf{1}]
    = 0.
\end{align*}
Then for any vector $\mathbf{z}$, we have
\begin{align*}
    \langle \mypsi,\tilde{\mathbf{z}} \rangle 
    = \langle \mypsi,\mathbf{z} - \frac{\langle \mathbf{z}, \mathbf{1}  \rangle}{n} \mathbf{1} \rangle
    = \langle \mypsi,\mathbf{z} \rangle- \frac{\langle \mathbf{z}, \mathbf{1}  \rangle}{n} \langle \mypsi, \mathbf{1} \rangle 
    = \langle \mypsi,\mathbf{z} \rangle.
\end{align*}
This also implies that
\begin{equation*}
    corr(\mypsi,\tilde{\mathbf{w}}_1) \propto \langle \mypsi, \tilde{\mathbf{w}}_1 \rangle 
    = \langle \mypsi, \mathbf{w}_1 \rangle
    = \sum_{i=2}^{n} \frac{\langle \mathbf{w}_i , \mathbf{1} \rangle^2}{1- \lambda_i/\lambda_1} > 0,
\end{equation*}
where the strict inequality follows from Assumption~\ref{assump.uniform} that $\mathbf{G}$ is not regular.

Now define the projection of $\tilde{\mathbf{w}}_1$ onto $\tilde{\ba}$ as
\begin{equation*}
    \hat{\ba} := \frac{\langle \tilde{\mathbf{w}}_1,\tilde{\ba} \rangle}{\langle \tilde{\ba},\tilde{\ba} \rangle} \tilde{\ba}.
\end{equation*}
Then we have
\begin{align*}
    ||\tilde{\mathbf{w}}_1 - \hat{\ba}||^2 =&
    \langle \tilde{\mathbf{w}}_1, \tilde{\mathbf{w}}_1\rangle + \langle \hat{\ba}, \hat{\ba}\rangle - 2\langle \tilde{\mathbf{w}}_1, \hat{\ba}\rangle
    = \langle \tilde{\mathbf{w}}_1, \tilde{\mathbf{w}}_1\rangle - \frac{\langle \tilde{\mathbf{w}}_1,\tilde{\ba} \rangle^2}{\langle \tilde{\ba},\tilde{\ba} \rangle} \\
    = &
    \langle \tilde{\mathbf{w}}_1, \tilde{\mathbf{w}}_1\rangle (1-corr(\tilde{\mathbf{w}}_1, \tilde{\ba})^2)
    < \epsilon(2-\epsilon)\langle \tilde{\mathbf{w}}_1, \tilde{\mathbf{w}}_1\rangle,
\end{align*}
where the last inequality follows from $corr(\tilde{\ba}, \tilde{\mathbf{w}}_1) > 1-\epsilon$ or $corr(\tilde{\ba}, \tilde{\mathbf{w}}_1) < -1+\epsilon$. Since $corr(\mypsi,\tilde{\mathbf{w}}_1) > 0$, by continuity, it follows that $corr(\mypsi, \hat{\ba}) > 0$ when $\epsilon$ is sufficiently small. Note that
\begin{equation*}
    \langle \mypsi,\hat{\ba} \rangle = \frac{\langle \tilde{\mathbf{w}}_1,\tilde{\ba} \rangle}{\langle \tilde{\ba},\tilde{\ba} \rangle} \langle \mypsi,\tilde{\ba} \rangle.
\end{equation*}
Then we have $sgn(corr(\mypsi,\tilde{\ba})) = sgn(corr(\tilde{\mathbf{w}}_1,\tilde{\ba}))$

Recall that $\langle \mypsi, \tilde{\mathbf{z}} \rangle = \langle \mypsi, \mathbf{z} \rangle$ for any vector $\mathbf{z}$, then $sgn(corr(\mypsi,\tilde{\ba})) = sgn(corr(\mypsi,\ba))$. Therefore, $corr(\mypsi, \ba) > 0$ (resp. $< 0$) when $corr(\tilde{\mathbf{w}}_1, \tilde{\ba}) > 1 - \epsilon$ (resp. $< -1 + \epsilon$). Applying Proposition~\ref{thm.uniform} then yields the proposition.
\hfill{$\square$}


\subsection{Proof of Proposition \ref{prop.twotype}}
Suppose $\mathbf{P}$ is a permutation matrix such that $\mathbf{P}\mathbf{G}\mathbf{P}^T = \mathbf{G}$, then
\begin{equation*}
    \lambda_1 \mathbf{P} \mathbf{w}_1 = \mathbf{P}\mathbf{G}\mathbf{w}_1
    = \mathbf{P}\mathbf{G}\mathbf{P}^T\mathbf{P}\mathbf{w}_1
    = \mathbf{G}\mathbf{P}\mathbf{w}_1.
\end{equation*}
This implies that $\mathbf{P}\mathbf{w}_1$ is an eigenvector associated with $\lambda_1$. Since $\mathbf{G}$ is connected, the normalized eigenvector associated with $\lambda_1$ is unique. Therefore, $\mathbf{P}\mathbf{w}_1 = \mathbf{w}_1$.

We also have
\begin{equation*}
    \mathbf{P}\mathbf{H}\mathbf{P}^T = \mathbf{P}[\mathbf{I}_n - \delta \mathbf{G}]^{-1}\mathbf{P}^T = \left\{ \mathbf{P}[\mathbf{I}_n - \delta \mathbf{G}]\mathbf{P}^T \right\}^{-1}
    = [\mathbf{P}\mathbf{P}^T - \delta \mathbf{P}\mathbf{G}\mathbf{P}^T]^{-1}
    = [\mathbf{I}_n - \delta \mathbf{G}]^{-1}
    = \mathbf{H}.
\end{equation*}
Moreover, it is clear that $\mathbf{P}\mathbf{1}=\mathbf{1}$, then
\begin{align*}
    \mathbf{P}\mypsi 
    =& \lim\limits_{\delta \rightarrow 1/\lambda_1}\mathbf{P}[\mathbf{w}_1\mathbf{1}^T\mathbf{H}-\mathbf{H}\mathbf{1}\mathbf{w}_1^T]\mathbf{1}
    = \lim\limits_{\delta \rightarrow 1/\lambda_1}[\mathbf{P}\mathbf{w}_1\mathbf{1}^T\mathbf{H}-\mathbf{P}\mathbf{H}\mathbf{P}^T\mathbf{P}\mathbf{1}\mathbf{w}_1^T]\mathbf{1}\\
    =& \lim\limits_{\delta \rightarrow 1/\lambda_1}[\mathbf{w}_1\mathbf{1}^T\mathbf{H}-\mathbf{H}\mathbf{1}\mathbf{w}_1^T]\mathbf{1}
    = \mypsi.
\end{align*}

Therefore, $\mathbf{w}_1$ and $\mypsi$ are invariant under graph automorphism. In a network with two types of nodes, for any $i, j \in V_k$ ($k = 1, 2$), we have $\mypsi(i) = \mypsi(j) \equiv \mypsi^{(k)}$ and $\mathbf{w}_1(i) = \mathbf{w}_1(j) \equiv \mathbf{w}_1^{(k)}$. By definition of the demeaned vector, we have $\langle \tilde{\mathbf{w}}_1, \mathbf{1} \rangle = 0$.
The previous section shows that $\langle \mypsi, \mathbf{1} \rangle = 0$, it then follows that:
\begin{equation*}
    \frac{\tilde{\mathbf{w}}_1^{(1)}}{\tilde{\mathbf{w}}_1^{(2)}} = \frac{\mypsi^{(1)}}{\mypsi^{(2)}} =  -\frac{|V_2|}{|V_1|}.
\end{equation*}
This also implies $\mypsi \propto \tilde{\mathbf{w}}_1$. Indeed, if $\mypsi \propto -\tilde{\mathbf{w}}_1$, then $\langle \mypsi, \tilde{\mathbf{w}}_1 \rangle < 0$, which contradicts the fact derived in the previous section that $\langle \mypsi, \tilde{\mathbf{w}}_1 \rangle > 0$. 

Since we assume $\mathbf{w}_1^{(1)} > \mathbf{w}_1^{(2)}$, it then follows that $\tilde{\mathbf{w}}_1^{(1)}, \mypsi^{(1)} > 0$ and $\tilde{\mathbf{w}}_1^{(2)}, \mypsi^{(2)} < 0$. Then we have:
\begin{align*}
    \langle \mypsi,\mathbf{a} \rangle 
    =& \mypsi^{(1)}(\sum_{i\in V_1}a_i) + \mypsi^{(2)}(\sum_{i\in V_2}a_i) 
    \propto |V_2|(\sum_{i\in V_1}a_i) - |V_1|(\sum_{i\in V_2}a_i) 
    = |V_1| |V_2| \left( \overline{a}(V_1) - \overline{a}(V_2)\right).
\end{align*}
Then $\langle \mypsi,\mathbf{a} \rangle >0$ is equivalent to $\overline{a}(V_1)-\overline{a}(V_2) > 0$. Applying Proposition~\ref{thm.uniform} then yields the proposition.
\hfill{$\square$}



\section{Additional Results}\label{appsec.online}

\subsection{An alternative definition for consumer surplus}\label{appsec.v}
In our main text, there are $n$ markets and each market has a representative consumer. Here we use an alternative definition where there is one representative consumer consuming in $n$ markets with utility:
\begin{equation}\label{appeq.u}
    U_{AV}(\mathbf{x},\bp) := \ba^T \mathbf{x} - \frac{1}{2} \mathbf{x}^T \mathbf{x} + \frac{\delta}{2} \mathbf{x}^T \mathbf{G} \mathbf{x}^T - \bp^T \mathbf{x}.
\end{equation}
The demand system under the first order condition is the same as the main text, so the analysis of equilibrium prices are the same as before.

For the welfare analysis, $\mathbf{x}(\bp) = [\mathbf{I}_n - \delta \mathbf{G}]^{-1} (\ba-\bp)$, then the consumer surplus has an alternative definition:
\begin{equation}\label{appeq.v}
    V_{AV}(\bp) := U_{AV}(\mathbf{x}(\bp),\bp) =
    \frac{1}{2} (\ba-\bp)^T [\mathbf{I}_n - \delta \mathbf{G}]^{-1} (\ba-\bp).
\end{equation}
Now we also consider problem (\ref{eq.intervention}) and get the following parallel results for efficient prices.

\begin{lemma}\label{applemma.pfamily}
    The solution to problem (\ref{eq.intervention}) can be expressed as:
    \begin{equation}\label{appeq.pfamily}
        \begin{aligned}
            \bp(\tau)
            =& \frac{\ba+\bc}{2} -  \sqrt{1-\tau} \frac{\ba-\bc}{2}.
        \end{aligned}
    \end{equation}
	
\end{lemma}
The consumption level is given by $\mathbf{x}(\tau) = (1+ \sqrt{1-\tau}) \mathbf{x}^{ur}$, consistent with the implication from \cite{armstrongMultiproductPricingMade2018} that relative quantities remain unchanged. Furthermore, we can characterize the bound for the equilibrium outcome.
\begin{proposition}
    If the Assumptions \ref{assump.k} and \ref{assump.spectrum} hold, the equilibrium consumer surplus ratio must locate in the range $[(1-\sqrt{1-\tau^*})^2,(1+\sqrt{1-\tau^*})^2]$ where $\tau^* := R_\Pi (\bp^*)$.
\end{proposition}
Here the bounds are independent of $\delta$, in fact, they equal to the bound when $\delta$ approaches $1/\lambda_1$. Then the Pareto frontier can be parameterized as:
\begin{equation}\label{appeq.pareto}
	\mathscr{P} = \{(1-t^2, (1-t)^2): t \in [-1,0]\}.
\end{equation} 

We can also derive a parallel result to Lemma \ref{lemma.pareto} to determine the types of price regulations that achieve the Pareto frontier. Specifically, we define
\begin{align}
   &\nabla R_{\Pi}(\bp(\tau)) \propto \boldsymbol{\iota} :=  [\mathbf{I}_n - \delta\mathbf{G}]^{-1} (\ba-\bc) \geq \mathbf{0}, \label{appeq.weight}
\end{align}
Here the weight $\boldsymbol{\iota}$ is parallel to $\mathbf{x}^{ur}$ and remains independent of $\tau$. Additionally, as $\delta$ approaches $1/\lambda_1$, it converges to $\mathbf{w}_1$.
\begin{proposition}\label{applemma.pareto}
    If Assumptions \ref{assump.k} and \ref{assump.spectrum} hold, a price regulation $K$ is Pareto efficient if and only if $\exists \tau \in [0,1]$ such that $\bp(\tau) \in K$ and $\langle \boldsymbol{\iota}, \bp-\bp(\tau) \rangle \leq 0$ for any $\bp \in K$.
\end{proposition}
In other words, the set $K$ must intersect the segment connecting $\bc$ and $(\ba+\bc)/2$ at $\bp(\tau)$, and the average price induced by any $\bp \in K$, weighted by $\boldsymbol{\iota}$, must not exceed the corresponding average price for $\bp(\tau)$. 

Our analysis in Section \ref{sec.large} for large $\delta$ applies in this section.


\subsection{Regular graph under uniform pricing}\label{appsec.regular}

If the network is a regular graph, the uniform price can be simplified as
\begin{equation*}
    \bp^0 = \frac{\mathbf{1}^T \bp^{ur}}{n} \mathbf{1}
\end{equation*}
which is irrelevant with the network structure and spillovers. Then by spectral decomposition, we have
\begin{align*}
    R_V(\bp^0) -1 =&  
    \sum_{i=1}^{n} \frac{\langle \mathbf{w}_i , \ba-\bp^{0} \rangle^2}{(1-\delta \lambda_i)^2} 
	\left/
	\sum_{i=1}^{n} \frac{\langle \mathbf{w}_i , \frac{\ba-\bc}{2} \rangle^2}{(1-\delta \lambda_i)^2} \right. -1 \\
    \propto &
    \sum_{i=1}^{n} \frac{1}{(1- \delta\lambda_i)^2} \left[\langle \mathbf{w}_i , \ba - \bp^0 \rangle^2 - \langle \mathbf{w}_i , \frac{\ba-\bc}{2} \rangle^2\right]\\
    =&
    \sum_{i=2}^{n} \frac{1}{(1- \delta\lambda_i)^2} \left[\langle \mathbf{w}_i , \ba \rangle^2 - \langle \mathbf{w}_i , \frac{\ba-\bc}{2} \rangle^2\right].
\end{align*}
The third equation follows from the facts that $\langle \mathbf{w}_1 , \bp^{0} \rangle = \frac{1}{\sqrt{n}} \langle \mathbf{1},\bp^{0} \rangle = \frac{1}{\sqrt{n}} \langle \mathbf{1},\bp^{ur} \rangle = \langle \mathbf{w}_1 , \bp^{ur} \rangle$ and $\langle \mathbf{w}_i , \bp^{0} \rangle = 0$ for $i \geq 2$.

\begin{figure}[H]
	\centering
	\subfloat[Case (i) with complete graph]{
		\includegraphics[width=0.45\textwidth]{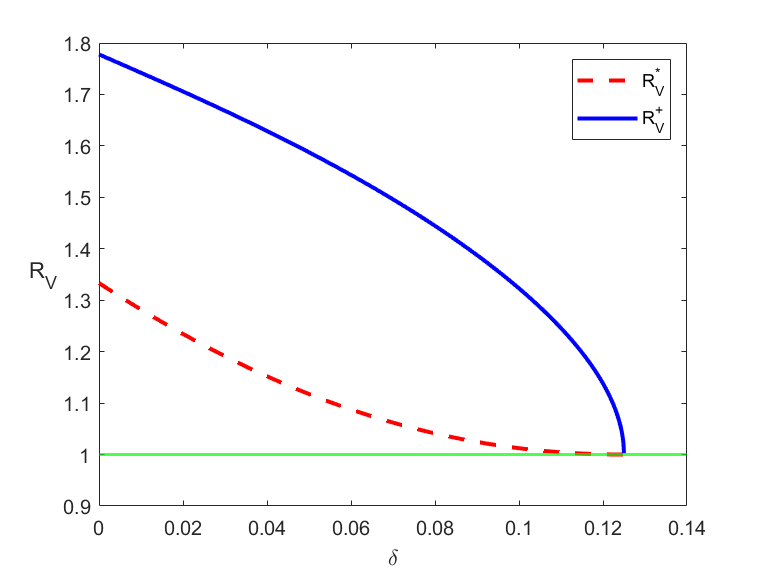}
		\label{fig.complete}
	}\hfill
	\subfloat[Case (ii) with complete graph]{
		\includegraphics[width=0.45\textwidth]{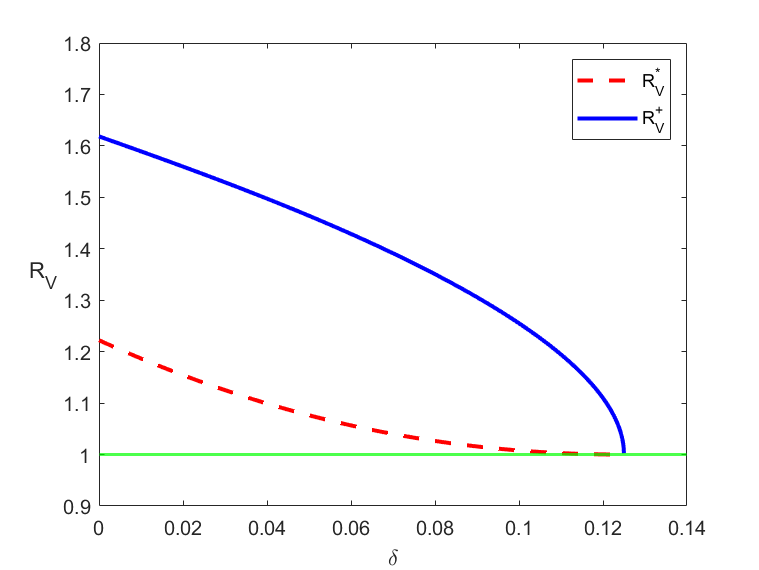}
		\label{fig.complete2}
	}
	\caption{Changes of $R_V^*$ and $R_V^+$ as $\delta$ varies from $0$ to $1/\lambda_1$\\
	}
	\label{fig.example2}
\end{figure}

Note that when $\bc$ are sufficiently close to 0 or $\ba$, the equation above implies $R_V(\bp^0) -1 > 0$ and thus consumers benefit from the ban on price discrimination. Furthermore, when $\delta$ is close to $1/\lambda_1$, the increase in the consumer surplus ratio is of the same order as $(\delta-\frac{1}{\lambda_1})^2$. Figure \ref{fig.example2} illustrates the simulation results using the same parameters as in Example \ref{example.uniform}, applied to a complete graph with 9 nodes.

\subsection{Regulation on price difference}\label{appsec.pdiff}

In this section, we use the same core-periphery network and parameters as in Example \ref{example.uniform}, but the price regulation is $\myd_{\boldsymbol{\Delta}}$ where $\Delta_{ij} \equiv \Delta \in \{0, 2.5, 5\}$. In this setting, a smaller $\Delta$ indicates tighter regulation. Specifically, when $\Delta = 0$, it corresponds to uniform pricing, while the regulation becomes ineffective when $\Delta = 5$.\footnote{Since $\max(\bp^{ur})-\min(\bp^{ur}) = 10-5=5$, the regulation becomes ineffective if the regulated price difference is equal to or greater than 5. Therefore, we set the maximum value of $\Delta$ as 5.} 
    
    \begin{figure}[htbp]
        \centering
        \par
        \subfloat[Case (i): $(\theta_1, \theta_2) = (20,10)$]{
            \includegraphics[width=0.45\textwidth]{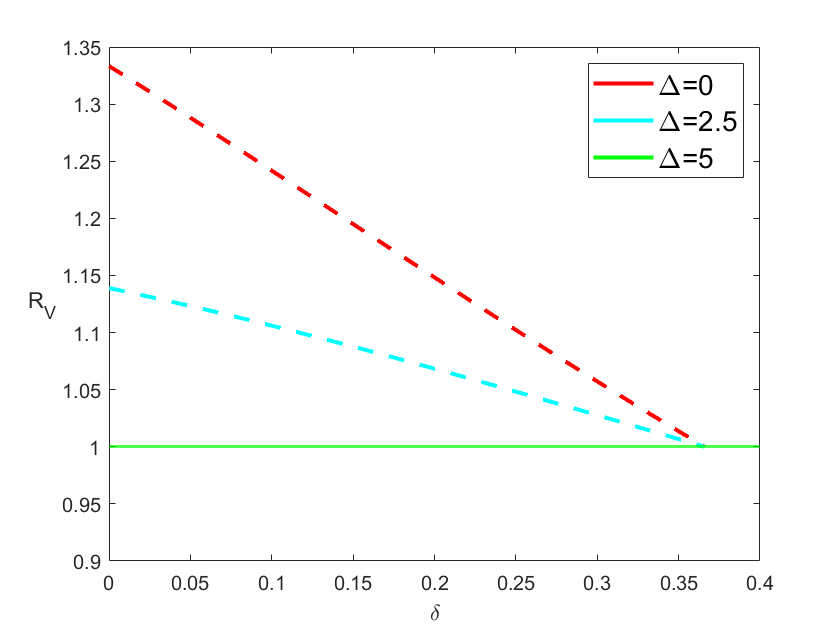}
            \label{fig.Delta}
        }\hfill
        \subfloat[Case (ii): $(\theta_1, \theta_2) = (10,20)$]{
            \includegraphics[width=0.45\textwidth]{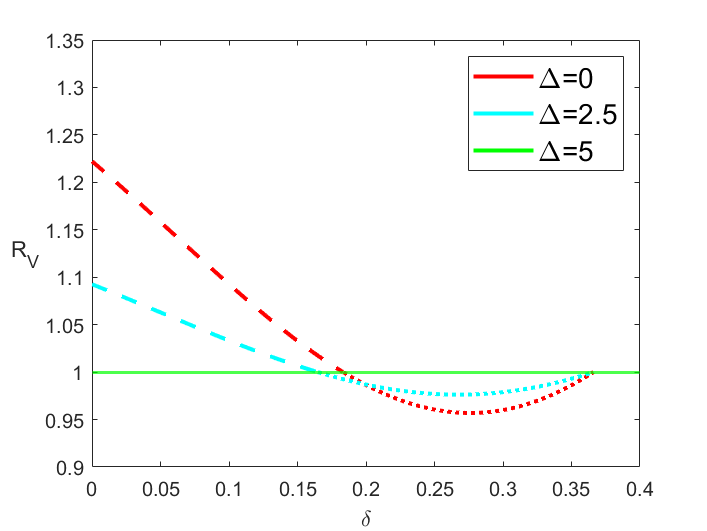}
            \label{fig.Delta2}
        }
        \caption{Changes of $R_V^*$ as $\delta$ varies from $0$ to $1/\lambda_1$ and $\Delta\in \{0, 2.5, 5\}$} 
        \label{fig.exampleDelta}
    \end{figure}
    
    As shown in Figure \ref{fig.exampleDelta}, a looser price regulation mitigates the welfare effect, either positive or negative. However, in case (ii), this may expand the range of $\delta$ in which consumers suffer.

\subsection{Complete bipartite network}\label{appsec.bipartite}
Here we conduct a numerical simulation of a complete bipartite network, which has two types of nodes and thus satisfies Definition \ref{def.twotype}. In particular, we set $|V_1| = 2$ and $|V_2| = 10$. The other parameters are the same as in Example \ref{example.uniform}, where $\bc = \mathbf{0}$ and $a_i = \theta_k$ if $i \in V_k$. We set $(\theta_1, \theta_2) = (20,10)$ for case (i) and $(\theta_1, \theta_2) = (10,20)$ for case (ii). For the regulation $\myd_{\boldsymbol{\Delta}}$, we also set $\Delta_{ij} \equiv \Delta \in \{0, 2.5, 5\}$ as in Example \ref{appsec.pdiff}.

\begin{figure}[H]
    \centering
    \par
    \subfloat[Case (i): $(\theta_1, \theta_2) = (20,10)$]{
        \includegraphics[width=0.45\textwidth]{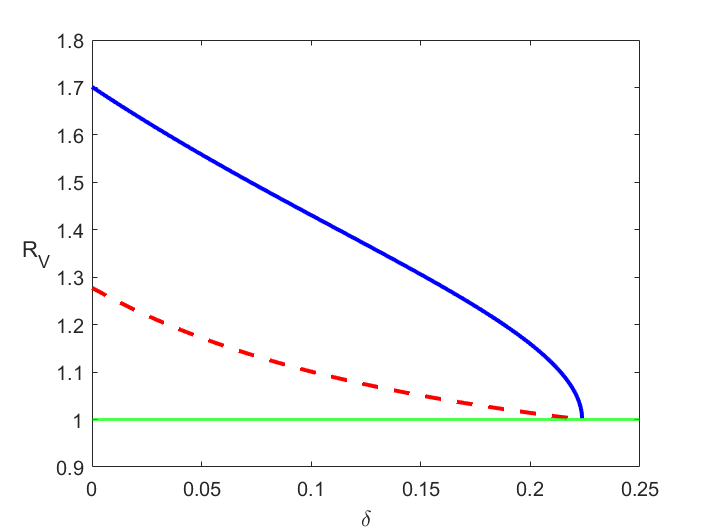}
    }\hfill
    \subfloat[Case (ii): $(\theta_1, \theta_2) = (10,20)$]{
        \includegraphics[width=0.45\textwidth]{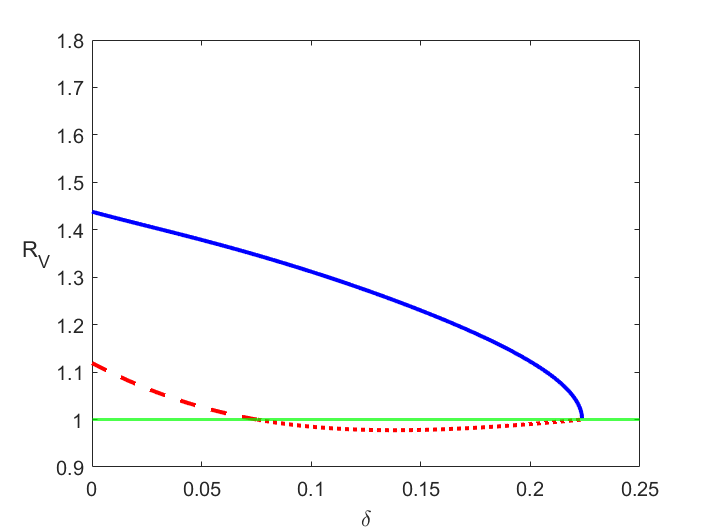}
    }
    \caption{Changes of $R_V^*$ (\textcolor{red}{Red curve}) and $R_V^+$ (\textcolor{blue}{Blue curve}) as $\delta$ varies from $0$ to $1/\lambda_1$  }
    \label{fig.cbexample}
\end{figure}

\begin{figure}[htbp]
    \centering
    \par
    \subfloat[Case (i): $(\theta_1, \theta_2) = (20,10)$]{
        \includegraphics[width=0.45\textwidth]{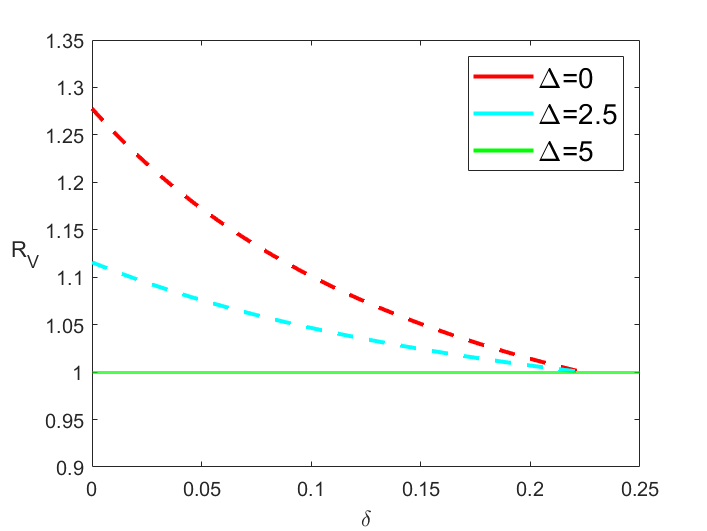}
        \label{fig.cbDelta}
    }\hfill
    \subfloat[Case (ii): $(\theta_1, \theta_2) = (10,20)$]{
        \includegraphics[width=0.45\textwidth]{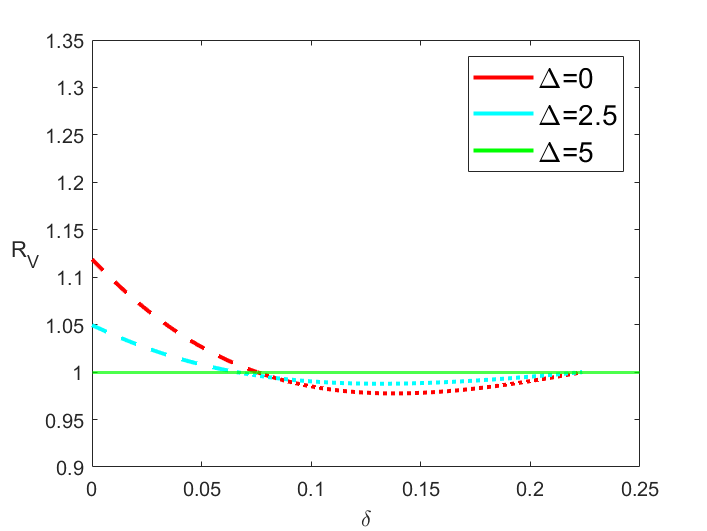}
        \label{fig.cbDelta2}
    }
    \caption{Changes of $R_V^*$ as $\delta$ varies from $0$ to $1/\lambda_1$ and $\Delta\in \{0, 2.5, 5\}$} 
    \label{fig.cbexampleDelta}
\end{figure}

These results are qualitatively the same as in Examples \ref{example.uniform} and \ref{appsec.pdiff}.

\newpage
\nocite{*}
\setlength\bibsep{6pt}
\bibliographystyle{apalike}
\bibliography{ref.bib}

@article{aguirreMonopolyPriceDiscrimination2010,
	title = {Monopoly {{Price Discrimination}} and {{Demand Curvature}}},
	shorttitle = {Monopoly {{Price Discrimination}} and {{Demand Curvature}}},
	author = {Aguirre, Inaki and Cowan, Simon and Vickers, John},
	year = {2010},
	month = sep,
	journal = {American Economic Review},
	volume = {100},
	number = {4},
	pages = {1601--1615},
	issn = {00028282},
	doi = {10.1257/aer.100.4.1601},
	urldate = {2024-05-02}
}

@article{aoyagiBertrandCompetitionNetwork2018,
	title = {Bertrand Competition under Network Externalities},
	author = {Aoyagi, Masaki},
	year = {2018},
	month = nov,
	journal = {Journal of Economic Theory},
	volume = {178},
	pages = {517--550},
	issn = {00220531},
	doi = {10.1016/j.jet.2018.10.006},
	urldate = {2024-07-19},
	langid = {english}
}

@article{armstrongCompetitionTwosidedMarkets2006,
  title = {Competition in Two-Sided Markets},
  author = {Armstrong, Mark},
  year = {2006},
  journal = {The RAND Journal of Economics},
  volume = {37},
  number = {3},
  pages = {668--691},
  issn = {1756-2171},
  doi = {10.1111/j.1756-2171.2006.tb00037.x},
  urldate = {2023-05-25},
  langid = {english}
}

@article{armstrongMultiproductPricingMade2018,
	title = {Multiproduct {{Pricing Made Simple}}},
	author = {Armstrong, Mark and Vickers, John},
	year = {2018},
	month = aug,
	journal = {Journal of Political Economy},
	volume = {126},
	number = {4},
	pages = {1444--1471},
	issn = {0022-3808, 1537-534X},
	doi = {10.1086/697902},
	urldate = {2024-10-23},
	langid = {english}
}

@article{armstrongWelfareEffectsPrice1991,
	title = {Welfare {{Effects}} of {{Price Discrimination}} by a {{Regulated Monopolist}}},
	author = {Armstrong, Mark and Vickers, John},
	year = {1991},
	journal = {The RAND Journal of Economics},
	volume = {22},
	number = {4},
	eprint = {2600990},
	eprinttype = {jstor},
	pages = {571--580},
	publisher = {[RAND Corporation, Wiley]},
	issn = {0741-6261},
	doi = {10.2307/2600990},
	urldate = {2024-08-17}
}

@inproceedings{arthurPricingStrategiesViral2009,
	title = {Pricing {{Strategies}} for {{Viral Marketing}} on {{Social Networks}}},
	booktitle = {Internet and {{Network Economics}}},
	author = {Arthur, David and Motwani, Rajeev and Sharma, Aneesh and Xu, Ying},
	editor = {Leonardi, Stefano},
	year = {2009},
	pages = {101--112},
	publisher = {Springer},
	address = {Berlin, Heidelberg},
	doi = {10.1007/978-3-642-10841-9_11},
	isbn = {978-3-642-10841-9},
	langid = {english}
}

@article{ballesterWhosWhoNetworks2006,
	title = {Who's {{Who}} in {{Networks}}. {{Wanted}}: {{The Key Player}}},
	shorttitle = {Who's {{Who}} in {{Networks}}. {{Wanted}}},
	author = {Ballester, Coralio and {Calv{\'o}-Armengol}, Antoni and Zenou, Yves},
	year = {2006},
	journal = {Econometrica},
	volume = {74},
	number = {5},
	pages = {1403--1417},
	issn = {1468-0262},
	doi = {10.1111/j.1468-0262.2006.00709.x},
	urldate = {2024-08-05},
	langid = {english}
}

@article{banerjiLocalNetworkExternalities2009,
	title = {Local Network Externalities and Market Segmentation},
	author = {Banerji, A. and Dutta, Bhaskar},
	year = {2009},
	month = sep,
	journal = {International Journal of Industrial Organization},
	volume = {27},
	number = {5},
	pages = {605--614},
	issn = {01677187},
	doi = {10.1016/j.ijindorg.2009.02.001},
	urldate = {2024-05-21},
	copyright = {https://www.elsevier.com/tdm/userlicense/1.0/},
	langid = {english}
}

@article{biscegliaFairGatekeepingDigital2024,
	title = {Fair {{Gatekeeping}} in {{Digital Ecosystems}}},
	author = {Bisceglia, Michele and Tirole, Jean},
	year = {2024},
	journal = {mimeo},
	langid = {english}
}

@article{blochPricingSocialNetworks2013,
	title = {Pricing in Social Networks},
	author = {Bloch, Francis and Qu{\'e}rou, Nicolas},
	year = {2013},
	month = jul,
	journal = {Games and Economic Behavior},
	volume = {80},
	pages = {243--261},
	issn = {0899-8256},
	doi = {10.1016/j.geb.2013.03.006},
	urldate = {2024-08-22}
}

@article{boiteux1956gestion,
  title={Sur la gestion des monopoles publics astreints {\`a} l'{\'e}quilibre budg{\'e}taire},
  author={Boiteux, Marcel},
  journal={Econometrica, Journal of the Econometric Society},
  pages={22--40},
  year={1956},
  publisher={JSTOR}
}

@article{bonacich1972factoring,
	title={Factoring and weighting approaches to status scores and clique identification},
	author={Bonacich, Phillip},
	journal={Journal of mathematical sociology},
	volume={2},
	number={1},
	pages={113--120},
	year={1972},
	publisher={Taylor \& Francis}
}

@article{brezaUsingAggregatedRelational2020,
  title = {Using {{Aggregated Relational Data}} to {{Feasibly Identify Network Structure}} without {{Network Data}}},
  author = {Breza, Emily and Chandrasekhar, Arun G. and McCormick, Tyler H. and Pan, Mengjie},
  year = 2020,
  month = aug,
  journal = {American Economic Review},
  volume = {110},
  number = {8},
  pages = {2454--2484},
  issn = {0002-8282},
  doi = {10.1257/aer.20170861},
  urldate = {2025-11-24},
  langid = {english}
}

@article{brezaConsistentlyEstimatingNetwork2023,
  title = {Consistently Estimating Network Statistics Using Aggregated Relational Data},
  author = {Breza, Emily and Chandrasekhar, Arun G. and Lubold, Shane and McCormick, Tyler H. and Pan, Mengjie},
  year = 2023,
  month = may,
  journal = {Proceedings of the National Academy of Sciences},
  volume = {120},
  number = {21},
  pages = {e2207185120},
  issn = {0027-8424, 1091-6490},
  doi = {10.1073/pnas.2207185120},
  urldate = {2025-11-24},
  langid = {english}
}

@article{candoganOptimalPricingNetworks2012,
	title = {Optimal {{Pricing}} in {{Networks}} with {{Externalities}}},
	author = {Candogan, Ozan and Bimpikis, Kostas and Ozdaglar, Asuman},
	year = {2012},
	month = aug,
	journal = {Operations Research},
	volume = {60},
	number = {4},
	pages = {883--905},
	issn = {0030-364X, 1526-5463},
	doi = {10.1287/opre.1120.1066},
	urldate = {2024-03-05},
	langid = {english}
}

@article{chenCompetitivePricingStrategies2018,
	title = {Competitive Pricing Strategies in Social Networks},
	author = {Chen, Ying-Ju and Zenou, Yves and Zhou, Junjie},
	year = {2018},
	journal = {The RAND Journal of Economics},
	volume = {49},
	number = {3},
	pages = {672--705},
	issn = {1756-2171},
	doi = {10.1111/1756-2171.12249},
	urldate = {2023-11-06},
	langid = {english}
}

@article{chenImpactNetworkTopology2022,
	title = {The Impact of Network Topology and Market Structure on Pricing},
	author = {Chen, Ying-Ju and Zenou, Yves and Zhou, Junjie},
	year = {2022},
	month = sep,
	journal = {Journal of Economic Theory},
	volume = {204},
	pages = {105491},
	issn = {0022-0531},
	doi = {10.1016/j.jet.2022.105491},
	urldate = {2024-08-17}
}

@article{cohenPriceDiscriminationFairness2022,
	title = {Price {{Discrimination}} with {{Fairness Constraints}}},
	author = {Cohen, Maxime C. and Elmachtoub, Adam N. and Lei, Xiao},
	year = {2022},
	month = dec,
	journal = {Management Science},
	volume = {68},
	number = {12},
	pages = {8536--8552},
	publisher = {INFORMS},
	issn = {0025-1909},
	doi = {10.1287/mnsc.2022.4317},
	urldate = {2024-05-21}
}

@article{contreras2017non,
  title={Non-discrimination and FRAND commitments},
  author={Contreras, Jorge L and Layne-Farrar, Anne},
  journal={The Cambridge Handbook of Technical Standardization Law},
  volume={1},
  year={2017}
}

@article{cowan2016welfare,
  title={Welfare-increasing third-degree price discrimination},
  author={Cowan, Simon},
  journal={The RAND Journal of Economics},
  volume={47},
  number={2},
  pages={326--340},
  year={2016},
  publisher={Wiley Online Library}
}

@article{dasarathaIncentiveDesignSpillovers2024,
	title = {Incentive {{Design With Spillovers}}},
	author = {Dasaratha, Krishna and Golub, Benjamin and Shah, Anant},
	year = {2024},
	month = jun,
	journal = {Available at SSRN 4853054},
	doi = {10.2139/ssrn.4853054},
	urldate = {2024-07-05},
	langid = {english}
}

@article{debreuNonnegativeSquareMatrices1953,
  title = {Nonnegative {{Square Matrices}}},
  author = {Debreu, Gerard and Herstein, I. N.},
  year = 1953,
  journal = {Econometrica},
  volume = {21},
  number = {4},
  eprint = {1907925},
  eprinttype = {jstor},
  pages = {597--607},
  publisher = {[Wiley, Econometric Society]},
  issn = {0012-9682},
  doi = {10.2307/1907925},
  urldate = {2024-08-26}
}

@article{degiorgiConsumptionNetworkEffects2020,
	title = {Consumption {{Network Effects}}},
	author = {De Giorgi, Giacomo and Frederiksen, Anders and Pistaferri, Luigi},
	year = {2020},
	month = jan,
	journal = {The Review of Economic Studies},
	volume = {87},
	number = {1},
	pages = {130--163},
	issn = {0034-6527},
	doi = {10.1093/restud/rdz026},
	urldate = {2025-01-08}
}

@article{demangeOptimalTargetingStrategies2017a,
  title = {Optimal Targeting Strategies in a Network under Complementarities},
  author = {Demange, Gabrielle},
  year = 2017,
  month = sep,
  journal = {Games and Economic Behavior},
  volume = {105},
  pages = {84--103},
  issn = {0899-8256},
  doi = {10.1016/j.geb.2017.07.004},
  urldate = {2025-12-17}
}

@article{fainmesserPricingNetworkEffects2016,
	title = {Pricing {{Network Effects}}},
	author = {Fainmesser, Itay P. and Galeotti, Andrea},
	year = {2016},
	month = jan,
	journal = {The Review of Economic Studies},
	volume = {83},
	number = {1},
	pages = {165--198},
	issn = {0034-6527, 1467-937X},
	doi = {10.1093/restud/rdv032},
	urldate = {2024-03-05},
	langid = {english}
}

@article{galeottiLawFew2010,
  title = {The {{Law}} of the {{Few}}},
  author = {Galeotti, Andrea and Goyal, Sanjeev},
  year = {2010},
  month = sep,
  journal = {American Economic Review},
  volume = {100},
  number = {4},
  pages = {1468--1492},
  issn = {0002-8282},
  doi = {10.1257/aer.100.4.1468},
  urldate = {2025-05-12},
  langid = {english}
}

@article{galeottiTargetingInterventionsNetworks2020,
	title = {Targeting {{Interventions}} in {{Networks}}},
	author = {Galeotti, Andrea and Golub, Benjamin and Goyal, Sanjeev},
	year = {2020},
	journal = {Econometrica},
	volume = {88},
	number = {6},
	pages = {2445--2471},
	issn = {0012-9682},
	doi = {10.3982/ECTA16173},
	urldate = {2024-07-05},
	copyright = {https://creativecommons.org/licenses/by/4.0/},
	langid = {english}
}

@article{galeotti2024robust,
  title={Robust Market Interventions},
  author={Galeotti, A and Golub, B and Goyal, S and Talamas, E and Tamuz, O},
  year={2024},
  journal = {mimeo}
}

@inproceedings{hartlineOptimalMarketingStrategies2008,
	title = {Optimal Marketing Strategies over Social Networks},
	booktitle = {Proceedings of the 17th International Conference on {{World Wide Web}}},
	author = {Hartline, Jason and Mirrokni, Vahab and Sundararajan, Mukund},
	year = {2008},
	month = apr,
	series = {{{WWW}} '08},
	pages = {189--198},
	publisher = {Association for Computing Machinery},
	address = {New York, NY, USA},
	doi = {10.1145/1367497.1367524},
	urldate = {2024-07-17},
	isbn = {978-1-60558-085-2}
}

@article{huangValuePriceDiscrimination2022,
	title = {The {{Value}} of {{Price Discrimination}} in {{Large Social Networks}}},
	author = {Huang, Jiali and Mani, Ankur and Wang, Zizhuo},
	year = {2022},
	month = jun,
	journal = {Management Science},
	volume = {68},
	number = {6},
	pages = {4454--4477},
	issn = {0025-1909, 1526-5501},
	doi = {10.1287/mnsc.2021.4108},
	urldate = {2025-01-22},
	langid = {english}
}

@article{kohPricesSymmetries2025,
	title = {Prices and {{Symmetries}}},
	author = {Koh, Andrew and {Martinez-Bruera}, Pedro},
	year = {2025},
	journal = {Available at SSRN 5132200},
	doi = {10.2139/ssrn.5132200},
	urldate = {2025-02-21},
	langid = {english}
}

@article{liLimitTargetingNetworks2022,
	title = {The Limit of Targeting in Networks},
	author = {Li, Jian and Zhou, Junjie and Chen, Ying-Ju},
	year = {2022},
	month = apr,
	journal = {Journal of Economic Theory},
	volume = {201},
	pages = {105418},
	issn = {0022-0531},
	doi = {10.1016/j.jet.2022.105418},
	urldate = {2024-08-30}
}

@article{ramsey1927contribution,
  title={A Contribution to the Theory of Taxation},
  author={Ramsey, Frank P},
  journal={The economic journal},
  volume={37},
  number={145},
  pages={47--61},
  year={1927},
  publisher={JSTOR}
}

@article{rochetTwosidedMarketsProgress2006,
  title = {Two-Sided Markets: A Progress Report},
  shorttitle = {Two-Sided Markets},
  author = {Rochet, Jean-Charles and Tirole, Jean},
  year = {2006},
  journal = {The RAND Journal of Economics},
  volume = {37},
  number = {3},
  pages = {645--667},
  issn = {1756-2171},
  doi = {10.1111/j.1756-2171.2006.tb00036.x},
  urldate = {2024-12-06},
  langid = {english}
}

@article{schmalenseeOutputWelfareImplications1981,
	title = {Output and {{Welfare Implications}} of {{Monopolistic Third-Degree Price Discrimination}}},
	author = {Schmalensee, Richard},
	year = {1981},
	journal = {The American Economic Review},
	volume = {71},
	number = {1},
	eprint = {1805058},
	eprinttype = {jstor},
	pages = {242--247},
	publisher = {American Economic Association},
	issn = {0002-8282},
	urldate = {2025-03-05}
}

@article{shiScreeningNetworkInformation2025,
	title = {Screening {{Network Information}}: {{Optimal Network Interventions}} under {{Asymmetric Information}}},
	shorttitle = {Screening {{Network Information}}},
	author = {Shi, Fanqi and Xing, Yiqing and Chen, Litian},
	year = {2025},
	journal = {Available at SSRN 5033820},
	doi = {10.2139/ssrn.5033820},
	urldate = {2025-02-20},
	langid = {english}
}

@article{ushchevPriceCompetitionProduct2018,
	title = {Price Competition in Product Variety Networks},
	author = {Ushchev, Philip and Zenou, Yves},
	year = {2018},
	month = jul,
	journal = {Games and Economic Behavior},
	volume = {110},
	pages = {226--247},
	issn = {08998256},
	doi = {10.1016/j.geb.2018.04.002},
	urldate = {2024-06-27},
	langid = {english}
}

@article{varianPriceDiscriminationSocial1985,
	title = {Price {{Discrimination}} and {{Social Welfare}}},
	author = {Varian, Hal R.},
	year = {1985},
	journal = {The American Economic Review},
	volume = {75},
	number = {4},
	eprint = {1821366},
	eprinttype = {jstor},
	pages = {870--875},
	publisher = {American Economic Association},
	issn = {0002-8282},
	urldate = {2025-03-05}
}

@article{yangFairnessRegulationPrices2024,
	title = {Fairness {{Regulation}} of {{Prices}} in {{Competitive Markets}}},
	author = {Yang, Zongsen and Fu, Xingyu and Gao, Pin and Chen, Ying-Ju},
	year = {2024},
	month = sep,
	journal = {Manufacturing \& Service Operations Management},
	volume = {26},
	number = {5},
	pages = {1897--1917},
	publisher = {INFORMS},
	issn = {1523-4614},
	doi = {10.1287/msom.2022.0552},
	urldate = {2024-09-25}
}

@article{zhangOptimalNonlinearPricing2020,
  title = {Optimal {{Nonlinear Pricing}} in {{Social Networks Under Asymmetric Network Information}}},
  author = {Zhang, Yang and Chen, Ying-Ju},
  year = {2020},
  month = may,
  journal = {Operations Research},
  volume = {68},
  number = {3},
  pages = {818--833},
  issn = {0030-364X, 1526-5463},
  doi = {10.1287/opre.2019.1915},
  urldate = {2025-03-22},
  langid = {english}
}

@book{ziegler2012lectures,
	title={Lectures on polytopes},
	author={Ziegler, G{\"u}nter M},
	volume={152},
	year={2012},
	publisher={Springer Science \& Business Media}
}

@book{goyal2023networks,
	title={Networks: An economics approach},
	author={Goyal, Sanjeev},
	year={2023},
	publisher={MIT Press}
}

@book{bertsekas2003convex,
	title={Convex analysis and optimization},
	author={Bertsekas, Dimitri and Nedic, Angelia and Ozdaglar, Asuman},
	volume={1},
	year={2003},
	publisher={Athena Scientific}
}

@book{boyd2004convex,
  title={Convex optimization},
  author={Boyd, Stephen and Vandenberghe, Lieven},
  year={2004},
  publisher={Cambridge university press}
}

@book{brouwer2011spectra,
  title={Spectra of graphs},
  author={Brouwer, Andries E and Haemers, Willem H},
  year={2011},
  publisher={Springer Science \& Business Media}
}

@book{axler2024linear,
	title={Linear algebra done right},
	author={Axler, Sheldon},
	year={2024},
	publisher={Springer Nature}
}

@book{jackson2008social,
	title={Social and economic networks},
	author={Jackson, Matthew O},
	volume={3},
	year={2008},
	publisher={Princeton university press Princeton}
}

@book{robinson1969economics,
	title={The economics of imperfect competition},
	author={Robinson, Joan},
	year={1969},
	publisher={Springer}
}


\end{document}